\documentclass[preprint]{aastex} 

\righthead{FUSE OVI survey}
\def\dex#1{10$^{#1}$}
\def\tdex#1{$\times$10$^{#1}$}
\def\cmm#1{~cm$^{-#1}$}
\def\kms{~km\,s$^{-1}$}
\def\kpc{~kpc}
\def\Mpc{~Mpc}
\def\fu{~erg\,cm$^{-2}$\,s$^{-1}$\aa$^{-1}$}
\def\aa{~\AA}  
\def\maa{~m\AA}
\def\vlsr{$v_{\rm LSR}$}
\def\vlsra{$\vert$$v_{\rm LSR}$$\vert$}
\let\deg=\arcdeg

\def\ref{\par\noindent}
\def\sig#1{$\sigma_{\rm W,#1}$}

\def\SNres{S/N$_{\rm res}$}
\def\SNbin{S/N$_{\rm bin}$}

\def\FUSE{{\it FUSE}}
\def\IUE{{\it IUE}}
\def\HST{{\it HST}}
\def\GHRS{{\it GHRS}}
\def\STIS{{\it STIS}}
\def\FOS{{\it FOS}}
\def\NED{{\it NED}}
\def\SIMBAD{{\it SIMBAD}}
\def\MAST{{\it MAST}}
\def\STScI{{\it STScI}}

\def\l{\ $\lambda$}
\def\ll{\ $\lambda\lambda$}

\def\H#1{H$_#1$}
\def\HI{\protect\ion{H}{1}}
\def\Lya{Ly$\alpha$}
\def\Lyb{Ly$\beta$}
\def\HII{\protect\ion{H}{2}}

\def\HeII{\protect\ion{He}{2}}

\def\CII{\protect\ion{C}{2}}
\def\CIIl{\CII\l1036.337}
\def\CIII{\protect\ion{C}{3}}
\def\CIIIl{\CIII\l977.020}
\def\CIV{\protect\ion{C}{4}}
\def\NI{\protect\ion{N}{1}}

\def\NIII{\protect\ion{N}{3}}

\def\NV{\protect\ion{N}{5}}
\def\OI{\protect\ion{O}{1}}
\def\OIl{\OI\l1039.230}
\def\OIlb{\OI\l988.773}
\def\OII{\protect\ion{O}{2}}
\def\OIII{\protect\ion{O}{3}}

\def\OVI{\protect\ion{O}{6}}
\def\OVIll{\OVI\ll1031.926, 1037.617}
\def\OVIla{\OVI\l1031.926}
\def\OVIlb{\OVI\l1037.617}

\def\SiII{\protect\ion{Si}{2}}
\def\SiIIla{\SiII\l1020.699}
\def\SiIIlb{\SiII\l989.873}

\def\SII{\protect\ion{S}{2}}
\def\SIII{\protect\ion{S}{3}}

\def\ClI{\protect\ion{Cl}{1}}
\def\ClII{\protect\ion{Cl}{2}}
\def\ArI{\protect\ion{Ar}{1}}
\def\ArIl{\ArI\l1048.220}

\def\FeII{\protect\ion{Fe}{2}}
\def\FeIII{\protect\ion{Fe}{3}}

\def\Fspectra{1}
\def\Fovishift{2}
\def\Fcompare{3}
\def\Fmagflux{4}
\def\Fvlim{5}
\def\Ferrhist{6}
\def\Ferrcorr{7}
\def\Foviratio{8}
\def\Fvelhist{9}
\def\Flsrmap{10}
\def\FHVCmap{11}

\def\Tbasic{1}
\def\Tderived{2}
\def\Tdetratio{3}

\def\Sect{Sect.~}
\def\Sects{Sects.~}

\def\Sdefsample{2}
\def\Sdefine{2.1}
\def\Ssample{2.2}
\def\Sdata{3}
\def\Sdatasum{3.1}
\def\Spipeline{3.2}
\def\SHI{3.3}
\def\Svshift{3.4}
\def\Scombine{3.5}
\def\Scontinuum{3.6}
\def\SSoverN{3.7}
\def\SHtcontam{3.8}
\def\SAGNcontam{3.9}
\def\Sgroupcontam{3.10}
\def\Smeasure{4}
\def\Svrange{4.1}
\def\Seqvwidth{4.2}
\def\Scoldens{4.3}
\def\Sresults{5}
\def\SMWYdet{5.1}
\def\SHVCdet{5.2}
\def\Sveldist{5.3}
\def\SLocGrp{5.4}
\def\Smaps{5.5}

\begin{document}


\title{THE FUSE SURVEY OF \OVI\ ABSORPTION IN AND NEAR THE GALAXY}
\author{
B.P. Wakker\altaffilmark{1},
B.D. Savage\altaffilmark{1},
K.R. Sembach\altaffilmark{2},
P.   Richter\altaffilmark{1,3},
M.   Meade\altaffilmark{1},
E.B. Jenkins\altaffilmark{4},
J.M. Shull\altaffilmark{5},
T.B. Ake\altaffilmark{6,7},
W.P. Blair\altaffilmark{6},
W.V. Dixon\altaffilmark{6},
S.D. Friedman\altaffilmark{6},
J.C. Green\altaffilmark{8},
R.F. Green\altaffilmark{9},
J.W. Kruk\altaffilmark{6},
H.W. Moos\altaffilmark{6},
E.M. Murphy\altaffilmark{10},
W.R. Oegerle\altaffilmark{6},
D.J. Sahnow\altaffilmark{6},
G.   Sonneborn\altaffilmark{11},
E.   Wilkinson\altaffilmark{5},
D.G. York\altaffilmark{12}
}

\altaffiltext{1}{Department of Astronomy, University of Wisconsin, 475 N Charter St, Madison, WI 53706}
\altaffiltext{2}{Space Telescope Science Institute, 3700 San Martin Drive, Baltimore, MD 21218}
\altaffiltext{3}{Osservatorio Astrofisico di Arcetri, I-50125, Firenze, Italy}
\altaffiltext{4}{Department of Astrophysical Sciences, Princeton University, Princeton, NJ 08544}
\altaffiltext{5}{Center of Astrophysics and Space Astronomy, University of Colorado, Boulder, CO 80309}
\altaffiltext{6}{Department of Physics and Astronomy, Johns Hopkins University, Baltimore, MD 21218}
\altaffiltext{7}{Science Programs, Computer Sciences Corporation, Lanham-Seabrook, MD 20706}
\altaffiltext{8}{Center for Astrophysics and Space Astronomy, Department of Astrophysical and Planetary Sciences, University of Colorado, Campus Box 389, Boulder, CO 80309}
\altaffiltext{9}{National Optical Astronomy Observatories, P.O.\ Box 26732, Tucson, AZ 85726}
\altaffiltext{10}{Department of Astronomy, University of Virginia, P.O. Box 3818, Charlottesville, VA 22903}
\altaffiltext{11}{Laboratory for Astronomy and Solar Physics,, NASA Goddard Space Flight Center, Code 681, Greenbelt, MD 20771}
\altaffiltext{12}{Department of Astronomy and Astrophysics, University of Chicago, 5640 South Ellis Av., Chicago, IL 60637}


\begin{abstract}
We present {\it Far Ultraviolet Spectroscopic Explorer} (\FUSE) observations of
the \OVIll\ absorption lines associated with gas in and near the Milky Way, as
detected in the spectra of a sample of 100 extragalactic targets and 2 distant
halo stars. We combine data from several \FUSE\ Science Team programs with guest
observer data that were public before 2002 May 1. The sightlines cover most of
the sky above galactic latitude $\vert$$b$$\vert$$>$25\deg\ -- at lower latitude
the ultraviolet extinction is usually too large for extragalactic observations.
We describe the details of the calibration, alignment in velocity, continuum
fitting, and manner in which several contaminants were removed -- Galactic \H2,
absorption intrinsic to the background target and intergalactic \Lyb\ lines.
This decontamination was done very carefully, and in several sightlines very
subtle problems were found. We searched for \OVI\ absorption in the velocity
range $-$1200 to 1200\kms. With a few exceptions, we only find \OVI\ in the
velocity range $-$400 to 400\kms; the exceptions may be intergalactic \OVI. In
this paper we analyze the \OVI\ associated with the Milky Way (and possibly with
the Local Group). We discuss the separation of the observed \OVI\ absorption
into components associated with the Milky Way halo and componet at
high-velocity, which are probably located in the neighborhood of the Milky Way.
We describe the measurements of equivalent width and column density, and we
analyze the different contributions to the errors. We conclude that low-velocity
Galactic \OVI\ absorption occurs along all sightlines -- the few non-detections
only occur in noisy spectra. We further show that high-velocity \OVI\ is very
common, having equivalent width $>$65\maa\ in 50\% of the sightlines and
equivalent width $>$30\maa\ in 70\% of the high-quality sightlines. The
high-velocity \OVI\ absorption has velocities relative to the LSR of
$\pm$(100--330)\kms; there is no correlation between velocity and absorption
strength. We discuss the possibilities for studying \OVI\ absorption associated
with Local Group galaxies, and conclude that \OVI\ is probably detected in M\,31
and M\,33. We limit the extent of an \OVI\ halo around M\,33 to be $<$100\kpc\
(at a 3$\sigma$ detection limit of log $N$(\OVI)$\sim$14.0). Using the measured
column densities, we present 50\kms\ wide \OVI\ channel maps. These show
evidence for the imprint of Galactic rotation. They also highlight two known
\HI\ high-velocity clouds (complex~C and the Magellanic Stream). The channel
maps further show that \OVI\ at velocities $<$$-$200\kms\ occurs along all
sightlines in the region $l$=20\deg--150\deg, $b$$<$$-$30\deg, while \OVI\ at
velocities $>$200\kms\ occurs along all sightlines in the region
$l$=180\deg--300\deg, $b$$>$20\deg.
\end{abstract}

\newpage
\section{INTRODUCTION}
The {\it Far Ultraviolet Spectroscopic Explorer} (\FUSE) provides high
resolution spectra in the wavelength regime between 905 and 1187\aa, making it
one of the few observatories (past or present) that allows observations
shortward of 1150\aa\ down to the Galactic Lyman edge. This spectral region
contains the resonance absorption lines of the most abundant atoms and molecules
including, for example, \HI, \H2, \OI, \OVI, \CII, \CIII, \FeII\ and \FeIII.
Interstellar absorption line observations with \FUSE\ enable studies of all
phases of the interstellar gas, including the cold neutral and molecular medium,
the warm neutral medium, the warm ionized medium and the hot ionized medium.
\FUSE\ was launched in June 1999. The capabilities of \FUSE\ are described in
detail by Moos et al.\ (2000) and Sahnow et al.\ (2000).
\par \FUSE\ data consist of 8 separate $\sim$90\aa\ wide spectra, identified as
LiF1A, LiF1B, LiF2A, LiF2B, SiC1A, SiC1B, SiC2A and SiC2B. The LiF1A, LiF2B,
SiC1A and SiC2B spectra cover the region near 1030\aa, which allows the study of
absorption by five-times ionized oxygen (O$^{+5}$), a good diagnostic of gas at
temperatures near 3\tdex5~K. This ion has a high enough ionization potential
(113.9~eV is required to convert O$^{+4}$ to O$^{+5}$) that it is not produced
by photoionization caused by extreme ultraviolet radiation from normal stars,
which is suppressed above energies of 54.4~eV because of the strong He$^+$
absorption edge in the stellar atmosphere. O$^{+5}$ has two strong resonance
absorption lines at 1031.9261\aa\ and 1037.6167\aa, with oscillator strengths of
0.133 and 0.066 (Morton 1991).
\par \FUSE\ is the first instrument with sufficient sensitivity {\it and}
spectral resolution to observe \OVI\ absorption using large numbers of
extragalactic objects as background targets. The \FUSE\ Science Team designed a
number of programs to systematically map the distribution and amount of Galactic
\OVI. Several programs concentrate on the Galactic disk, while others emphasize
the Galactic halo. The halo program is divided into two parts -- a study of the
vertical distribution of the \OVI\ using a set of stars located at different
heights above the Galactic plane, and a study of the integrated column density
using a sample of extragalactic targets. This paper describes the basic target
information, the technical details of the data handling, and derived parameters
for the extragalactic study.
\par We present a catalog and some general analyses of this dataset, with an
emphasis on discriminating the different phenomena that are present. Detailed
interpretations, with more emphasis on a physical understanding, are presented
by Savage et al.\ (2002) and Sembach et al.\ (2002b). The former paper analyzes
the \OVI\ absorption that is clearly associated with the Milky Way, which we
also refer to as the ``thick disk'' \OVI. It discusses in depth the angular and
spatial distributions, the kinematics, the relation of \OVI\ to other components
of the Galaxy and the implications for understanding the Galaxy as a whole. As
we find in this catalog paper, high-velocity \OVI\ absorption is present along
many sightlines. This turns out to sample varied phenomena, some of which may
not be directly associated with the Milky Way. The physical interpretation of
this aspect of the \OVI\ sky is discussed in detail by Sembach et al.\ (2002b).
\par This paper is constructed as follows. First, we discuss the manner in which
the sample of extragalactic objects was assembled (\Sect\Sdefsample).
Table~\Tbasic\ lists the object and observation information. In \Sect\Sdata\ we
first summarize the reduction steps (\Sect\Sdatasum) and then describe the
calibration (\Sect\Spipeline), the absolute alignment and relative alignment
between different spectrograph segments (\Sects\SHI\ and \Svshift) and finally
how different observations and segments were combined (\Sect\Scombine). Next we
describe how the continuum was fit (\Sect\Scontinuum), how the final binning was
determined (\Sect\SSoverN), and how contamination by Galactic molecular hydrogen
(\H2), intrinsic AGN lines, and intergalactic absorption was removed
(\Sects\SHtcontam--\Sgroupcontam). Section~\Smeasure\ details the process of
measuring the \OVI\ lines. First, a velocity range of integration is determined
(\Sect\Svrange), then we measure the equivalent width and study the distribution
of the statistical and systematic errors in the equivalent width
(\Sect\Seqvwidth). Next we discuss the column densities and compare the values
derived from the \OVIla\ and \OVIlb\ lines to determine whether saturation could
be important (\Sect\Scoldens). Scientific results are presented in
\Sect\Sresults\ -- the detection rate for Galactic and high-velocity \OVI\
(\Sects\SMWYdet--\Sveldist), a check on the possibilities for detecting \OVI\
absorption from other Local Group galaxies using our sample (\Sect\SLocGrp) and
\OVI\ channel maps (\Sect\Smaps). Finally, we add an Appendix, in which notes
are presented concerning each sightline.


\section{OBJECT SELECTION}

\subsection{Observing Program Summary}
\par \FUSE\ Science Team targets were chosen through a process that involved
considerations of expected flux and sky coverage. For most objects in programs
P101, P107, P108 and P207, the selection process was similar. First we searched
the \HST\ and \IUE\ archives for ultraviolet observations of extragalactic
sources. Typically, this involved checking low dispersion \IUE\ data or Faint
Object Spectrograph ({\it \FOS}) data. We fitted a power law continuum to all
objects with detectable flux, making sure to avoid obvious stellar absorption
features and emission lines. This fit was typically performed in the
1200--1600\aa\ spectral region. Once the fit was performed, we extrapolated the
continuum fit to shorter wavelengths and estimated a continuum flux at 1000\aa.
Differential extinction was not taken into account, but most sightlines are at
high latitude and thus have E(B$-$V)$<$0.07 (see Table~\Tbasic). The predicted
fluxes are compared with the observed values in Sect.~\Scontinuum.2.
\par We then constructed a catalog of the 200 objects having the highest
1000\aa\ fluxes. The cutoff at the low flux end was typically \dex{-14}\fu. The
brightest targets were then selected for inclusion in the main \OVI\ halo survey
of extragalactic objects (28 objects in programs P101 and P108, as well as two
objects observed during the commissioning phase; programs X017 and I904). When
choosing from among these objects, we tried to choose diverse locations on the
sky as well as to select objects for which high quality \HST\ observations of
\CIV\ and \NV\ were already available for comparison with the \OVI. Many of the
remaining objects were chosen for short snapshots in the D/H program to search
for extragalactic Lyman-limit systems (37 objects in program P107). Since early
operations with \FUSE\ had to be performed with limited maneuvering, the
intended 2~ks snapshots were often extended to 10~ks or longer. Thus, the data
quality of the P107 snapshot program observations was often higher than
originally anticipated.
\par We excluded most known Seyfert 2 AGNs from the halo and D/H surveys since
these objects tend to have strong absorption features due to the stellar
populations of the host galaxies. Thus, most of the objects in the sample in
these programs are either quasars or type 1 Seyferts.
\par In the second cycle of \FUSE\ observing, 29 objects were included in the
P207 snapshot program based on their optical and X-ray fluxes as derived from
the ROSAT point source catalog. These objects had no previously known
ultraviolet data and were of unknown source type, although most were expected to
be QSOs. Eleven were found to have sufficient flux to be included in the final
sample.
\par For five six \FUSE\ Science Team programs that were geared toward
observations of Galactic \OVI\ (P101, P107, P108, X017, I904), a total of
3267~ks of exposure time was obtained for 70 objects, of which 3042~ks resulted
in useful spectra for 57 objects. The remaining 225~ks were not suitable for our
analysis because either the object was too faint (7 objects), had a Lyman limit
system covering the \OVI\ lines (3C\,351.0), had \H2\ or IGM lines covering the
\OVI\ lines (4 objects, see \Sects\SHtcontam\ and \SAGNcontam), or was
mispointed (IRAS\,F07546+3928). For six objects some observations were not used
because much longer observations were available.
\par Several other Science Team programs had different goals (e.g., measuring
D/H or intrinsic absorption in AGNs), but also produced spectra useful for
measuring Galactic \OVI\ (P110, P111, P190, P191, P198, P207, Q105, Q106, Q223
and Q224). A total of 1817~ks of exposure time for 58 objects was obtained for
these programs, of which 466~ks for 24 objects resulted in spectra useful for
our program. Of the 1351~ks that we do not use, 886~ks was useful for other
purposes, while 465~ks were spent on 30 objects that turned out to be too faint
to be useful.
\par In addition to the objects observed as part of the \FUSE\ Science Team
effort described above, we decided to include public data from the \FUSE\ guest
observer program. The rationale for this was twofold. First, many of these
observations were not taken for the purpose of studying Galactic \OVI, but
usually for the purpose of studying either the intergalactic medium or the AGN
being observed. Further, by including these objects, we expanded our sample
considerably, especially in some regions of the sky where the Science Team
program had only sparse coverage. We checked all guest observer datasets that
became public before 1 May 2002. Of these 96 sources (1749~ks total exposure
time), 57 were too faint and 12 had a continuum that was too complicated in the
1030\aa\ region to allow a measurement of Galactic \OVI\ absorption. Two objects
(Mrk\,153 and PG\,1011-040) yielded good spectra, but were excluded from our
sample because the Guest Observer's science aims overlapped ours. The remaining
25 objects (555~ks total exposure time) were included in the sample. These are
objects from programs A023 (PI Buat), A035 (PI Mulchaey), A036/B022 (PI Thuan),
A046 (PI Heckman), A060 (PI Koratkar), A068 (PI Bregman), A086 (PI Keel), A088
(PI Brown), A121 (PI Gibson), B062 (PI Mathur) and B087 (PI Prochaska). We thank
these investigators for their permission to use their data.
\par Since there are 5 objects that are in more than one of the categories
listed above, the final sample consists of 290 observations of 219 objects
observed for a total of 6834~ks (1898 hours or 79.0 days). A spectrum with
sufficiently high S/N ratio and reliable information for our study of Galactic
\OVI\ was obtained for 102 objects (100 extragalactic and 2 distant halo stars,
excluding Mrk\,153 and PG\,1011$-$040), amounting to 4214~ks (1171 hours or 48.7
days) of exposure time.
\par All objects that were part of the P, Q, X and I series of programs were
calibrated by us (see Sect.~\Spipeline\ below for details). Guest Observer data
were first retrieved from the ``Multi-Mission Archive at Space Telescope''
(\MAST) archive at the \STScI\ (Space Telescope Science Institute), which
provides reasonably-well calibrated datasets. If the observation was deemed
possibly useful, we calibrated it more carefully, using the same procedure as
for the Science Team data, described in \Sect\Spipeline\ below.


\subsection{Object Data}
Table~\Tbasic\ gives object and observation information for all objects in our
sample. We checked the basic data in \NED, the {\it NASA-IPAC Extragalactic
Database; http://nedwww.ipac.caltech.edu}. Using our preferred names (see below)
in a \NED\ search will yield information for the object. We discuss some objects
with troublesome names below, so that the reader can find these also.
\par Column~1 of Table~\Tbasic\ gives the object name. This is usually, but not
always, the same name as was used in the \FUSE\ observation log. Many objects
have multiple names. Column~16 of Table~\Tbasic\ lists the alternative ``Mrk'',
``PG'', ``Ton'', ``UGC'' or Zwicky name for the object, if it exists. We
established the following order of preference for the prefix indicating the
source catalogue: ``NGC'', ``3C'', ``Mrk'', ``PG'', ``Ton''. This led us to
prefer Mrk\,116 over I\,Zw\,18, Mrk\,771 over PG\,1229+024, Mrk\,1095 over
Akn\,120, Mrk\,1502 over I\,Zw\,1, Mrk\,1513 over II\,Zw\,136, PG\,0844+349 over
Ton\,951 and PG\,1302$-$102 over PKS\,1302$-$102. We further used the full name
as specified in \NED\ for some objects for which an abbreviated 12-character
version was used in the \FUSE\ observation log. This left 10 cases where we
found that the name used in the \FUSE\ observation log was highly irregular. We
then used \NED\ to find an alternative name. For 7 of these the choice was easy,
as other sources with the same prefix are present in the sample. In two cases
both the original name and the alternative name were possibilities. We chose
PHL\,1811 over FIRST\,J2155$-$0922, Tol\,1924$-$416 over IRAS\,19245$-$4140 and
UGC\,12163 over Akn\,564, since these preferred names are more indicative that
the object is a blue galaxy.
\par Four objects had some name complications. First, both the \FUSE\
observation log and \NED\ give PG\,1114+445 instead of PG\,1114+444. The B1950
declination of PG\,1114+444 is +44\deg29\arcmin56\arcsec, and the IAU naming
convention calls for cutting, not rounding the coordinates. Second, in the
original PG catalogue, the right ascension of PG\,1544+489 is given as
15$^h$44$^m$00$^s$, whereas a more precise determination gives
15$^h$43$^m$59.93$^s$. Since the IAU rule is that a name should be based on the
original catalogue, this implies that the name PG\,1543+489 used in

\noindent  {\it NED} is
in error. Third, the object named QSO\,0045+3926 in the \FUSE\ observation log
is only recognized by \NED\ as IO\,And, S\,10785 or RX\,J0048.3+3941; we used
the last of these. Finally, we note that \NED\ stores HS\,1102+3441 under the
name PG\,1102+347, even though the PG catalogue lists no object at this
position.
\par The J2000 equatorial coordinates of the objects are given in Cols.~2 and 3
of Table~\Tbasic, with galactic longitude and latitude in Cols.~4 and 5. For
most objects these were found from the V\'eron-Cetty \& V\'eron (2000) catalogue
of QSOs and Seyferts. Thirty-five objects (including thirteen elliptical
galaxies and 5 \HII\ regions) are not present in that catalogue, so we used the
coordinates given in \NED.

\par We used the \NED\ database to find the velocity or redshift for each
object. Column~6 gives the object's radial velocity, if it is $<$6000\kms, or
the redshift, if it is $>$0.02. The object HS\,1549+1919 is unknown within
either \NED\ or \SIMBAD, and its redshift is unknown. For RX\,J0042.0+3641 and
RX\,J1306.3+3917 \NED\ does not provide a redshift. In three cases
(PG\,0804+761, PG\,1211+143, PHL\,1811) the redshift in \NED\ is incorrect, and
an updated value is provided (see the notes in Appendix~A).
\par Column~7 of Table~\Tbasic\ gives the classification, also from \NED. This
can be one of the following: ``QSO'' or ``BLLac'' (for quasars), ``Sey\#''
(where a number between 1 and 2 gives the numerical subclassification),
``Gal:{\it mmm}'' (with {\it mmm} the morphological classification), `Gal:HII''
(for the three Tololo objects, which are galaxies whose light is dominated by
\HII\ regions), ```BCG'' (Blue Compact Galaxy), ``BCD'' (Blue Compact Dwarf),
``StrBrst'' (a starburst galaxy), ``HII({\it ggg})'' (for Mrk\,59, NGC\,588,
NGC\,592 and NGC\,5461, which are bright \HII\ regions inside another galaxy,
whose name is given by {\it ggg}), ``sdO'' (for PG\,0832+675 and vZ\,1128). Note
that for QSOs \NED\ gives ``QSO'' on its summary line, but in the body of the
information each QSO is classified as either ``Sey1'', ``Opt.var.'' or (in a few
cases) ``E2'' or ``E4''. We use the ``QSO'' classification in these cases.
\par In Col.~8 the blue magnitude of the object is given, taken from the
catalogue of V\'eron-Cetty \& V\'eron (2000). \NED\ gives a preferred magnitude,
but does not specify for which band it is; B magnitudes are not always given
either. Column~9 shows the value of the reddening that \NED\ gives for the
direction to the target. These are based on the {\it IRAS/DIRBE} measurements of
diffuse infra-red emission (Schlegel et al.\ 1998), but may not be free of
systematic errors (see explanations in \NED).
\par Columns 10--13 give the basic \FUSE\ observation information -- the year,
month and day of the observation, the identification code of the \FUSE\
observing program, the observation ID within that program and the nominal
exposure time. The observation ID has two parts -- a target ID within the
observing program and a visit number. Within each observation there are a
varying number of single orbit exposures. For a few observations the information
is preceded by ``['' and followed by ``]''. These are observations that were not
used in the final combined spectrum for the object (see \Sect\Svshift). Except
for the observation of NGC\,5253 all spectra were taken through the LWRS
aperture.
\par The nominal exposure time given in Col.~13 is the time that the observation
log specifies. However, parts of the observation are often lost because of
``event bursts'' or other problems (see \Sect\Spipeline), and the usable time is
less than the nominal time; this can differ for the different spectrograph
segments. We also decided not to use the SiC data due to the higher noise in
that channel. Therefore, Cols.~14 and 15 give the useful on-object time for both
the LiF1A and the LiF2B segments. In a few early observations there is no signal
in the LiF2B segment, as indicated by the dash in Col.~15 (3C\,249.1,
Fairall\,9, Mrk\,352, Mrk\,817, NGC\,985, PG\,0052+251, PG\,1302$-$102,
PKS\,2155$-$304, Ton\,S180, Ton\,S210). For many faint guest observer sources,
we did not run the \FUSE\ pipeline again (described in \Sect\Spipeline), but
instead just used the \MAST\ data, in which bursts have not been removed. This
is indicated by ellipses in the table.


\section{DATA PREPARATION}

\subsection{Summary of Steps}
In this section we discuss the steps necessary to create a final spectrum in the
region around the \OVI\ absorption lines. The resulting spectra are shown in
Fig.~\Fspectra. Their construction is described in \Sects\Spipeline\ to
\Sgroupcontam.
\par The top panel of Fig.~\Fspectra\ presents a wide wavelength range centered
on 1032\aa, showing the general behavior of the continuum. For this and the
other panels, the vertical scale gives the absolute flux in units of
\dex{-14}\fu. The top level is set to 1.4 times the flux level at 1032\aa, with
a minimum value of 2.8\tdex{-14}\fu. The strong line near 1025.722\aa\ is
Galactic \Lyb\ absorption, in the center of which geocoronal \Lyb\ emission is
always seen. Geocoronal \OI\ and \OI*\ emission may also be present, at
1027.431, 1028.157, 1039.230, 1040.942 and 1041.688\aa. Several interstellar
lines are always seen in the 1015--1050\aa\ wavelength range, these are the
slightly broad features at 1020.699\aa\ (\SiII), 1026.476\aa\ (\OI),
1036.337\aa\ (\CII), 1037.018\aa\ (\CII*), 1039.230\aa\ (\OI) and 1048.220\aa\
(\ArI). In addition many \H2\ lines fall in this wavelength range (3 with $J$=0,
5 with $J$=1, 5 with $J$=2, 6 with $J$=3 and 7 with $J$=4). These show up as
very narrow lines.
\par The second panel presents the \OVIla\ absorption line, on the calibrated
LSR velocity scale (see \Sect\Svshift). The positions of nearby \H2\ lines are
indicated by the labels P3 and R4. The line showing the continuum includes the
expected \H2\ lines, whose parameters were determined in the manner described in
\Sect\SHtcontam. \H2\ wavelengths and oscillator strengths were taken from
Abgrall \& Roueff (1989) and Abgrall et al.\ (1993a, 1993b). If a feature has
been identified with intergalactic absorption or if absorption is associated
with the background object (see \Sect\SAGNcontam), this is indicated by the
label. For features that may be \Lyb, but which are not positively identified, a
``?'' is added. If the possible \Lyb\ is not confirmed, but very likely exists
because of the presence of a galaxy group in the sightline, a ``:'' is added.
The small bar in the lower left corner shows the $\pm$1$\sigma$ noise level (at
10-pixel binning, i.e., per 20\kms\ resolution element). The thick vertical bars
connected to the top axis show the velocity ranges over which the \OVIla\ line
was integrated to derive equivalent widths and column densities (see
\Sect\Svrange).
\par The third panel shows the \OVIlb\ line. There are several other absorption
lines near this line, which are indicated by the labels \CII\ and \CII* for the
\CIIl\ and \CII*\l1037.018 lines, and R0, R1, P1 and R2 for the four \H2\ lines
at 1036.546, 1037.146, 1038.156 and 1038.690\aa, respectively. The continuum is
also shown in this panel, except for a few sightlines where the continuum fit
near 1037\aa\ is too uncertain.
\par The fourth panel contains the apparent column density profile for both
\OVI\ lines (see \Sect\Scoldens). The thick line is for the \OVIla\ line, the
thin one for \OVIlb. The latter was shifted to positive velocities by 10\kms\ to
correct for an apparent discrepancy in the wavelength calibration present in
v1.8.7 of the \FUSE\ pipeline (see \Sect\Svshift\ for a description of the
apparent column density and the justification for the shift). Note that we used
the nominal shift for the \OVIlb\ absorption profile in the panel above. The
labels identify the other absorption lines near both \OVI\ lines. The profile
for the \OVIlb\ line is cut off below a velocity of $-$110\kms\ and above a
velocity of 280\kms\ in order to avoid clutter associated with the \CII, \CII*
and \H2\ lines. This panel is omitted for the objects for which the spectrum was
too noisy or the \OVI\ line was not measureable for other reasons.
\par The fifth, sixth and seventh panels from the top present the low-ionization
absorption lines (\CIIl, \SiIIla\ and \ArIl) that were used to align the \FUSE\
spectrum with the \HI\ emission data (see \Sect\Svshift). These lines also serve
as a guide for showing the properties of strong, medium and weak low-ionization
species in neutral and weakly-ionized gas.
\par The bottom two panels show the \HI\ spectrum (see \Sect\SHI) -- first with
a vertical scale emphasizing the brightest component, then with a vertical scale
emphasizing the low-intensity higher-velocity components. The label in the top
left corner gives the telescope used to obtain the \HI\ spectrum -- ``LDS''
means the Leiden-Dwingeloo Survey, ``VE'' means Villa Elisa, ``GB'' means the
Green Bank 140-ft, ``Pks'' means Parkes and ``Eff'' means the Effelsberg
telescope, See \Sect\SHI\ for more details. The bottom of the figure lists the
parameters of a gaussian decomposition of the \HI\ spectrum. Five numbers are
given. First a component identification, next the central velocity in \kms, then
the amplitude, $A$, of the gaussian in K, the FWHM, $\Gamma$, of the line in
\kms, and the \HI\ column density, in units of \dex{18}\cmm2. $N$(\HI) is
calculated as $\sqrt{\pi/4\ln2}$ $A$ $\Gamma$ $\times$ 1.82\tdex{18}.
\par Table~\Tderived\ summarizes the derived parameters for each object. In this
table the objects are sorted by object name, divided into two groups. The first
group contains all the objects for which the measured signal-to-noise ratio per
resolution element is $>$3 (see \Sect\SSoverN), toward which we can measure the
Galactic \OVI\ column density. The remaining, fainter, objects follow.
\par Below, we summarize the contents of the columns of Table~\Tderived, but we
refer to later sections for the detailed descriptions of the meaning of the
specified values.
\par Column~1 of the table gives the object name (see \Sect\Ssample).
\par Column~2 gives the number of LiF1A and LiF2B segments used to construct the
final spectrum (see \Sect\Scombine). For objects where the LiF2B segment was not
used a preceding ``f'' indicates that the reason is that there is no recorded
flux in the segment, while a ``d'' indicates that the LiF1A and LiF2B segments
disagree. For spectra with high S/N ratio ($>$14) only LiF1A was used.
\par Column~3 gives the ``effective exposure time''. This is the total exposure
time in the LiF1A segment plus half that in the LiF2B segment (if it was used),
after removing bursts (see \Sect\Ssample). The LiF2B channel is weighted half as
much as the LiF1A channel since it produces half the number of counts for the
same exposure time. Some values are given within parentheses. These are for
guest observer objects that we did not recalibrate because either the continuum
is too complicated to measure Galactic \OVI, or because the source is too faint.
\par Columns~4--9 give the flux, the rms noise, the signal-to-noise (S/N) ratio
in a 10-pixel bin (corresponding to 20\kms, or one resolution element), the
signal-to-noise ratio at the final rebinning, the final rebinning near the \OVI\
line, and a ``quality factor''; see \Sect\SSoverN\ for details.
\par Columns~10 and 11 describe the continuum fit (see \Sect\Scontinuum).
\par Column~12 summarizes the intergalactic absorption (see \Sect\Sgroupcontam).
Columns~13 and 14 list the galaxy groups that the sightline intersects, as
determined using the catalogues of Geller \& Huchra (1983, 1984) and Tully
(1988).
\par Columns~15 and 16 give the highest rotational level of H$_2$ that is
detected, and whether the \H2\ lines contaminate the \OVIla\ absorption (see
\Sect\SHtcontam).
\par Columns~17--23 give the LSR velocity range and equivalent width of the
Milky Way and HVC \OVI\ absorption (see \Sect\Svrange). The ``x''-es and the
``[, ]'' pairs before or after the minimum or maximum velocity of integration
(Cols.~17/21 and 18/22) indicate sightlines with a difficult separation between
Milky Way and high-velocity \OVI\ (see \Sect\Svrange\ for the full description.
In Col.~20 a classification is given for the high-velocity \OVI\ component. The
corresponding phenomena are summarized in \Sect\Smaps. Sembach et al.\ (2002b)
present a full discussion.

\subsection{Calibration}
A general description of the \FUSE\ mission and its characteristics was given by
Moos et al.\ (2000) and Sahnow et al.\ (2000). \FUSE\ consists of four aligned
telescopes and spectrographs, two of which have an Al+LiF coating to produce
spectra between 1000 and 1187\aa, and two of which have a SiC surface optimized
for the 905--1105\aa\ wavelength range. There are two detectors, one for each
LiF/SiC channel pair, called Side 1 and 2, which are composed of two
microchannel plates, called segment A and B. The segments are separated by a
10\aa\ gap and the detectors are aligned such that most wavelengths are covered
at least twice.
\par Since all our objects produce low data rates, all observations were done in
the ``time-tag'' mode, in which the ($x$,$y$) address of each photon is saved,
with a time tag inserted once a second. One such photon list is produced for
each segment for each orbital visibility period, during which on the order of
2000 seconds of data are taken. Most observations take longer than a single
orbit, so the first step in our procedure is to use the program
``ttag\_combine'' (v1.0.4) to combine all of the raw data of the different
one-orbit exposures into a single photon list.
\par For a few sightlines of special interest (PG\,1259+593, NGC\,1705, and
NGC\,3310) a slightly more elaborate procedure was applied: all individual
exposures were separately processed and Doppler-corrected before combining. For
long observations ($>$30~ks) of bright objects (flux $>$3\tdex{-14} \fu) it
should in general be possible to align the data for each individual orbit, and
combine the aligned exposures. As the pointing may vary slightly from orbit to
orbit this can in principle improve the resolution of the data. For three
objects (Mrk\,153, Mrk\,279 and Mrk\,817) we compared the spectra obtained using
this method with those obtained from the standard ``ttag\_combine''. Indeed,
narrow lines such as \ArIl\ become slightly narrower (by 1--2\kms), but the
difference is small. We conclude that in general it is not necessary to apply an
exposure-by-exposure alignment when the goal is to measure the broad \OVI\
lines.
\par A complication of \FUSE\ data is that they contain intermittent increases
in the count rate (see Sahnow et al.\ 2000). The duration of these ``bursts'' is
between a few and several hundred seconds. They are still unexplained. Since the
bursts are isolated in time, they can easily be removed. We used the IDL utility
``fuse\_scan'' that is provided by the \FUSE\ software group to identify and
remove the bursts by hand, separately for each segment.
\par Next we ran the \FUSE\ calibration pipeline. We used version 1.8.7, which
was available from the \FUSE\ web site at the Johns Hopkins University as of
November 2000. This process does the following: a) it screens out the times that
the satellite passed through the South Atlantic Anomaly (SAA), and times that it
pointed near the Earth limb; b) it corrects for detector drift and geometric
distortions on the detector; c) it corrects for satellite motion and grating
rotation during the observation; d) it subtracts the background; e) it applies a
wavelength calibration; f) it converts the counts to a flux, taking into account
dead time. No flat fielding or astigmatism correction is applied; see Moos et
al.\ (2000) and Sahnow et al.\ (2000) for discussions of these effects on the
data. The background correction that is applied is based on assuming a uniform
rate of 1 count\,\cmm2\,s$^{-1}$ across the detector, and ignores the scattered
light differences between observations during orbital day and night. This
implies that there may be slight offsets for long integrations.
\par A newer version (v2.0.5) of the pipeline software became available as we
were finishing the data handling. We did a few comparisons between the two
versions and decided not to recalibrate the data, as the largest difference is
in the wavelength calibration, and we already addressed this difference (see
\Sect\Svshift). If one retrieves calibrated data from the \MAST\ archive at the
Space Telescope Science Institute, most of these calibration steps are also
done, except that bursts (and, in a few cases, bad exposures) are not removed.
This adds noise to the spectra, and also implies that the flux calibration may
be in error because the on-source exposure time is no longer the nominal
exposure time.
\par The \FUSE\ pipeline was applied separately to each of the 8 segments. We
ran this process twice, once for all data, and once selecting only the data that
were taken during orbital night. In the latter dataset airglow lines are much
fainter. The airglow lines that lie closest to the \OVI\ lines are those of
\Lyb\ at 1025.722\aa\ and \OI\ and \OI*\ at 1027.431, 1028.157, 1039.230,
1040.942 and 1041.688\aa. These do not overlap the \OVI\ lines, and for the
current study we therefore used the combined orbital day and night data.
\par For two objects (Ton\,S210 and HE\,0226$-$4110), an earlier short
observation was followed by a second, longer observation, which was taken using
a ``focal-plane split'' (i.e., the spectrum was deliberately shifted on the
detector from orbit to orbit in order to reduce the effects of fixed-pattern
noise in the combined spectrum). As the \FUSE\ pipeline assumes that the
detector position stays constant from exposure to exposure within a multi-orbit
observation, we ran the pipeline separately for each of the 15 (for Ton\,S210)
or 19 (for HE\,0226$-$4110) orbits and combined the data later.
\par After running the pipeline, we were left with one file for each detector
segment, each covering about 90\aa, with a pixel spacing of about 0.0068\aa, or
about 2\kms. Not all these detector pixels are independent, however. The
properties of the optics and the detector combine so that a photon at a given
wavelength may produce signal in any one of about 3--4 (near 1030\aa) or even
5--6 (near segment edges) adjacent detector pixels. We therefore always binned
the data over at least 5 pixels. For fainter sources we created larger bins.
\par We note that \FUSE\ produces a very low background signal. For long
integrations this usually remains below $\sim$2\tdex{-15}\fu, and it should have
been corrected for by the calibration software. We checked this in each spectrum
by examining the flux level in the \CIIl\ line. This line is always strongly
saturated over the wavelength range where $N$(\HI)$>$\dex{18}\cmm2\ and the flux
in the center of the line should be zero. There were just two objects for which
this was not true: HE\,0226$-$4110 and Mrk\,771. These discrepancies may reflect
non-uniformities in the intrinsic or scattered-light components of the
background. Since these are not precisely known, we did not apply a correction.


\subsection{\HI\ Spectra}
In order to better interpret and measure the absorption spectra, and in order to
properly align the different segments, we collected an \HI\ 21-cm spectrum for
each of our objects. For objects with declination $>$$-$35\deg, we first
extracted a spectrum from the Leiden-Dwingeloo Survey (LDS, Hartmann \& Burton
1997); the Dwingeloo telescope has a 35\arcmin\ beam. For 97 objects only LDS
data were available. For the 16 objects with more southern declinations we
requested a spectrum from the survey by Arnal et al.\ (2000), which was done
using the Villa Elisa telescope (34\arcmin\ beam).
\par It has been shown previously that the smaller the beam the better the
approximation to the \HI\ column density in the pencil-beam toward a background
source (e.g., Wakker \& Schwarz 1991, Wakker et al.\ 2001, 2002). Therefore,
whenever possible, we also used data from several other telescopes with smaller
beams. For 22 objects, we have Effelsberg data available; the Effelsberg
telescope has a beam of 9\farcm7. These spectra were taken in the same runs as
were described by Wakker et al.\ (2001), and are mostly for objects projected on
known \HI\ high-velocity clouds. For one object (Fairall\,9), a Parkes
(16\arcmin\ beam) spectrum was available, courtesy Gibson et al.\ (2001). For 80
objects we took the spectra from Murphy et al.\ (1996), who used the Green Bank
140-ft Telescope (21\arcmin\ beam) to measure accurate \HI\ spectra in the
direction of 220 QSOs and AGNs. The Green Bank spectra also are more sensitive
and have a better baseline subtraction than the LDS data.
\par All \HI\ spectra have a velocity resolution of $\sim$1\kms, and all spectra
were corrected for stray-radiation. For the objects where we have data from more
than one radio telescope available, we cross-checked the component structure to
see whether or not weak components were present in both spectra. The \HI\
spectrum with the smallest beam is included in Fig.~\Fspectra, as is the result
of a gaussian fit to the \HI\ components; the telescope is indicated by a label
(LDS, VE, GB, Pks, or Eff). See also the description in \Sect\Sdatasum.


\subsection{Alignment in Velocity}
The spectra that come out of the \FUSE\ calibration pipeline are nominally
calibrated in wavelength. However, in practice there are offsets between
segments and the absolute wavelength scale is usually slightly offset. This
problem was particularly severe before February 2001, at which time an improved
wavelength calibration algorithm was provided by the \FUSE\ software group 
although this was not included in our version of the calibration pipeline and
needed to be applied later. Also, in v1.8.7 of the pipeline there was a sign
error in the heliocentric to LSR conversion. Furthermore, we discovered that at
first we did not use the proper orbital parameter file (we did not recalibrate
these spectra). All these problems resulted in minor distortions of the
wavelength scale (20\kms\ over 50\aa), but mostly in large unpredictable shifts
(up to 100\kms). After we finished determining the channel alignments, version
2.0.5 of the \FUSE\ calibration pipeline became available. This version has an
improved wavelength calibration, except that minor offsets (up to 50\kms) in the
absolute alignment of the wavelength scale still occur.
\par Our calibrations were complete before v2.0.5 became available. We knew
about the problems with the wavelength scale, so we therefore determined the
zero-point of the wavelength scale in the region around the \OVIll\ doublet from
the data, rather than from using the wavelength calibration. This was done by
aligning the \SiIIla\ and \ArIl\ lines with the \HI-21 cm emission observed in
the direction of the extragalactic object. Below, we first discuss the rationale
for using \ArIl\ and \SiIIla\ (point 1). Next we discuss our treatment of the
\HI\ spectrum (point 2), followed by some comments on the resulting alignments
(point 3). Then we comment on differences between the velocity scales of the two
versions of the calibration pipeline (point 4), followed by a discussion of the
intrinsic difficulty of aligning absorption and emission lines originating in
different kinds of gas (point 5). Finally, we summarize the sources of alignment
errors and present a conclusion on the accuracy of the alignment (point 6).
\bigskip\bigskip
\par {\it 1) Rationale for using \ArIl\ and \SiIIla.} We used the \ArIl\ and
\SiIIla\ lines since a) they lie on opposite sides of the \OVIla\ line and b)
the \HI\ column density usually is on the order of a few \dex{20}\cmm2, in which
case these lines are strong but usually not highly saturated, so that the
position of the deepest absorption usually corresponds to the strongest \HI\
component. For a few sightlines we also checked several of the \H2\ lines
between 1028 and 1043\aa, but these were only used to confirm the alignment, as
it is not a-priori clear with which \HI\ component they correspond. For
sightlines with low FUV flux we further checked the \CIIl\ line, and demanded
that the full extent of the \HI\ profile fell within the region where \CIIl\ is
saturated.
\par {\it 2) The \HI\ spectrum.} To find the alignment, we first fitted a set of
gaussians to each \HI\ spectrum, in the manner described by Wakker et al.\
(2001). We then noted the velocity of the strongest component. In the majority
of cases (185 out of 219) this velocity lies within about 15\kms\ from
\vlsr=0\kms. A small positive offset (\vlsr=19\kms) is seen toward NGC\,1705. In
31 directions lying in the area of sky containing the Intermediate-Velocity Arch
and Spur (see Wakker 2001), the strongest component in the \HI\ profile lies at
a velocity between $-$60 and $-$15\kms. Finally, toward PG\,1259+593 HVC
complex~C at \vlsr=$-$128\kms\ is the strongest \HI\ component. These cases
clearly show that blindly assuming that the peak of the \HI\ is located near
0\kms\ might lead to an erroneous velocity scale.
\par For 32 sightlines no single \HI\ component dominates the spectrum, and
instead there are two (or even three) components of similar strength. In this
case the interstellar absorption lines may be broad, or if the different
components have different element depletions due to the presence of dust, one
absorption component may be stronger, although it would not be a-priori clear
which one. Some of these objects have very low FUV flux and the alignment cannot
be adequately determined. In other sightlines we checked the individual
absorption features carefully to determine the most likely alignment.
\par {\it 3) Results.} To do the alignment, two of us (Wakker/Richter)
independently determined the apparent velocity of the \ArIl\ and \SiIIla\ lines
in the LiF1A segment and visually estimated the shift necessary to align these
lines with the \HI. Subsequently, we overlaid the LiF2B spectrum and determined
its shift relative to that of LiF1A for the same observation by visually
aligning them as well as possible. In almost all cases this gives an unambiguous
result for the differential alignment, accurate to about 3--5\kms\ at high S/N
ratios, and to about 10--15\kms\ at low S/N ratios.
\par In most cases, the shifts implied by the \ArIl\ and \SiIIla\ lines appeared
to be nearly the same, although in a few cases they differed by up to 20\kms.
Such ambiguities often made it necessary to simply adopt an average shift.
In a limited number of cases we later compared our shifts with those implied by
v2.0.5 of the pipeline. This showed that there may be a systematic difference of
15\kms\ between the two lines (see point 4 below). We did not revisit the
velocity alignment for all 119 sightlines after we found this out. Also, because
the \ArIl\ and \SiIIla\ lines differ in shape, and because we had not noted down
for each individual sightline whether we gave the highest weight to \ArIl\ or to
\SiIIla, we concluded that we could not correct for this systematic difference
by applying a blanket average shift of 8\kms.
\par The final shifts we determined for the LiF1A segment from the v1.8.7
calibrations have a bi-modal distribution: there is a peak at $-$85\kms\ with an
rms of 14\kms, and another at $-$8\kms\ with an rms of 19\kms. These shifts may
be due to a misalignment between the LiF1A channel and the Fine Error Sensor
(the camera that determines the pointing). This should not happen, but an
analysis by Bowen \& Jenkins (priv.\ comm.) suggests that it does. For our
project this does not matter because we align the spectra with the \HI\ data
a-posteriori. Of more interest is the difference in the shift between the LiF1A
and LiF2B segments. We find that this difference has a gaussian distribution
with an average value of $v$(LiF1A)$-$$v$(LiF2B)=$-$18\kms\ and a dispersion of
18\kms, showing that even after applying the proper wavelength calibration one
should still carefully check the alignment between the different segments.
\par {\it 4. Comparison between v1.8.7 and v2.0.5.} For six high signal-to-noise
objects (3C273.0, H\,1821+633, Mrk\,153, Mrk\,335, Mrk\,509 and vZ\,1128), we
also ran v2.0.5 of the pipeline and compared the positions of several absorption
lines. This shows that it is still necessary to introduce a data-based shift in
the velocities in order to align the absorption with the \HI. However, these
shifts are smaller than for v1.8.7. If we align the \OVIla\ lines produced by
v1.8.7 and v2.0.5 of the pipeline, there appears to be a systematic change in
the relative position of the other absorption lines. For the \H2\ line at
1008.553\aa\ $v$(v1.8.7)=$v$(v2.0.5)$-$10\kms; for the \SIII\,\l1012 line the
difference is $-$5\kms, for \SiIIla\ it is 0\kms, for \CIIl, \OVIlb, \OIl\ it is
$-$10\kms, for \ArIl\ it is $-$15\kms, for \FeII\,\l1063.176 it is $-$25\kms,
and for \H2\,\l1077.138 it is $-$30\kms.
\par However, the implied velocities of the center of the absorption in each of
these lines may still differ; for 3C273.0, Mrk\,153 and vZ\,1128 the velocities
of the deepest absorption in the \ArIl\ and \SiIIla\ lines agree better in the
v1.8.7 calibration; for H\,1821+643, Mrk\,335, Mrk\,509 they agree better in
v2.0.5. This effect may be due to intrinsic differences in the absorption lines
(see point 5 below), or it may be due to residual errors in the wavelength
calibration. In any case, there is still some uncertainty in the absolute
velocity of the different lines.
\par We note that we also checked the data used to actually define the
wavelength calibration for v2.0.5. These data are measurements of the raw pixel
positions of many lines in several spectra with very high signal-to-noise
ratios. Distortions of up to 20\kms\ are removed by fitting the average line
positions, but the residuals still have a dispersion of 5\kms.
\par The systematic wiggle in the wavelength scale of v1.8.7 is particularly
important for comparing the two \OVI\ absorption lines. Even before v2.0.5 was
available, we had already noticed that the \OVIlb\ line seemed systematically
shifted by $-$10\kms\ relative to the \OVIla\ line. This is very obvious in some
sightlines (e.g., vZ\,1128 and 3C\,273.0, see Sembach et al.\ 2001b).
Figure~\Fovishift\ presents four sightlines that illustrate this effect. On the
left hand side we show apparent column density ($N_a(v)$) profiles (see
\Sect\Scoldens\ for the definition), using the nominal wavelength scale produced
by the v1.8.7 pipeline, while on the right hand side a +10\kms\ shift for the
\OVIlb\ line is included. For each of the four sources the two $N_a(v)$ profiles
are shown in the top panel, while the ratio of the two is shown in the bottom
panel. For vZ\,1128 the shift is clearly visible in the profiles, unlike what is
the case for the other three sightlines. For these one needs to look at the
ratio plot, which is clearly much flatter after the correction is applied, in
the velocity range defined by the heavy vertical lines on the $N_a(v)$ plot. The
conclusion we draw from this comparison is that the wavelength scale used in
v1.8.7 of the calibration pipeline has a systematic offset of 10\kms\ between
the two \OVI\ lines.
\par {\it 5. Intrinsic difficulties with aligning absorption lines with \HI.} We
note that even if the \HI\ profile shows a sharp peak, the absorption lines do
not necessarily peak at exactly the same velocity since the \HI\ spectrum
measures the average emission over a 10\arcmin, 21\arcmin\ or 36\arcmin\ beam,
within which there may be velocity gradients. The absorption, however, is seen
on the pencil-beam toward the background source. This is illustrated by the
complications found for the sightline toward 3C\,273.0 (Sembach et al.\ 2001b).
They found that in the 21\arcmin\ Green Bank 140-ft beam the \HI\ shows peaks at
$-$6 and 23\kms, while the \SII\ absorption seen with the \GHRS\ on \HST\
centers at $-$15 and 23\kms, although there is an extra, narrow, feature at
$-$6\kms. Arguing that there could be ionization differences between the
different ions in the 23\kms\ positive-velocity gas, they align the two features
seen in the \ArIl, \SiIIla\ and \FeII\,\l1063.176\aa\ lines with the smoothed
\SII\ spectrum, resulting in velocities of $-$15 and 18\kms\ for \SII. The \H2\
absorption would then be at 16\kms. However, an argument could be made that the
ionization differences occur in the gas near 0\kms\ and that the \ArIl\ and \H2\
absorption should align with the 23\kms\ \HI\ component. Using v2.0.5 of the
pipeline clarifies some of these issues, especially since the resolution seems
slightly better. If we align the \H2\ absorption with the 23\kms\ component,
then a second weak \H2\ absorption component lies at $-$14\kms, the \ArIl\ line
centers at $-$15 and 16\kms, the \SiIIla\ line at $-$16 and 11\kms, and the
\FeII\,\l1063.176 line at $-$14 and 10\kms. These differences in alignment
between different versions of the pipeline and between different ions illustrate
the difficulty of aligning the complex wide-beam \HI\ spectrum with the complex
pencil-beam absorption spectra.
\par {\it 6. Summary of alignment errors.} In summary, there are many sources of
error in determining the alignment of the \FUSE\ spectra. The internal accuracy
of the wavelength scale is about 5\kms, as determined by the fit to many lines
in some spectra with high signal-to-noise ratios. For v2.0.5 the relative
alignment between different absorption lines seems correct to within this error,
but for v1.8.7 there may be a systematic stretch of 15\kms\ between \ArIl\ and
\SiIIla. Since we did not correct for this stretch, on average there may be an
offset of $\sim$8\kms\ in out final alignments. For many sightlines the \HI\
spectrum has a well defined peak that can be aligned with either \ArIl\ or
\SiIIla\ to within one rebinned pixel (5\kms\ in spectra with high S/N ratio,
10\kms\ at low S/N ratio, see \Sect\Scontinuum). In some cases, the \HI\ peak is
broad and the alignment becomes more uncertain. However, the alignment between
the peak of the \HI\ emission and the peak absorption may differ by 10\kms\ (as
illustrated by the 3C\,273.0 sightline). Also, the gas observed in \HI, \ArI\
and \SiIIla\ may sample different parts of interstellar space under different
physical conditions. Taking into account all these effects we conclude that for
spectra with high signal-to-noise ratios our alignments are accurate to within
15\kms\ for 80\% of the sightlines (corresponding to a 1-$\sigma$ error of
$\sim$10\kms, but that in some cases a future study may show that our alignment
is in error by as much as 25\kms. 


\subsection{Combining Segments}
For the \OVI\ doublet, the most sensitive data are obtained from the LiF1A
segment (which covers 987--1082\aa), while the LiF2B segment (978--1074\aa)
yields a spectrum with a S/N that is $\sim$1.4 times lower. The SiC1A segment
(1004--1091\aa) and SiC2B segment (1016--1105\aa) also cover the \OVI\ lines,
but these data are much noisier, and we did not use them. The resolution in the
LiF1A segment is on the order of 20\kms\ (0.070\aa), while that in the LiF2B
segment is slightly worse.
\par Since the LiF1A segment has a better S/N ratio and resolution than the
LiF2B segment, we prefer to use it. In fact, for objects with sufficiently high
S/N ratio (see Sect.~\SSoverN), combining the two segments would actually lead
to a reduction in the quality of the spectrum because of the lower resolution
and the uncertainties in aligning different channels. In these cases we still
used the LiF2B segment as a check on the features in the spectrum, and in all
cases we found that within the errors the two segments give the same answers for
quantities like the equivalent width. For objects with a very low S/N ratio (see
Sect.~\SSoverN) we deemed it too difficult to determine the alignment between
the two segments, so we also used just the LiF1A segment. For objects with
intermediate S/N ratios we decided to combine the LiF1A and LiF2B segments. Then
the factor $\sqrt{1.5}$ increase in the S/N ratio more than offsets the problems
associated with combining two segments which have slightly different
resolutions.
\par To combine the spectra we used the count arrays provided by the \FUSE\
pipeline. Using the flux arrays we reconstructed the conversion from count to
flux for each segment. The LiF1A pixels were then shifted appropriately. Next,
the LiF2B counts and conversion factors were shifted, regridded to the LiF1A
wavelength grid and added to the LiF1A data. The final counts were then
converted back to flux. A similar procedure is used when combining multiple
observations.
\par For a number of sightlines more than one observation was obtained, on
different dates. These were also combined, after applying the shifts, which were
determined separately for each observation. In a few cases (ESO\,265$-$G23,
Mrk\,79, Mrk\,618, Mrk\,817, Mrk\,1095, VII\,Zw\,118) one of the 2 or 3
observations had a much shorter integration time than the other(s). We then
decided that adding the extra counts did not compensate for the uncertainties in
determining the proper alignment for data with low S/N.
\par For most of the sightlines the spectrum in the LiF1A segment is nearly
identical to that in the LiF2B segment, within the noise. However, there are
cases where this is not so. For 3C249.1, HE\,0450$-$2958, IRAS09149$-$6206,
Mrk\,59, Mrk\,357, NGC\,595 and Ton\,S210 the differences are fairly sutble and
certainly within the noise. For HS\,0624+6907, Mrk\,106, Mrk\,618, PG\,1415+451
and Tol\,1924$-$416, the differences are quite easily noticeable, but since
these spectra have low S/N ratios, it can still be argued that the differences
lie within the noise. Since the weight of the LiF2B segment is half that of the
LiF1A segment, we still combined the LiF1A and LiF2B segments. For
MRC\,2251$-$178, NGC\,5548, PG\,1001+291 and PG\,1211+143, however, the
difference is particularly striking, and we decided that we should discard the
LiF2B segment.
\par Another kind of problem occurs for 6 bright objects for which multiple
observations are available (Mrk\,279, Mrk\,509, Mrk\,817, PG\,0804+761,
PG\,0953+414, PKS\,2155$-$304). For these the calibrated flux differs between
the different observations, by up to a factor two, although the spectra have the
same shape and show the same features. For the early observations of Mrk\,509
and PKS\,2155$-$304 (prior to November 1999), the LiF1A and LiF2B fluxes also
disagree, but this might be explained by misalignments in the telescope/slit or
mirror/detector systems. However, since for the other objects the LiF1A and
LiF2B segments in a single observation do agree, it is likely that the
differences in flux between different observations are due to intrinsic
variations in the extragalactic sources.


\subsection{Continuum Fitting}
\subsubsection{Fitting Procedure}
After combining different segments, the resulting spectrum was considered to be
the final one. The next step was to fit a continuum. To do this, we first
selected two or more narrow (a few \AA) wavelength regions in the wavelength
range between 1020 and 1050\aa. The selected regions lie away from known
Galactic absorption lines and are visually free of other redshifted lines. We
then fitted a Legendre polynomial to the selected regions, using the algorithm
described by Sembach \& Savage (1992). This method also provides an estimate for
the error in the placement of the continuum for each pixel.
\par The fit described above provides a global continuum. It is the fit shown in
Fig.~\Fspectra. If the fits only made sense in part of this range, only that
part is displayed. However, such a fit is not appropriate for determining the
error in the equivalent width and column density associated with the continuum
placement, as for this purpose the continuum near \OVIla\ should not be
constrained by what happens 20\aa\ away. Therefore, we also fitted a local
continuum, between 1029 and 1035\aa, using the global continuum as a guide, and
making sure that at the \OVIla\ line the local continuum level differs by
$<$0.5$\sigma$ from the global level. This local fit usually can be accomplished
using a lower polynomial order. However, it produces substantially larger (and
more realistic) continuum placement errors.
\par The orders of both the global and local polynomials are listed in Col.~10
of Table~\Tderived. In those cases where a local fit was not possible because
the continuum had too much curvature, or there were interfering features, no
separate local fit was done. This is indicated by having an ``x'' as the order
of the local continuum fit. We also assigned a one-word description of the shape
of the continuum to each spectrum (Col.~11). We further noted the continuum flux
level at the center of the \OVIla\ line. This is given in Col.~4 of
Table~\Tderived. For objects where the \OVIla\ line sits in the wing of \Lyb\
absorption associated with the background object, or where the spectrum shows
large fluctuations, this number is lower than the continuum flux elsewhere in
the spectrum. Then the continuum value over most of the spectrum is also given
in Table~\Tderived, in Col.~11, between square brackets, after the word
describing the continuum shape.

\subsubsection{Continuum Characteristics}
\par As described in Sect.~\Sdefine, \IUE\ spectra were used to estimate the FUV
flux for all AGNs observed by that satellite in order to define a sample of
objects for the \FUSE\ Science Team Projects. Figure~\Fcompare\ shows the
comparison of the extrapolated flux and the flux actually observed at 1030\aa,
for the 72 observed objects with predicted flux $>$2\tdex{-14}\fu. The
correlation coefficient of this collection of points is just 0.69. The ratio
averages to 1.09 with a dispersion of 0.81. For 75\% of the objects the
prediction is correct to within a factor of 2, for 10\% the ratio
observed/predicted is $>$2, and for 15\% it is $<$0.5. Clearly, a source that is
bright at $\lambda$$>$1200\aa\ will generally be bright at 1030\aa, but the
actual flux remains uncertain. This uncertainty is of similar magnitude for
galaxies, Seyferts and quasars, and it is independent of the shape of the
continuum. Some of the difference between observed and predicted flux may be due
to intrinsic variability. 
\par Not all AGNs that \FUSE\ might observe were already observed by \IUE. To
allow an estimate of the flux at 1030\aa\ (F(1030)) for any AGN, we therefore
also compared the visual and blue magnitude from the V\'eron-Cetty \& V\'eron
catalogue of AGNs with the observed flux -- visual magnitudes are available in
this catalogue for 151 targets, blue magnitudes for 83. Figure~\Fmagflux\ shows
the results of this comparison. Panels a) and b) show that for QSOs there is a
fairly good correlation between magnitude and F(1030) (correlation coefficient
$\sim$$-$0.75). This correlation does not depend on the reddening; correcting
for the reddening does not change it substantially, as the corrections are
generally small. The faintest objects for which the flux level can be somewhat
reliably determined have F(1030)$\sim$5\tdex{-15}\cmm2. Including just objects
brighter than this, the correlation coefficient is also $\sim$$-$0.75. Almost
all QSOs with B$<$15 have F$>$\dex{-14}\fu\ at 1030\aa, while those with
B$>$16.5 are too faint to observe with \FUSE. On the other hand, the correlation
for Seyferts (panels c and d) is poor (correlation coefficient $\sim$$-$0.45).
For nearby galaxies with V$<$12 (panel e) the correlation is bad. However, these
galaxies usually are larger than the \FUSE\ aperture, and thus the visual/blue
magnitude refers to a different part of the galaxy than the \FUSE\ observation.
For the smaller (fainter) galaxies ($V$$>$12), the correlation coefficient is
$\sim$$-$0.6 (panels e and f).
\par The shape of the continuum is characterized as ``Flat'', and fit by a
straight line (polynomial order 1) for 60 objects with flux $>$4.5\tdex{-15}\fu\
(and for one fainter one with a long observation, KUG\,1031+398). Usually the
slope of the polynomial is $\sim$0. For 58 objects with flux
$<$4.5\tdex{-15}\fu\ we list a shape of ``Zero'' in Table~\Tderived, as there
really is not enough flux to determine the continuum shape. In 31 cases the
Galactic \OVI\ sits in the damping wing of \Lyb\ absorption associated with the
background galaxy. The continuum shape then is listed as ``\Lyb[\#]'', where the
\# gives the average flux over most of the spectrum. In these cases the
line-free regions selected to make the polynomial fit only cover the spectrum to
the side of the wing nearest to \OVIla.
\par For the remaining 66 objects the continuum shows varying degrees of
curvature. The least troublesome are the 16 we loosely classify as ``Curved'',
in which case there appears to be a slight curvature near the \OVI\ lines. Then
the global continuum can be fit with a polynomial of order 2 or 3. Some
sightlines with high S/N were classified as ``Curved'', although with lower data
quality we probably would have classified them as ``Flat''. For the 32 objects
in which the continuum fluctuates considerably in the 
\noindent wavelength range
1020--1050\aa\ the classification ``Wavy'' is used and a polynomial of order 4
or 5 was needed to fit the global continuum. The difference between ``Curved''
and ``Wavy'' is sometimes subjective.
\par Troublesome continua are provided by the 18 objects for which we classify
the continuum as ``Wobbly'' and which need a polynomial with order $>$5. These
split into the 11 sightlines where we still think that the continuum is
trustworthy enough to measure \OVI\ (IRAS\,F11431$-$1810, MRC\,2251$-$178,
Mrk\,36, Mrk\,79, Mrk\,335, Mrk\,1502, Mrk\,1513, NGC\,595, NGC\,3504, NGC\,4151
and NGC\,7714) and the 7 sightlines for which we decided that the continuum was
too uncertain, and which were therefore discarded (NGC\,3690, NGC\,3783,
NGC\,4214, NGC\,5236, NGC\,5461, NGC\,7496 and NGC\,7673).


\subsection{Measurement of S/N Ratio}
\subsubsection{Flux and rms Measurement}
We measured the flux in the \OVIla\ line as the value of the fitted continuum at
a wavelength of 1031.926\aa. This is listed in Col.~4 of Table~\Tderived. Next,
we measured the rms fluctuations around the fitted continuum in the line-free
regions used to define the continuum. The resulting values depend on the binning
that is used. We measured the rms at 10 pixel binning, i.e.\ pixels that are
about 20\kms\ wide, corresponding to one resolution element. These rms values
are listed in Col.~5 of Table~\Tderived. Column~6 gives the ratio of the
continuum flux at 1031.926\aa\ to the rms, i.e.\ the signal-to-noise ratio in
the continuum per resolution element (\SNres).
\par For data with S/N$>$10 we combine 5 detector pixels. This yields bins with
a velocity width of about 10\kms, or about half a resolution element. For \OVI\
measurements at low S/N ratios we needed to use larger bins. This is possible
because the FWHM of the \OVIll\ lines is larger than 50\kms. The choice for the
final binning depends on the measured S/N ratio per resolution element. We
decided to rebin to 10 pixels if the S/N ratio at 10-pixel rebinning was between
6 and 10, while we rebinned to 15 pixels if this number was between 3 and 6, and
to 20 pixels for even lower S/N ratios. We then remeasured the rms. In Col.~7 of
Table~\Tderived\ we list the S/N ratio in the continuum at the final rebinning
(\SNbin). This value is always $>$4 for the objects for which we decided that a
measurement was possible.
\par For faint objects (flux $<$4.5\tdex{-15}\fu), the flux that is listed may
be affected by inaccuracies in the background subtraction, and thus the value we
list is unreliable. That this is the case is shown by a few faint objects for
which the measured flux differs between an early short observation and a later
long observation. However, without a detailed analysis of every exposure it is
not possible to improve the calibration and determine a more precise background
correction for each individual object.

\subsubsection{Binning}
\par Col.~8 of Table~\Tderived\ shows that there are 47 objects for which we
could use the smallest bins (5 pixels or 10\kms). These objects all have a
\SNres\ ratio in the continuum $>$10. For the 30 objects with \SNres\ ratios
between 6 and 10 we decided that rebinning over 10 pixels was appropriate. At
lower S/N ratios (for the 42 objects that still have \SNres$>$3) even that was
insufficient and we rebinned over 15 pixels, increasing \SNbin\ to lie between
4.0 and 7.3. For the 97 faintest objects even rebinning to 20 pixels did not
increase \SNbin\ above 4, which is required to measure the \OVI\ absorption.
\par We note that for spectra that need heavy rebinning, their appearance on the
plots of Fig.~\Fspectra\ depends strongly on the pixel at which the binning is
started. That is, one can combine pixels 1--15, 2--16, 3--17,  4--18 or 5--19,
etc. We used this multiplicity to get a feel for which features are real versus
which features are just noise. For the displayed spectra in Fig.~\Fspectra\ we
chose a starting pixel such that the absorption profile looks smooth, while the
noise fluctuations look realistic. For the targets with high signal-to-noise
ratios it hardly matters which starting pixel is chosen, as the pixel-to-pixel
fluctuations are small. At the lowest signal-to-noise ratios the pixel-to-pixel
fluctuations are large, and the bins wide, and the rebinned spectra always look
noisy. It is at intermediate signal-to-noise ratios that the precise choice of
starting pixel makes the most difference in the appearance of the spectrum. The
measurements of equivalent width and column density are not substantially
affected, however as they differ by $<<$1$\sigma$.

\subsubsection{Quality Factor}
\par We also assigned each spectrum an ``\OVI-quality factor'', Q, which is
listed in Col.~9 of Table~\Tderived. This does not exactly correlate with the
chosen final binning. A value Q=4 was given to spectra in which the \OVIla\ line
can be clearly measured. This pertains to most spectra with \SNres$>$14. Quality
3 is for good data where a confident measurement can be obtained (\SNres=9--14).
Quality 2 (\SNres=5--9) is for spectra where we think that the measurement is
acceptable, but where some problems start to appear -- such as uncertainty about
the range of integration or even about whether certain features are or are not
\OVI. Quality 1 (\SNres=3--5) is for spectra for which the measurement is
unreliable, but for which we still list the results; these should be treated
carefully. Finally, quality 0 pertains both to spectra with very low S/N
(\SNres$<$3) and spectra in which the continuum near \OVIla\ was too uncertain
or where intergalactic absorption overlapped the Galactic \OVI\ lines (see
Sect.~\Sgroupcontam). In the end, out of the 217 objects in the sample, there
are 26 objects of quality 4, 23 of quality 3, 30 of quality 2, 23 of quality 1
and 115 of quality 0.
\par The quality factor was used to determine when to combine the LiF1A and
LiF2B segments (see also Sect.~\Scombine). For objects with Q=4 and Q=0 only
LiF1A was used, while for objects with Q=1, 2 or 3 the two segments were added
together. For seven objects this criterion led to a contradiction
(ESO\,141$-$G55, H\,1821+643, HE\,0226$-$4110, NGC\,7469, PG\,0844+349,
PKS\,0405$-$12 and VII\,Zw\,118). Using just the LiF1A segment would give
\SNres$<$14 and Q=3, while with LiF1A+LiF2B they have \SNres$>$14 and Q=4; we
decided to adopt the higher S/N ratio for these sightlines.


\subsection{Removing Contamination by \H2}
\subsubsection{Contaminating Lines}
A complication in measuring the Galactic \OVI\ lines is the presence of other
absorption lines in the same spectral region. Near \OVIlb\ the \H2 5--0 R(1)
\l1037.146, 5--0 P(1) \l1038.156 and 5--0 R(2) \l1038.690 lines are often
strong, and lie at $-$135, 155 and 310\kms\ on the \OVIlb\ velocity scale.
Further, \CII*\l1037.018 is usually present at $-$173\kms. In many cases these
lines make the \OVIlb\ line only visible over the velocity range between
$\sim$$-$90 and 110\kms. Whenever possible, we used the \OVIlb\ line to check
the results for the \OVIla\ line, but we did not attempt to decontaminate it.
\par Four lines may contaminate \OVIla. \ClI\l1031.507 is at $-$122\kms\
relative to \OVI. \ClI\ is only found when $N$(\H2) is very high, as it forms
when \ClII\ interacts with \H2 (Jura 1974). Then the HD 6--0 R(0) line at
1031.912\aa\ ($-$4\kms\ relative to \OVI) is also expected to occur. In fact,
toward a number of stars with high $N$(\H2), one finds that when HD is weakly
present, \ClI\ is still absent. In our sample there is only one sightline
showing HD so the \ClI\ line should not be a problem.
\par NGC\,7469 shows surprisingly strong \H2\ absorption
($N$(\H2)$>$\dex{19}\cmm2). Features with equivalent widths of 21$\pm$8,
12$\pm$6, 29$\pm$11, 19$\pm$7, 15$\pm$7 and 15$\pm$7\maa\ can be noticed at the
wavelengths of the HD 3--0 R(0) through 8--0 R(0) lines at 1066.271, 1054.433,
1042.847, 1031.912, 1021.456 and 1011.457\aa\ (Dabrowski \& Herzberg 1976).
Since all these lines have similar oscillator strengths, we concluded that the
narrow feature in the \OVIla\ profile at 1031.91\aa\ is due to HD 6--0 R(0).
This line was removed from the calculation for \OVI.
\par Two \H2\ absorption lines often contaminate the \OVIla\ line -- the
6--0 P(3) line at 1031.191\aa\ and the 6--0 R(4) line at 1032.356\aa. On the
\OVI\ velocity scale these lie at velocities of $-$214 and 125\kms. Although the
column density of local interstellar \H2\ is expected to be small along
sightlines at high latitude, blending with \H2\ has to be considered when
interpreting the \OVIla\ absorption line profiles at absolute velocities
$>$100\kms.
\par For sightlines where we find that these \H2\ lines contaminate the \OVIla\
line, we removed the \H2\ absorption by parametrizing the lines. It is not
necessary to model the entire \H2\ absorption spectrum for these two rotational
states since there are a number of other $J$=3 and $J$=4 lines with similar
oscillator strengths. These lines therefore have absorption profiles and
equivalent widths similar to the 6--0 P(3) and 6--0 R(4) lines and thus can be
used to model them without constructing curves of growth. In particular, we used
7--0 P(3) \l1019.506, 5--0 P(3) \l1043.498, 3--0 R(3) \l1067.478, 8--0 P(4)
\l1012.261, 5--0 R(4) \l1044.546, and 4--0 R(4) \l1057.379. To these lines we
fitted gaussian profiles, deriving FWHMs and absorption depths. We averaged
these parameters (weighted by S/N) for the three $J$=3 and the three $J$=4
lines, respectively, and reconstructed the \H2\ absorption as a gaussian in the
vicinity of the \OVIla\ absorption with these averaged parameters. We
cross-checked the validity of our model by comparing the shape and strength of
the \H2\ absorption with other $J$=3 and $J$=4 lines that have higher and lower
oscillator strengths. The \H2\ absorption then was removed from the \OVI\
profile by converting the \H2\ absorption depth into a peak optical depth
($\tau_0$=$-$ln[1$-$depth]), and assuming a gaussian optical depth profile
($\tau(v)=\tau_0 \exp[-(v-v_0)^2/b^2]$). The original continuum, $c$, was thus
replaced by $c\exp-\tau(v)$.
\par Columns 15 and 16 of Table~\Tderived\ list the basic \H2\ information for
each of the sightlines. In Col.~15, we show the maximum value of the rotational
state J for which absorption is found, where a dash indicates those sightlines
where no \H2\ was found. Column~16 contains a ``y'' for those sightlines where
\H2\ contaminates the \OVI\ absorption and an ``n'' otherwise.

\subsubsection{Discussion of Contaminated Sightlines}
\par There are 77 sightlines in which the \OVIla\ line is not contaminated by
\H2\ absorption. In 19 of these there is no discernable \H2. For the cases with
high S/N ratio (Q=4; Mrk\,279, Mrk\,817, NGC\,1705, PKS\,2155$-$304 and
vZ\,1128) the 3$\sigma$ detection limit on $N$(\H2) in each rotational level is
about \dex{14}\cmm2. For lower quality sightlines (Q=2, 3), the limit is more
like 4\tdex{14}\cmm2. At even lower S/N the quality of the spectrum may just not
be sufficient to discern the \H2\ that may still be present. The sightlines with
little or no \H2\ concentrate in the regions $l$=0--120\deg, $b$$>$40\deg,
$l$=210--60\deg, $b$$<$$-$40\deg. In 58 sightlines low-velocity \H2\ absorption
is detected that does not contaminate the \OVIla\ absorption. In 15 of these the
highest rotational level we can discern is $J$=1, in 6 it is $J$=2, and in 24 it
is $J$=3. Finally, in 13 sightlines $J$=4 is present, but neither $J$=3 or $J$=4
contaminate the \OVI. For the latter 37 sightlines, instead of fitting other
$J$=3 or $J$=4 lines to find the parameters of the \H2\ lines near \OVIla, we
measured the parameters of the 6--0 P(3) and 6--0 R(4) lines by directly fitting
a gaussian to the absorption profile.
\par In 5 sightlines (Mrk\,357, Mrk\,876, PG\,0832+675, PG\,1116+215 and
PG\,1351+640) the $J$=3 and $J$=4 lines show two components. One is at a
velocity near 0\kms, and is associated with the low-velocity \HI\ component. The
other lies near $-$50\kms\ and is associated with intermediate-velocity \HI\
(having \vlsr=$-$90 to $-$40\kms). In fact, in about 50\% of the sightlines
toward which intermediate-velocity \HI\ is seen, intermediate-velocity \H2\
absorption is detected. This is described in a separate paper (Richter et al.\
2002).
\par We now discuss the 25 contaminated sightlines. First we note that for the
sightline to NGC\,3783 the contamination by \H2\ is so severe that we cannot use
this sightline for measuring Galactic \OVI\ (see Sembach et al.\ 2001a).
\par Five different kinds of \H2\ contamination occur.
\par (a) Toward Mrk\,116, NGC\,3310 PG\,1004+130 and PHL\,1811 the
positive-velocity edge of the Galactic absorption abuts the 6--0 R(4) line at
125\kms. In these cases the contamination is minor and mostly limits our ability
to precisely determine the extent of the \OVI\ absorption.
\par (b) Toward 3C\,273.0, ESO\,141$-$G55, Mrk\,509, Mrk\,734, and PG\,0844+349
the 6--0 R(4) line is present in the middle of a wide high-positive velocity
\OVI\ wing that merges with the Milky Way thick disk absorption. This \H2\ line
is always sufficiently weak that we are confident that the \OVI\ measurements
are still reliable.
\par (c) Toward Mrk\,290, Mrk\,506, Mrk\,876, Mrk\,1502 and PHL\,1811 there is a
HVC component between velocities of $\sim$$-$200 and $-$100\kms, and the
negative-velocity edge of that component merges with the \H2\ $J$=3 absorption
at $\sim$$-$214\kms, making the velocity range of the HVC component somewhat
uncertain. The column density measurement is only weakly affected, however.
\par (d) Toward Mrk\,304, Mrk\,1513, NGC\,7714, as well as UGC\,12163 the
positive-velocity edge of a HVC component at \vlsr$<$$-$250\kms\ gets somewhat
confused with \H2. However, this has only a minor influence on the measurement
of \OVI.
\par (e) Toward Mrk\,335, Mrk\,352, Mrk\,357, Mrk\,509, NGC\,595, NGC\,7469,
PG\,0052+251, and PG\,2349$-$014 a strong $J$=3 line at $\sim$$-$214\kms\ sits
between two HVC components with central velocities of $\sim$$-$300 and
$\sim$$-$180\kms. In these sightlines the fit to the other $J$=3 lines is
reliable enough to conclude that the absorption that is apparently centered near
the $-$214\kms\ is a mix of the 6--0 P(3) line and \OVI\ absorption. After
dividing out the \H2\ line, the \OVI\ absorption profile looks simple, as can be
seen from the apparent column density plots shown in Fig.~\Fspectra\ for these
sightlines.
\par In this category (e), an especially difficult case of contamination is
presented by H\,1821+643 (see its spectrum in Fig.~\Fspectra). Absorption at
velocities of $\sim$$-$100 to $-$160\kms\ is interpreted as associated with the
Outer Arm, a distant Galactic spiral arm (e.g., Kepner 1970, Haud 1992). For the
H$_2$ $J$=3 line we derive a wide (FWHM 35\kms) and deep (32\% absorption)
profile, which separates two \OVI\ features. The \OVI\ absorption extends up to
$-$285\kms\ on the negative-velocity side, and to about $-$160 \kms\ on the
positive-velocity side, where the \OVI\ absorption suddenly gets deeper.
However, because of the uncertainties in deriving the H$_2$ parameters, the
systematic errors in these components are large. \par The spectrum of
H\,1821+643 is further contaminated by a strong (saturated) \OVI\ line near
0\kms, which is associated with the environs of the planetary nebula K1-16,
which lies about 96\arcsec\ away (Savage et al.\ 1995). This absorption was
fitted and removed from the calculation of the Galactic $N$(\OVI).


\subsection{Extragalactic Absorption Lines}
A source of contamination other than that by Galactic H$_2$ is provided by
redshifted lines from the neutral and low-ionization ISM in some of the
background targets or from intervening objects. We refer to the former as
``intrinsic absorption'' and to the latter as ``intergalactic absorption''. We
searched the velocity range $-$1200 to 1200\kms. The lower limit is set by the
presence of geocoronal \OI* emission lines near 1028\aa, which make absorption
hard to detect at more negative velocities. The upper limit is set by the
low-velocity interstellar \CIIl\ line. In this velocity range we find 169
absorption features in 101 of the 119 sightlines with sufficient S/N ratio
(excluding \OVI\ and \H2\ lines). Of these, 37 can be identified with intrinsic
\Lyb\ absorption, 13 with intrinsic \OVI\ lines, 29 with other intrinsic
absorption lines, and 18 with interstellar \CII. In addition, toward NGC\,7673
we find two high-velocity \OVI\ features, which are clearly present, but whose
measurement is too uncertain because of the intrinsic wide \SiIIla\ line. This
leaves 61 features which are confirmed, probably or possibly intergalactic.
\par For ten features we did not manage to find a reasonable identification, for
several reasons. (a) Some are weak, are only seen in one channel and/or may not
be real, while at the same time they are not at the velocity expected for
intergalactic \Lyb\ based on the presence of galaxy groups in the sightlines or
we find no \Lya\ at the same velocity (toward HE\,0238$-$1904, Mrk\,304,
Mrk\,506, Mrk\,926 (2 features), Mrk\,1095, PG\,0947+396 and PG\,1211+143). (b)
The absorption at 290\kms\ toward HE\,1228+0131 is a confusing blend, of which
only some parts can be identified. (c) The feature at 425\kms\ toward
PG\,1211+143 cannot be \Lyb, and while it may be \OVI, this is impossible to
confirm. No other intergalactic lines fall near the velocity of the feature.
These ten unidentified features are indicated by a ``?'' in Fig.~\Fspectra.
\par We discuss the intrinsic lines in the next subsection, the intergalactic
\Lyb\ in the one after that, and then we summarize the remaining intergalactic
features.

\subsection{Removing Contamination by the AGN}
Here we discuss the expected and detected intrinsic absorptions in the velocity
range between $-$500 and 500\kms\ relative to \OVI. To check whether such
absorption can occur, we calculated the redshifted position for \SiIIla,
\OI\l988.773, \CIIIl, \SIII\l1012.501, \Lyb\l1025.722, Ly$\gamma$\l972.537,
Ly$\delta$\l949.743 and Ly$\epsilon$\l937.804, and we noted the objects for
which any of these lines would lie within $\pm$500\kms\ from the Galactic \OVI\
absorption. This is the case for the sightlines discussed below. For all other
sightlines confusion with intrinsic absorption does not occur. All velocities
listed below refer to those on the Galactic \OVIla\ velocity scale.
\par Intrinsic \SiII\ absorption is expected near \OVIla\ in 6 sightlines. It is
indeed seen near the expected velocity (but not contaminating \OVI) toward
NGC\,3690, NGC\,3783, NGC\,7714 and Tol\,1924$-$416. For NGC\,3991 the intrinsic
\SiIIla\ line lies at \vlsr=$-$110\kms, but is relatively weak and shows a
simple profile; we thus fit and remove it from the Galactic \OVI\ line. Toward
NGC\,7673, however, the intrinsic \SiIIla\ line is broad and strong. We also
detect very broad and strong intrinsic \OI\l988.773 and \SiIIlb\ lines. Using
that the ratio of optical depths of the two \SiII\ lines is 10.1, the expected
optical depth for the intrinsic \SiIIla\ line is on the order of 0.1--0.5.
Therefore, the continuum near \OVIla\ is too uncertain, leading us to discard
NGC\,7673 from the Galactic \OVI\ sample.
\par The intrinsic \OI\l988.773 line is expected to lie near Galactic \OVI\ in
three AGNs. A broad line, centered at 290\kms\ is seen toward Mrk\,54; this
causes sufficiently bad contamination of the Galactic \OVI\ that we discarded
Mrk\,54 from the final sample. The possible \OIlb\ line is not detected at
$-$175\kms\ toward Mrk\,506 or at $-$145\kms\ toward NGC\,985.
\par Intrinsic \CIIIl\ is seen at 500\kms\ toward 3C\,382.0, where it does not
contaminate Galactic \OVI. Toward ESO\,265$-$G23 intrinsic \CIII\ is expected at
$-$55\kms, but not seen. However, toward IRAS\,09149$-$6206 the intrinsic \CIII\
line is centered at 320\kms, very broad and confuses the positive-velocity wing
of Galactic \OVI\ in such a manner that we decided to discard this object from
our sample.
\par Toward ESO\,350$-$IG38 a wide intrinsic \SIII\l1012.51 overlaps most of the
Galactic \OVI\ absorption. Corresponding features are also seen in many other
low, medium and high-ionization lines. We therefore did not include
ESO\,350$-$IG38 in the final sample.
\par Intrinsic \Lyb\ occurs at $-$165\kms\ in the spectrum of NGC\,7496. This
line is so strong that it completely covers Galactic \OVI\ and makes the
spectrum unusable for the current project. Toward NGC\,3504 the intrinsic \Lyb\
line was expected at $-$277\kms, but there only seems to be weak \Lyb\
absorption centered near $-$440\kms. In many other cases we see strong intrinsic
\Lyb\ with broad damping wings, centered several \aa\ from 1032\aa. Except for 
NGC\,1395, NGC\,1404 and NGC\,1407 the flux near 1032\aa\ remains sufficiently
high to measure \OVIla.
\par Intrinsic Ly$\gamma$ might occur at 20\kms\ toward Mrk\,1502, and at
260\kms\ toward Ton\,S180, but no absorption is seen, nor are there
corresponding \Lyb\ lines.
\par Intrinsic Ly$\delta$ is not present at $-$15\kms\ toward Mrk\,1383, as is
confirmed by the absence of Ly$\gamma$\ and \Lyb. Toward PG\,1351+640, however,
two intrinsic Ly$\delta$ lines are clearly present at velocities of $-$560 and
$-$340\kms, covering the \OVI\ velocity range between $-$700 and $-$260\kms.
This almost abuts the high-negative velocity \OVI\ absorption associated with
HVC complex~C, but there does not appear to be contamination. A similar
Ly$\delta$ line stays to the short wavelength side of the Galactic \OVI\ line
toward PG\,1411+442 (centered at $-$575\kms, extending to $-$150\kms).
\par Intrinsic Ly$\epsilon$ is seen at $-$410\kms\ toward PG\,1404+226.
\par A special case is presented by Tol\,0440$-$381, where no known intrinsic
line overlaps Galactic \OVI, but where there is nevertheless an apparent
emission feature located between $-$400 and 0\kms\ on the \OVI\ velocity scale.
This unidentified feature strongly contaminates Galactic \OVI, and we therefore
do not include Tol\,0440$-$381 in the final sample.


\subsection{Removing Contamination by Nearby Galaxy Groups}
\subsubsection{Identifying Nearby Galaxy Groups}
To identify the intergalactic \Lyb\ lines associated with nearby galaxy groups
we used two checks. First, we searched for matching \Lya\ absorption for those
objects for which a longer wavelength UV spectrum was also available, taken with
one of the spectrographs on the {\it Hubble Space Telescope} -- either the
\GHRS\ {\it (Goddard High Resolution Spectrograph}, the \FOS\ {\it (Faint Object
Spectrograph)} or the \STIS\ {\it (Space Telescope Imaging Spectrograph)}.
Second, we checked whether the line of sight passes through a galaxy group, in
which case the likelihood of detecting \Lyb\ associated with that group is
increased.
\par To find galaxy groups, we overlaid the position of each object on a map of
galaxies from Tully's (1988) ``Nearby Galaxy Catalog''. This lists 2367 galaxies
with systemic velocities less than 3000\kms. Each of these was assigned to one
of 36 galaxy ``galaxy groupings''. Tully's terminology for these was ``clouds'',
but that could lead to confusion with interstellar features. These ``galaxy
groupings'' are much bigger than the more concentrated galaxy groups and
clusters usually discussed -- they may encompass several groups. A list of
intersected ``galaxy groupings'' is given in the notes to Table~\Tderived. In
the overlays we also included a circle showing the position and size of the 176
groups catalogued by Geller \& Huchra (1983, 1984). These authors listed the
central position, velocity and diameter of galaxy groups, defined as regions
with a galaxy density enhancement greater than 20, in a magnitude-limited sample
of galaxies brighter than m=13.2; this sample goes out to a velocity of
$\sim$8000\kms.
\par Columns 12, 13 and 14 of Table~\Tderived\ summarize the results. Column~12
gives a code classifying the \Lyb\ absorption line near \OVIla, as described
below; a detailed discussion of each individual case is given in the Appendix.
If the sightline intersects a galaxy group, Col.~13 of Table~\Tderived\ lists
the group number as given by Geller \& Huchra (1983, 1984), and Col.~14 the
``galaxy grouping'' number given by Tully (1988).

\subsubsection{Detected Intergalactic \Lyb}
\par The codes used in Column~12 of Table~\Tderived\ indicate the kind of
detection that was made. ``NoGrp'' implies that the sightline does not pass
through a nearby galaxy group (38 objects). ``GrpNoIGM'' is for sightlines that
do pass through a group, but where no extra features are seen in the spectrum
(57 objects), while ``Grp'' is used for the 46 sightlines passing through a
group but for which the spectrum has insufficiently high S/N ratio. We note that
since the detection limit strongly varies with the quality of the spectrum, it
is quite likely that \Lyb\ absorption associated with nearby galaxy groups could
be present in many of the objects given code ``GrpNoIGM''.
\par The entry ``Intr\Lyb??' is used for six objects that are part of a group in
the Tully catalogue, and where intrinsic \Lyb\ might be expected, but where the
S/N ratio is too low to see this. ``Intr\Lyb'' is given for the 39 objects where
an intrinsic \Lyb\ line with damping wings is present.
\par There are 40 intergalactic \Lyb\ absorptions detected toward 30 objects, at
varying levels of confidence. Three entries are coded ``NoGrp\Lyb??'' (3C\,382.0
-- 2 possible \Lyb, HE\,0226$-$4110 and PG\,2349$-$014). Here a fairly strong
feature is present, and it is possible that it is \Lyb, although the sightline
does not intersect a galaxy group, nor is a spectrum for \Lya\ available.
Entries ``Grp\Lyb??'' are given for PG\,1302$-$102 and PG\,1352+183). In these
low S/N sightlines there is a feature at a velocity similar to that of the
galaxy group that is intersected, but the spectrum is too noisy to conclude with
confidence that the feature is \Lyb\ (or sometimes even whether it is real).
Toward Mrk\,478 and PG\,1444+407 such a feature occurs in addition to a positive
identification. ``Grp\Lyb?' is given for four objects (Mrk\,106, Mrk\,304 -- 2
features, PG\,0832+251 and PG\,0947+396). These sightlines are projected onto a
nearby group, and a feature that is probably \Lyb\ is seen in the correct
velocity range, but there is no spectrum that would allow confirmation using
\Lya.
\par A code of ``Grp\Lyb'' is used for the 13 objects where intergalactic \Lyb\
absorption from a known intervening group is confirmed by a matching \Lya\ line
(3C\,273.0 -- 2 \Lyb; HS\,1543+5921; Mrk\,205; Mrk\,335 -- 2 \Lyb; Mrk\,478;
Mrk\,734 -- 2 \Lyb; Mrk\,817; Mrk\,876; NGC\,7469; PG\,1004+130; PG\,1211+143;
PG\,1259+593 -- 2 \Lyb; and PG\,1444+407). For several of these objects, the
\Lya\ absorption line was previously shown by Penton et al.\ (2000). Finally,
confirmed \Lyb\ not associated with a group is found toward Mrk\,509, NGC\,4670
and VII\,Zw\,118 (code ``\Lyb'').
\par Toward five objects the confirmed intergalactic \Lyb\ is at a velocity such
that it contaminates the \OVIla\ line (3C\,232, HE\,1228+0131 -- 3 \Lyb,
Mrk\,771, PG\,1048+342 and PG\,1216+069) -- code ``Contam\Lyb''. Only in the
case of Mrk\,771 was it possible to use the \OVIlb\ line to measure $N$(\OVI).
\par In addition to the unidentified features, the identified intrinsic
absorption lines, and the identified or probable \Lyb\ absorptions, 21 other
features are found in 18 sightlines. Three of these are high-redshift Ly$\gamma$
(toward HS\,0624+6907, PG\,1211+143 and PG\,1415+451) or \CIIIl\ (toward
Mrk\,1513 and PG\,1211+143). Another three are high-redshift \CII\ll903.624,
903.962 (toward PG\,1116+215 and PG\,1302$-$102). These identifications are based
on the presence of many other Lyman lines at the same redshifts. The remaining 13
features are confirmed or probable low-redshift ($v$$<$1200\kms) \OVI\
absorbers. These will be described in a separate paper.


\section{MEASURING THE \OVI}

\subsection{Velocity Range}
\subsubsection{Determining Velocity Limits}
Of the 219 objects in the sample, 121 were deemed to have a signal-to-noise
ratio sufficiently high to be of interest (S/N$>$3 per resolution element after
combining the LiF1A and LiF2B segments). Of these 121 objects, Mrk\,153 and
PG\,1011$-$040 were eliminated because the data were obtained as part of guest
observer programs with science goals similar to those of our program. Six were
eliminated because the continuum placement was too uncertain (NGC\,604,
NGC\,3690, NGC\,4214, NGC\,5236, NGC\,5253, NGC\,5461). Finally, eleven more
were eliminated because Galactic, intrinsic or intergalactic absorption either
was too confusing or overlapped the \OVIla\ line (ESO\,350$-$IG38,
IRAS\,09149$-$6206, HE\,1228+0131, Mrk\,54, NGC\,592, NGC\,3783, NGC\,7496,
NGC\,7673, PG\,1048+342, PG\,1216+069, Tol\,0440$-$381). This left a sample of
100 extragalactic and 2 stellar objects for which we could measure the Galactic
\OVI\ lines.
\par Toward most objects in our final sample, we find a strong \OVIla\ line
centered near a velocity of 0\kms. This component clearly is associated with the
Milky Way. In about two-thirds of the sightlines a second component is also
found, at velocities \vlsra$>$120\kms. Differential Galactic rotation can
produce velocities up to about $\pm$150\kms\ in the Galactic plane, depending on
the Galactic longitude; lower maximum velocities are expected at high latitudes.
Therefore, any \OVI\ absorption at high velocities cannot be part of the
rotating Galactic disk. Fortunately, the high-velocity component and the Milky
Way absorption can usually be fairly cleanly separated from each other.

\subsubsection{Distribution of Velocity Limits}
\par The top panel of Fig.~\Fvlim\ shows the distribution of the minimum and
maximum velocity to which the Milky Way absorption extends. The thin solid line
gives the histogram for all sightlines, the thick line excludes the eleven
sightlines for which only an upper limit could be derived, as well as the 26
sightlines where it was difficult to separate the Milky Way and high-velocity
absorption. These distributions show that both at the negative and the positive
velocity edge the Galactic \OVI\ absorption typically extends to anywhere
between $\pm$50 and $\pm$120\kms, with $\pm$90\kms\ as the average. The spike at
100\kms\ is artificial and has to do with some of the sightlines where the
separation between the Milky Way and high-velocity absorption was difficult (see
below).
\par The bottom two panels of Fig.~\Fvlim\ show the velocity limits for the
high-velocity \OVI\ components. Again, the thin solid line includes all
sightlines, while the thick solid line excludes the difficult ones. This
distribution shows that \OVI\ absorption with \vlsra=100 to 400\kms\ is common,
and has no preferred velocity limits. We find only eight sightlines with \OVI\
at velocities between 400 and 1200\kms, and in all cases this can be identified
as intergalactic.
\par For the majority of sightlines (80 out of 102) the Galactic \OVIla\ line
extends no further than $\pm$120\kms\ and in 58 of these it is easily separated
from high-velocity absorption. In only five sightlines does it extend slightly
further (to $\pm$130--140\kms\ (HE\,0226$-$4110, Mrk\,209, NGC\,3310,
PG\,1004+130 and PKS\,0558$-$304). For the eleven sightlines where we can only
set an upper limit, a velocity range that seemed reasonable was chosen to
determine the detection limit (HE\,0450$-$2958, HE\,1115$-$1735,
HE\,1326$-$0516, Mrk\,205, Mrk\,926, NGC\,588, NGC\,595, NGC\,3504,
PG\,0052+251, PG\,2349$-$014 and SBS\,0335$-$052). Three sightlines lie
projected on the Outer Arm, a distant Galactic spiral arm (Habing 1966, Kepner
1970, Haud 1992). In the direction of H\,1821+643, differential Galactic
rotation may produce velocities for gas in the outer galaxy as high as the
$-$145\kms\ seen in the \OVI\ line. Similarly, toward 3C\,382.0 velocities as
high as $-$130\kms\ can easily be understood. Thus, for these two cases, the
absorption over the full observed velocity range was considered to be Galactic.
In the third sightline crossing the Outer Arm (HS\,0624+6907) the \OVI\
absorption does not extend as far as the \HI.

\subsubsection{Sightlines with a Difficult Separation}
\par For 26 objects the separation between the Galactic and high-velocity
component was not clear-cut. These fall into five categories.
\par (a) Nine sightlines lie projected onto the HVC complex~C, which is a large,
infalling cloud with a metallicity of $\sim$0.1 solar, lying more than 4\kpc\
above the Galactic plane (e.g., Wakker 2001). In \HI\ 21-cm emission, complex~C
is seen at velocities between $-$180 and $-$100\kms. In each of these
directions, the \OVI\ profile extends out to rather high negative velocities
($-$150 to $-$260\kms). Clearly, there is \OVI\ absorption associated with HVC
complex~C. In 5 of the 9 sightlines, there appear to be two components in the
\OVI\ profile (Mrk\,279, Mrk\,290, Mrk\,817, PG\,1259+593 and PG\,1626+554), one
of which more or less covers the velocity range of the \HI. We then placed a cut
between the Galactic and HVC absorption at about the velocity that separates the
\OVI\ components. In the remaining 4 directions (Mrk\,501, Mrk\,506, Mrk\,876
and PG\,1351+640) this cannot be done. Instead, we decided to cut at a velocity
of $-$100\kms, which seems a reasonable compromise, especially since this
velocity is among the most common ones for directions where the extent of the
Galactic \OVI\ absorption is clear (see Fig.~\Fvlim).
\par (b) Two sightlines intersect HVC complex~A, known to have a z-height of
5.7$\pm$1\kpc\ (Wakker 2001). In one of these sightlines (Mrk\,116), the \OVI\
extends to about $-$125\kms, whereas the \HI\ emission lies between $-$205 and
$-$125\kms. So, no separate HVC component was included. In the other sightline
(Mrk\,106), the \OVI\ goes to $-$150\kms, while the \HI\ lies between $-$190 and
$-$120\kms, as well as between $-$70 and 30\kms. In this case, the \OVI\
absorption between $-$150 and $-$100\kms\ is presumed to have some association
with complex~A.
\par (c) A group of eleven sightlines in the quadrant $l$$>$180\deg,
$b$$>$0\deg\ show extended positive-velocity wings in \OVI, with velocities up
to 275\kms. An \OVI\ component can be discerned in six of these (ESO\,572$-$G34,
HS\,1102+3441, Mrk\,734, Mrk\,1383, PG\,0947+396 and PG\,1116+215), separated
from the Galactic absorption at velocities of 100, 95, 140, 100, 100 and
115\kms, respectively. For the other five sightlines (IRAS\,F11431$-$1810,
Mrk\,421, PG\,0953+414, PG\,1001+291 and Tol\,1247$-$232), there is no such
separation, and instead we decided that the best we could do is to cut between
the Galactic and high-velocity absorption at a velocity of 100\kms.
\par (d) In part of the southern sky ($l$$<$180\deg, $b$$<$0\deg) many
sightlines exhibit \OVI\ absorption at high negative velocities. For two of
these (Mrk\,335, and PKS\,2155$-$304) the separation between Galactic and
high-velocity absorption is not clear-cut. We decided to separate the two at
velocities of $-$75 and $-$85\kms, respectively, since a) at these velocities
the absorption appears minimal and b) similarly low negative limits occur toward
other sightlines in this part of the sky: MRC\,2251$-$178 ($-$65), Mrk\,304
($-$40), Mrk\,1095 ($-$35), Mrk\,1502 ($-$35), Mrk\,1513 ($-$75), NGC\,7469
($-$65), NGC\,7714 ($-$60), PHL\,1811 ($-$65).
\par (e) Finally, in the directions of Fairall\,9 and PKS\,2005$-$489, there is
no separation between the Galactic and high-velocity component. The latter
extends out to velocities of 275 and 225\kms, respectively, which is clearly
incompatible with differential Galactic rotation. For PKS\,2005$-$489, we
decided to place a cut at 120\kms, where at least there is an inflection in the
profile. For Fairall\,9, we arbitrarily selected 100\kms\ as the place to cut.
\par For each object, Cols.~17 and 18 of Table~\Tderived\ list the final
velocity ranges within which we decided that \OVI\ absorption is associated with
the Milky Way. An ``x'' precedes or follows values where the separation between
the Galactic and high-velocity component is not clear, but where it appears to
be possible to discern separate components. A pair of ``['' and ``]'' precedes
or follows values where the separation between the Galactic and high-velocity
absorption was based on a rule rather than on features in the spectrum, as
described in detail above.
\par In Cols.~21 and 22 of Table~\Tderived\ we list the velocity ranges used to
measure the parameters of the high-velocity \OVI\ absorption. Column 20 gives
the classifications for the high-velocity gas, which will be discussed in detail
by Sembach et al.\ (2002b).


\subsection{Equivalent Width Measurement}
\subsubsection{The Equivalent Width and its Errors}
Having identified the velocity range of the absorption associated with the
Galaxy and with high-velocity gas, we next calculated equivalent widths by a
straight integration of the line profile: $$
  {\rm W} = \int_{\lambda_{\rm min}}^{\lambda_{\rm max}}\ \left( 1 - {F_\lambda\over C_\lambda} \right) d\lambda,
$$ where $F_\lambda$ is the observed flux, $C_\lambda$ is the continuum flux and
$\lambda_{\rm min/max}$ are the wavelengths corresponding to the velocity range
of the absorption. The resulting equivalent widths are listed in Cols.~19 and 23
of Table~\Tderived.
\par We calculate five contributions to the error in the equivalent width:
random noise, the continuum fit, fixed-pattern noise, uncertainties in the
integration range and uncertainties in removing \H2\ contamination. The
random-noise and continuum-fit error are combined in quadrature into a
``statistical error'', while the fixed-pattern error, the velocity-limits error
and the decontamination error are combined in quadrature into a ``systematic
error''. These are both included in Col.~19 of Table~\Tderived\ for the Milky
Way component and in Col.~23 for the high-velocity absorption. The statistical
error indicates how accurately the measurement can be made, while the systematic
error indicates how much the listed equivalent width could be offset from the
actual value.
\par The \FUSE\ calibration pipeline provides an error associated with random
noise in the flux for each pixel. This is mainly the poisson error (square root
of counts), but a

\noindent small background error is also included. This yields an error
in the equivalent width of: $$ 
\sigma_{\rm W,noise} = \sqrt{\left(  \Sigma \left[ {\delta F_\lambda\over C_\lambda} \,d\lambda \right]^2  \right)}.
$$ The second error contribution comes from the uncertainty in the placement of
the continuum. The method of fitting Legendre polynomials (Sembach \& Savage
1992) produces an error in the coefficients and this can be converted to an
error in the placement of the continuum for each pixel. Since these errors are
correlated, the associated error in the equivalent width becomes: $$
\sigma_{\rm W,cfit} = \Sigma\left[ {\delta C_\lambda \over C_\lambda}\ {F_\lambda\over C_\lambda} \right].
$$
\par The third error is from the detector ``fixed-pattern noise''. This takes
into account the possibility that there are pixel-to-pixel sensitivity
variations in the detector. However, because the placement of the spectrum on
the detector can vary, there can be variations in the correspondence between a
particular pixel on the detector and a particular wavelength. Thus, a
fixed-pattern feature is not always present at the same wavelength. By studying
the good-quality spectra and noting that sometimes there appear to be weak,
narrow features present (e.g., at $-$360\kms\ toward 3C\,273.0) and measuring
these features, we estimate that a reasonable fixed-pattern error is
$\sim$6.8\maa, which is the equivalent width of a 10\% absorption that is one
resolution element wide.
\par Fourth, there are multiple possible ways in which the velocity limits of
the integration can be uncertain, especially for noisy spectra and for
sightlines where the separation between the high-velocity and Milky Way
absorption is not clear. We estimate this contribution to the systematic error
as: $$
\sigma_{\rm W,vlim} = \vert {\rm W}(v_--15)-{\rm W}(v_-+15)\vert\ +
$$$$
\hskip1.3cm           \vert {\rm W}(v_+-15)-{\rm W}(v_++15)\vert,
$$ where $v_-$ gives the negative side of the integration and $v_+$ the positive
side.
\par Finally, for sightlines where the \OVI\ absorption is contaminated by \H2\
we systematically varied the parameters of the \H2\ lines (by 10\kms\ in width,
20\% in depth, 10\kms\ in velocity), and then recalculated the equivalent width.
We then find the error associated with decontamination as half of the difference
between the most extreme values. For sightlines where the Milky Way component or
the positive-velocity HVC are contaminated by the 6--0 R(4) line at 1032.356\aa,
this error is at most 5\maa. In the three sightlines (Mrk\,290, Mrk\,506 and
Mrk\,876) where the \OVI\ absorption associated with complex~C abuts the
6--0 P(3) line at 1031.191\aa, the decontamination error is also $<$5\maa.
However, the decontamination error is large for the high-negative velocity \OVI\
absorption on all twelve sightlines in the region $l$=30\deg\ to 130\deg,
$b$=$-$60\deg\ to $-$25\deg. Usually it lies in the range between 12 and 25\maa,
but it can be as high as 36\maa\ (for Mrk\,352).

\subsubsection{Error Distributions}
\par Figure~\Ferrhist\ shows the distributions of the measured equivalent
widths, random-noise errors, continuum-fit errors and velocity-limits errors.
Different panels are given for spectra of different quality, separately for
Galactic and high-velocity \OVI. For the thick disk the range in the measured
equivalent widths is independent of the spectral quality, giving confidence in
the measurements at low S/N ratios. Equivalent widths for the high-velocity
component are generally lower, and fainter components are detected in higher
quality spectra, suggesting that we are missing a substantial fraction of the
high-velocity \OVI\ in the lower S/N spectra. This is discussed in more detail
in \Sect\SHVCdet.
\par In Fig.~\Ferrcorr\ we plot the different errors against each other. The top
panel shows that the error associated with random noise (\sig{noise}) correlates
with that associated with the continuum fit (\sig{cfit}). The different symbols
indicate different orders for the local continuum (see \Sect\Scontinuum). If
this order is 1 (plus symbols in Fig.~\Ferrcorr) the continuum-fit error is
0.30$\pm$0.02 times the random-noise error (with correlation coefficient 0.86),
while if the fit order is 2 (filled circles in Fig.~\Ferrcorr) the ratio of the
errors is 0.66$\pm$0.04 (with correlation coefficient 0.89). For higher order
fits (open circles) the ratio is 0.34$\pm$0.11 with correlation coefficient
0.66.

\noindent Thus, on average, including the error in the continuum placement is
equivalent to increasing the random-noise error by a factor
$\sqrt{1+0.3^2}$=1.04 (order=1) or a factor $\sqrt{1+0.66^2}$=1.2 (order=2).
However, the ratio is not a constant, and there are ten sightlines where the
ratio is $>$0.8. The ratio also depends on the order of the fit. Therefore, the
continuum-fit error needs to be calculated separately for each spectrum.
\par The random-noise and continuum-fit errors correlate with spectral quality
(by definition). They range from $\sim$30--60\maa\ at Q=1 to $\sim$5--15\maa\ at
Q=4. As expected, the systematic error associated with the selection of the
velocity integration range is more or less independent of spectral quality. It
also has a much wider distribution, varying from low values of $\sim$10\maa\ to
as much as 60\maa.
\par Panel b of Fig.~\Ferrcorr\ compares the continuum-fit error determined from
fitting Legendre polynomials to the error that follows from an alternative
method that is often used. In this method the continuum is shifted up and down
by one-third of the rms fluctuations in the flux, and the continuum placement
error is found as half of the difference of the two resulting equivalent widths.
Figure~\Ferrcorr b shows that this error, \sig{rms/3} is on average 1.7 times
the Legendre fit error if the polynomial fit has order 1 (with correlation
coefficient 0.9), but also that it is on average only 0.8 times as large (with
correlation coefficient 0.9) for a fit order of 2. Thus, the $\pm$rms/3 method
of estimating continuum errors tends to overestimate these for flat continua,
but gives reasonable answers for continua fit with a second-order polynomial.
\par Finally, in Fig.~\Ferrcorr c we show that there is no correlation between
the random-noise error and the velocity-limits error. The correlation
coefficient is only 0.12. This is to be expected, as the velocity-limits error
is a measure of the difficulties with defining the velocity integration limits
intrinsic to the spectrum, while the random-noise error is simply a measure of
data quality.


\subsection{Column Density Measurement}
\subsubsection{Background Information}
To derive column densities, we used the apparent optical depth method (Savage
\& Sembach 1991), which is applicable if the spectral line to be measured is
unsaturated and nearly fully resolved. Using this method, the total column
density is measured by first converting each absorption profile to an apparent
optical depth profile:
$$ N_a(v)={m_e c\over \pi e^2}\,f\,\lambda\ \ln{C(v)\over F(v)}, $$
where $C(v)$ is the continuum as a function of $v$, $F(v)$ the observed flux,
and the oscillator strength $f$ of the \OVIla\ line is 0.133 (Morton 1991). This
is then integrated over the chosen velocity range. The error in the column
density is derived in a manner similar to that described above for the
derivation of the equivalent width error. The resulting column densities are not
listed in this paper, but in the companion papers that discuss in detail the
Galaxy's thick disk (Savage et al.\ 2002) and the high-velocity \OVI\ (Sembach
et al.\ 2002b).
\par In those papers we also list a measure of the position and width of the
\OVI\ absorption. These are calculated as: $$
\bar{v}_{\rm obs} =           \int{ {  v                     \,N_a(v)\,dv } \over N_a };\ \ \
                b = \left[ 2\ \int{ { (v-\bar{v}_{\rm obs})^2\,N_a(v)\,dv } \over N_a }\right]^{1/2}. $$
The average velocity is the column density weighted mean. The width parameter,
$b$, is numerically similar to the Doppler spread parameter. See Savage et al.\
(1997) for more discussion of the usefulness of $b$. The
\noindent ratio between the FWHM
and $b$ is FWHM=$\sqrt{4\ln2}b$. The values and distribution of
$\bar{v}_{\rm obs}$ and $b$ are discussed in the companion papers.

\subsubsection{Comparing the Two \OVI\ Lines}
\par If the $N_a(v)$ profiles of the two lines of the doublet match, one
concludes that there is no saturation. If they do not match, there probably is
saturation or some other problem with the observation or the placement of the
continuum. We checked the \OVIlb\ lines for all 60 objects with S/N ratio
$>$6.5. For 20 Milky Way and 2 HVC components it was possible to calculate the
ratio of the column density derived from the \OVIla\ line to that derived from
the \OVIlb\ line, although the velocity range over which the comparison is
possible is usually narrower than that used to calculate the total \OVI\ column
density. The velocity ranges and the ratio of the \OVIla\ and \OVIlb\ column
densities are given in the Appendix. For the other sightlines with S/N ratio
$>$6.5 the comparison between the two \OVI\ lines is strongly affected by the
uncertainties introduced by a curvature in the continuum or by the encroachment
of \H2\ lines. In two cases where saturation appears present (Mrk\,771 and
Mrk\,876) the \OVIlb\ line is in fact contaminated by weak \Lyb\ absorption at
3485\kms; the corresponding \Lya\ lines are clearly detected in the \STIS\
spectra of these objects.
\par Figure~\Foviratio a shows the ratio for the 22 components, with the error
calculated by combining the random-noise error and continuum fit errors, but
ignoring the systematic errors. For 17 of the 22 components, the ratio is 1
within the 1$\sigma$ error bar. In the one direction where the ratio is large
(1.60$\pm$0.18 for Mrk\,421) the profile comparison (see Fig.~\Fspectra) shows
that there is some saturation, though the \OVIlb\ line is rather noisy. For
PG\,1259+593, Mrk\,817, Mrk\,1383 and 3C\,273.0 the saturation is less obvious,
but the continua are defined well enough that the ratio calculation is
trustworthy, and thus we conclude that there is slight saturation toward these
sightlines.
\par In Fig.~\Foviratio b we plot the ratio against the column density derived
from the \OVIla\ line. This shows that the five sightlines with slightly
saturated \OVI\ do not seem to concentrate toward high \OVI\ column densities.
We therefore conclude that higher values of $N$(\OVI) are not systematically
more affected by saturation.
\par Since in the majority of cases there are no significant differences between
the column density derived from the weaker \OVIlb\ line with that derived from
the stronger \OVIla\ line, we conclude that it is justified to integrate just
the \OVIla\ line to derive column densities. For about 20\% of sightlines some
saturation may occur, which would result in corrections on the column density on
the order of a factor 1.2, or 0.08 in the log. This is about twice the typical
statistical error for high quality sightlines, but comparable to the typical
statistical error for sightlines with Q=2. In the subsample of 20 sightlines
where a check is possible, we found only one sightline where stronger saturation
appears to be present.
\par Thus, although there will be a (limited) number of sightlines where
saturation occurs, and where the derived column density is slightly in error,
the differences are sufficiently small that they should not affect any of the
scientific conclusions derived by Savage et al.\ (2002) and Sembach et al.\
(2002b).


\section{RESULTS}

\subsection{Detection Rate for Galactic \OVI}
We have detected Galactic \OVI\ absorption along almost all of our sightlines.
For the 26 objects with Q=4 the statistical error in the equivalent width lies
between 6 and 15\maa. The Galactic \OVI\ absorption is detected in every
sightline with an S/N ratio greater than 7\sig{stat} (the lowest value is
97$\pm$13\maa\ toward NGC\,7469). The 23 objects with Q=3 have a statistical
error between 14 and 25\maa, and \OVI\ absorption is detected at at least the
6\sig{stat} level toward each of these. The lowest value is 82$\pm$12\maa\
toward Mrk\,1095.
\par At the Q=2 level (30 objects, statistical error 22--49\maa) Galactic \OVI\
is detected above the 3\sig{stat} level in 24 cases. The 2.8\sig{stat}
(75$\pm$27\maa) detection toward NGC\,7714 appears real, but toward
HE\,0450$-$2958, NGC\,588, NGC\,595, PG\,0052+251 and SBS\,0335$-$052 we can
only set 3\sig{stat} upper limits of 126, 147, 126, 117 and 123\maa,
respectively. However, in 3 sightlines with Q=4, 1 with Q=3 and 4 with Q=2 an
equivalent width $<$125\maa\ is measured. Thus, although Galactic \OVI\ may
still be present in the three directions with upper limits, $N$(\OVI) must be
relatively low.
\par For the 23 objects with Q=1 the equivalent width statistical error is
29--65\maa. Galactic \OVI\ is detected above the 3\sig{stat} level in 16 of
these. The 2.6\sig{stat} detection toward Mrk\,290 also appears real. The
2.5\sig{stat} detection toward Mrk\,771 is based on the \OVIlb\ line. For the
remaining six objects (HE\,1115$-$1735, HE\,1326$-$0516, Mrk\,205, Mrk\,926,
NGC\,3504 and PG\,2349-014) we can only set 3\sig{stat} upper limits of 115,
174, 195, 216, 192 and 117\maa, respectively. Compared to the minimum values
detected in the sightlines with Q=3 and Q=4, the limits for HE\,1115$-$1735 and
PG\,2349$-$014 are still low enough that it is clear that W(\OVI) must be
relatively small. For the other four the error is so large that these
non-detections are not really significant, as there are 30 directions toward
which W(\OVI)$<$200\maa.
\par Of the directions with clear detections, the ratio of the equivalent width
to its systematic error is $<$4 in only six cases. In five of these
(IRAS\,F11431$-$1810, MRC\,2251$-$178, Mrk\,501, Mrk\,734 and NGC\,7714, it is
the velocity-limit error that dominates the systematic error, because the
separation between the high-velocity and Milky Way absorption is not sharp. For
UGC\,12163 the equivalent width itself is rather low.
\par In summary, we clearly detect Galactic \OVI\ in 91 out of 102 directions.
For six of the directions yielding an upper limit $N$(\OVI) must be low
($<$125\maa). Since there are eight sightlines with a significant detection
where the equivalent width lies below 125\maa, slightly better data might yield
detections. We thus conclude that \OVI\ is present in basically all directions.


\subsection{Detection Rate for High-Velocity \OVI}
High-velocity \OVI\ with \vlsra$>$100\kms\ is found along many sightlines. Here,
we discuss the detection rate as function of Q. Thirty five high-velocity
components occur in 22 of the 26 sightlines with Q=4, with equivalent widths
ranging from 14 to 200\maa. Only toward NGC\,1068, PG\,0804+761, VII\,Zw\,118
and vZ\,1128 is high-velocity \OVI\ not found. Fourteen of these high-velocity
\OVI\ components have a relatively small systematic error (ratio of equivalent
width to \sig{syst}$>$3). Another 15 are clearly significant (ratio of
equivalent width to statistical error $>$4), but with a relatively large
systematic error (W/\sig{syst}=1.2--3). The latter is usually dominated by
uncertainties in the velocity limits of the integration, sometimes by the
removal of the $J$=3 or 4 \H2\ line(s). Since these components are clearly wider
than a single resolution element and/or they form a clear wing that continues
the Galactic absorption, they are therefore real, but difficult to measure
accurately. Six components at high-positive velocities are both weak
(14--37\maa, 2--3.5\,\sig{stat}) and have a large systematic error (toward
3C\,273.0, H\,1821+643, Mrk\,421, Mrk\,876, PG\,1116+215 and PG\,1259+593) and
some of these components may not be real.
\par Seventeen high-velocity components occur in 12 of the 23 sightlines with
Q=3, with equivalent widths ranging from 30 to 265\maa. Ten of these are both
statistically highly significant and have a small systematic error
(W/\sig{syst}$>$3.0). Six are strong, but with a relatively large systematic
error (W/\sig{syst}=1.4--3.0), which is dominated by uncertainties in the
integration limits. The 125\kms\ component toward PG\,1351+640 is weak
(31$\pm$10\maa) and has a high systematic error (29\maa) because the lower
velocity limit is rather uncertain; it is probably real, however.
\par Twenty high-velocity components occur in 14 of the 30 sightlines with Q=2,
with equivalent widths ranging from 65 to 300\maa. All but three have
W/\sig{stat}$>$4, all have W/\sig{syst}$>$1.4, and all appear real. The ones
with the lowest ratio of W/\sig{syst} (Mrk\,357, Mrk\,734, Mrk\,501, NGC\,7714,
PG\,0052+251) are clearly present but hard to measure accurately because of \H2\
contamination.
\par Thirteen high-velocity components occur in 11 of the 23 sightlines with
Q=1, with equivalent widths ranging from 65 to 290\maa. Five of these are
clearly significant -- the ratios of both W/\sig{stat} and W/\sig{syst} are
$>$4. Another four are clearly present (W/\sig{stat}$>$3.5) but hard to measure
because of uncertainties in the integration limits or the removal of \H2\
contamination. Of the remaining five HVC components, one is strong ($>$100\maa)
and clearly real (toward NGC\,3991). The most negative-velocity HVC toward
PG\,2349$-$014 is weak, has low S/N and a large systematic error, mostly due to
uncertainties in \H2\ decontamination, but is probably real. The remaining two
ccomponents (toward Mrk\,106 and PG\,1001+291) may not be real, however.
\par Table~\Tdetratio\ lists the detection rate (number of components with
W$>$3\sig{stat}) of high-velocity \OVI\ for different minimum relative
systematic errors and different equivalent width limits. This table shows that
for a given equivalent width limit the detection rate is quite constant as
function of Q. Strong HVC absorption (EW$>$200\maa) occurs in 10\% of the
sightlines, whereas Galactic absorption of such strength occurs in 60\% of the
sightlines. And while 90\% of the sightlines show Galactic absorption stronger
than 65\maa\ (and \OVI\ is likely to exist at this level in the eleven
sightlines yielding an upper limit), only 47\% of the sightlines have
high-velocity \OVI\ that strong. However, in the best quality (Q=3 and 4)
spectra, the detection rate increases to 70\% (12+22 out of 23+26) for an
equivalent width limit of 30\maa. Even weaker components (down to 14\maa) are
detected in five sightlines (3C\,273.0, Mrk\,817, Mrk\,876, PG\,1116+215 and
PG\,1259+593). However, for components this weak it is sometimes no longer clear
whether they are real or whether they are a fixed-pattern noise feature.
Nevertheless, there are 13 sightlines in which high-velocity \OVI\ could have
been detected at the 20\maa\ level (S/N$>$18 per resolution element), and it is
found in 12 of these (92\%). It is notable that the only one of these sightlines
without a HVC is the direction toward vZ\,1128, which is a star at $z$=9.5\kpc.
This suggests that the high-velocity \OVI\ may be distant.
\par In summary, it appears that the detection rate for high-velocity \OVI\
strongly increases with the equivalent width limit. It may even reach 100\% at a
sufficiently low equivalent width limit (possibly $\sim$10\maa, or
$\sim$8\tdex{12}\cmm2).


\subsection{Velocity Distribution of \OVI\ Components}
In Fig.~\Fvelhist\ we show the distribution of the central velocities of all
measured components, including both the Galactic and the high-velocity \OVI. The
top panels plot the column density against the central velocity, and the
$b$-value codes the size of the symbols. This shows that the average and minimum
value of the thick disk absorption clearly are larger than those for the
high-velocity absorption, but also there are many high-velocity components that
are stronger than the Galactic absorption in some sightlines.
\par It is notable that the distribution of points between $\pm$100\kms\ is
qualitatively different than that at higher velocities. This may be due to the
blending of many weaker components in that velocity range; i.e.\ the Galactic
absorption is probably not due to a single absorber, but to a mixture of several
in the line of sight. This also implies that at low LSR velocities there
probably are many hidden absorption components that are related to the same
phenomenon that produces the high-velocity \OVI.
\par The average column density of the Galactic \OVI\ components is
14.33$\pm$0.20 (the median is 14.36), while for the high-velocity \OVI\ the
average is 13.95$\pm$0.34 (with median 13.97). If the column density
distribution of HVC components represents the real distribution of \OVI\ clouds
in the Milky Way, then on average the low-velocity absorption represents a
mixture of two to three absorbing clouds in the sightline.
\par Figures~\Fvelhist b to f present the histograms of the distribution of the
average velocity, $\bar{v}_{\rm obs}$, for several different equivalent width
ranges. There does not appear to be a relation between the strength of the
absorption and its velocity: both weak and strong high-velocity \OVI\ occur at
all velocities. Figures~\Fvelhist i to n present the same histograms, but
concentrating on the Milky Way absorption. Again, there does not appear to be a
correlation of strength with velocity, although in this case that was also not
expected.


\subsection{Detecting \OVI\ in Local Group Galaxies?}
For all Local Group galaxies in the list of Mateo (1998) we determined whether a
sightline in our sample passes within 50\kpc\ of such a galaxy, giving the
possibility of searching for \OVI\ in its extended halo.
\par In the case of M\,31, \FUSE\ program Z002 (PI Wannier) aimed at observing
several QSOs near it. These five sightlines (as well as HS\,0035+4405) all pass
within 60\kpc\ from M\,31, but there is insufficient flux for all but one object
(RX\,J0048.3+3941). The spectrum of the short (8.6~ks) observation of this faint
object has low S/N ratio (2.3 per resolution element), and we did not include it
in our sample. Nevertheless, we can see that the Milky Way absorption is weak,
and that a very strong (300$\pm$80\maa) high-velocity feature is present at
\vlsr=$-$180\kms\ (which is similar to the velocity of M\,31). However, a
high-negative velocity \OVI\ component is seen in all directions in this part of
the sky. Toward RX\_J0048.3+3951 it is much stronger than most of the other such
components, which typically are about 130\maa. Therefore, this component
probably is a mixture of the extended \OVI\ feature and absorption due to M\,31.
A longer observation of this QSO would clearly be useful.
\par Some of our sightlines also pass near one of the companions of M\,31
(NGC\,147, NGC\,185, NGC\,205, And\,I--VII, LGS\,3). Since these are dwarf
elliptical or dwarf spheroidal galaxies (diameter $<$1\kpc) without much gas,
none of the sightlines in our sample come close enough to be interesting (the
smallest impact parameter is 5.7\kpc).
\par Four of the objects in our sample are actually \HII\ regions within M33 --
NGC\,588, NGC\,595, NGC\,592 and NGC\,604. The flux we see is an amalgan of the
continua of many O stars. Nine individual O stars inside M\,33 were also
observed by \FUSE\ (PI Hutchings). Although each of these objects has a rather
difficult continuum, very strong \OVI\ absorption is seen in all sightlines at
velocities of $-$200\kms, and this absorption is probably associated with M\,33.
A future study of \OVI\ in M\,33 seems warranted.
\par There are four sightlines in our sample that pass within 150\kpc\ from
M\,33. The one with the smallest impact parameter (3C\,48, 2\fdg6 or 38\kpc) is
faint and the spectrum has low S/N ratio. A longer integration could be helpful.
The sightlines toward Mrk\,352, Mrk\,357 and PG\,0052+251 pass 7\fdg4 (90\kpc),
7\fdg9 (96\kpc) and 10\fdg9 (133\kpc), respectively, from the center of M\,33.
In all three, high-negative velocity \OVI\ absorption is seen, just as is the
case in all sightlines in this part of the sky. However, the rotation field of
M\,33 (e.g., Deul \& van der Hulst 1987) is such that velocities near or above
0\kms\ are expected at the positions of these background targets. No absorption
is seen at such velocities down to 3$\sigma$ limits of log $N$(\OVI)=14.08,
13.90 and 13.95, respectively. Considering this, it appears that any extended
\OVI-containing halo around M\,33 is smaller than 100\kpc.
\par A number of sightlines in our sample pass within 10\kpc\ from one of the
nearby dwarf ellipticals and dwarf spheroidals in the northern sky. 3C\,351.0
and PG\,1626+554 lie 3\fdg4 and 7\fdg6 (5 and 10\kpc) from Draco
($v$=$-$293\kms, distance $d$=82\kpc, full diameter as given in \NED\
$D$=0.8\kpc). Mrk\,876, Mrk\,279 and PG\,1351+640 pass $\sim$7\deg\ (8\kpc) from
Ursa Minor ($v$=$-$248\kms, $d$=66\kpc, $D$=0.5\kpc). In all these sightlines
high-velocity \OVI\ associated with complex~C is seen, but none is detected at
$v$$<$$-$200\kms.
\par At $l$$>$180\deg, PG\,1116+215 lies 1\fdg6 (5.6\kpc) from Leo\,II
($v$=75\kms, $d$=205\kpc, $D$=0.7\kpc) and NGC\,3115 lies 6\fdg4 (10\kpc) from
Sextans ($v$=227\kms, $d$=86\kpc, $D$=0.9\kpc). The smallest impact parameter is
found for PG\,1004+130, which lies 0\fdg6 (2.5\kpc) from Leo\,I ($v$=290\kms,
$d$=250\kpc, $D$=0.6\kpc). However, in this sightline there is an absorption
feature at velocities between 100 and 400\kms\ that is probably intrinsic
\OIII\l832.927, which hides any weak \OVI\ associated with Leo\,I.
\par Near Leo\,A ($v$=25\kms, $d$=690\kpc, $D$=1.0\kpc) lie the sightlines to
3C\,232 (1\fdg7, 20\kpc), PG\,1001+291 (2\fdg1, 25\kpc) and PG\,0946+301
(2\fdg2, 27\kpc). None of these three objects allow a determination of whether
this dwarf irregular has an extended \OVI\ halo. First, not only is 3C\,232 very
faint, it also samples the halo of NGC\,3067, and the \OVI\ line is confused by
\Lyb\ absorption associated with NGC\,3067. PG\,0946+301 just is too faint.
PG\,1001+291 could in principle be useful. However, the velocity of Leo\,A is
too low to separate any absorption from a Leo\,A halo from that due to the Milky
Way, and the low-velocity \OVI\ absorption toward PG\,1001+291 is not unusually
strong.
\par Impact parameters for the more distant ($d$=1.2--1.6\Mpc) irregulars in the
northern sky are generally $>$50\kpc. Among UGCA\,92, Sextans\,A, Sextans\,B,
Antlia, NGC\,3109 and GR\,8 the smallest impact parameter is 25\kpc\
(PG\,1004+054 lies 1\fdg1 from Sextans\,B). In any case, no matching lines of
\OVI\ were found.
\par There are three nearby dwarf ellipticals in the southern sky.
PKS\,0558$-$504 lies 6\fdg6 (12\kpc) from Carina ($v$=224\kms, $d$=100\kpc,
$D$=0.6\kpc). A high-velocity \OVI\ component is seen in the spectrum of
PKS\,0558$-$504, at 260\kms, but it is more likely that this is associated with
the small high-positive velocity HVCs nearby, which may be flecks of the
Magellanic Stream. HE\,0226$-$4110 lies 6\fdg9 (17\kpc) from Fornax ($v$=53\kms,
$d$=138\kpc, $D$=0.6\kpc) and Ton\,S210 lies 7\fdg1 (10\kpc) from Sculptor
($v$=108\kms, $d$=79\kpc, $D$=0.8\kpc). The velocities of these dwarfs are too
small -- any associated absorption component would be hidden by the Galactic
absorption.
\par Eight dwarf irregulars with distances 0.4--1.3\Mpc\ lie in the region where
high negative-velocity \OVI\ at $v$$\sim$$-$300 and $\sim$$-$180\kms\ is always
seen: NGC\,6822 ($v$=$-$53\kms), IC\,10 ($v$=$-$344\kms), IC\,1613
($v$=$-$237\kms), UGCA\,438 ($v$=62\kms), DDO\,210 ($v$=$-$137\kms), Pegasus
($v$=$-$182\kms), SagDIG ($v$=$-$75\kms) and WLM ($v$=$-$123\kms), Only the
sightline toward Mrk\,509 passes less than 50\kpc\ from one of these (31\kpc\
from DDO\,210). It is notable that many of the dwarf irregulars have $v$ between
$-$100 and $-$350\kms\ and that they are found in the same region of the sky
where high-negative velocity \OVI\ is prevalent.
\par Four more distant ($d$=0.4-1.6\Mpc) galaxies lie at $l$$>$270\deg,
$b$$<$0\deg\ (Phoenix, Tucana, NGC\,55 and IC\,5152). However, this region of
the sky is poorly sampled in our survey.


\subsection{Channel Maps}
\subsubsection{Construction of Channel Maps}
In this subsection we present maps of the \OVI\ column density. Savage et al.\
(2002) and Sembach et al.\ (2002b) show the total column density maps for the
Galactic and high-velocity components. Here (Fig.~\Flsrmap), we show a series of
maps created by calculating the \OVI\ column density in fixed, narrow velocity
ranges relative to the Local Standard of Rest (LSR). These maps are constructed
by stepping through galactic longitude and latitude, and determining which
object lies closest to each selected position. A value is then chosen from a
gaussian distribution that is centered on the column density toward that closest
sightline and has a dispersion equal to the 1-$\sigma$ measurement error. The
resulting column density determines the color of the pixel, as coded by the
wedge below the plot. If no object lies within 12\deg\ from the tested position,
no assignment is made. For an accurately measured value, the color is almost the
same across the whole 12\deg\ radius patch, but for values with relatively large
errors a strong mottling effect can be seen.
\par The velocity channels are 50\kms\ wide between $\pm$200\kms, but a wider
range (100\kms) was used from $\pm$(200--300)\kms\ because there is much less
\OVI\ at these velocities. The two edge channels are even wider
($\pm$(300--500)\kms). No integration was done outside the velocity limits where
we deemed \OVI\ to be present. For example, if we had previously determined that
the sightline showed \OVI\ between $-$85 and 45\kms, we integrated from $-$85 to
$-$50, $-$50 to 0 and 0 to 45\kms. We also calculated the error in each of these
channels, and we only included the sightline in the channel map if the ratio of
the resulting column density in the channel to its statistical error is $>$2.
For the sightlines with Q=4, significant upper limits can be set (2$\sigma$
limit for log $N$(\OVI) $<$13.20). To indicate these, we calculated the typical
error in a 50\kms\ wide bin, and then drew a small patch (2\fdg5 radius) around
these sightlines with a color corresponding to the 2$\sigma$ level. No
integrations were done in the $\pm$100\kms\ velocity range for the eleven
sightlines where we deemed the low-velocity measurement to be just an upper
limit.
\par The figure captions list the number of sightlines for which \OVI\ is
detected in absorption in each of the channels. This shows that the number of
sightlines in a channel is about the same as the number in the channel with
opposite sign, i.e.\ the velocity distribution of the \OVI\ is rather symmetric.
The 0 to 50\kms\ channel (Fig.~\Flsrmap g) shows the source identifications. To
help in understanding the content of these maps Fig.~\FHVCmap\ shows a map of
the velocity field of the \HI\ high-velocity clouds.

\subsubsection{Results -- Milky Way}
\par The four channels between velocities of $-$100 and 100\kms\ show the \OVI\
associated with the Milky Way (Fig.~\Flsrmap e--h). It is obvious that this gas
is widespread, and that it occurs at all of these velocities in almost all
directions. By comparing the maps centered at \vlsr=$\pm$75\kms\ (Figs.~\Flsrmap
e and h), the imprint of Galactic differential rotation can be discerned in the
northern sky. At $-$75\kms\ the column densities are higher at $l$$<$180\deg\
than at $l$$>$180\deg, while at 75\kms\ the opposite is the case. In the
southern sky this effect is not as visible. This may be partly due to the fact
that the region $l$$<$180\deg, $b$$<$0\deg\ has low total \OVI\ column density.

\subsubsection{Results -- High-Negative Velocity Gas}
\par The channel maps show that the \OVI\ at the highest negative velocities is
concentrated in the southern half of the first quadrant, or more precisely, in
the region $l$=20\deg--140\deg, $b$$<$$-$30\deg. In the velocity ranges $-$500
to $-$300 and $-$300 to $-$200\kms\ (Fig.~\Flsrmap a, b) the \OVI\ column
density in this part of the sky is considerable -- about \dex{14}\cmm2. The many
upper limits of log $N$(\OVI)$<$13.20 show that elsewhere in the sky \OVI\ with
such velocities could have been detected, but was not. A remarkable property of
the \OVI\ in this octant is that in most directions there are two components
with high-negative velocity, one near $\sim$$-$300\kms, and one with lower
velocities, typically $\sim$$-$150\kms. These may or may not represent different
phenomena. Figure 11 of Sembach et al.\ (2002b) shows the map of the central
velocities of the high-velocity \OVI, overlaid on the \HI\ HVC, clearly showing
the two negative-velocity components.
\par In the same region of the sky, the \HI\ map (Fig.~\FHVCmap) shows the
presence of the very-high-velocity clouds (VHVCs), which have velocities up to
$-$465\kms. These clouds are usually rather small. None of the sightlines in our
sample passes through a neutral VHVC, although some come within a few degrees.
It is possible that the VHVCs and the high-velocity \OVI\ are somehow related,
but more data is needed to investigate this. It would be useful to obtain \OVI\
column densities in the region of the Anti-Center HVCs ($l$=150\deg\ to 200\deg,
$b$=$-$45\deg\ to 0\deg), but unfortunately the extinction in this part of the
sky is high because of nearby star-forming complexes, and no sufficiently bright
extragalactic targets are known.
\par A possible explanation for the high-negative velocity component of the
\OVI\ sky is that this gas is related in some fashion to the Local Group. The
main concentration of Local Group galaxies is in the region
$l$=100\deg--150\deg, $b$=$-$60\deg--0\deg, and these galaxies all have negative
velocities. The high-negative velocity \OVI\ is also seen in the directions near
and in M\,31 and M\,33, where it blends with the absorption due to those
galaxies. However, a more detailed analysis is required to determine the origin
of the high-negative velocity \OVI; see Sembach et al.\ (2002b).

\subsubsection{Results - Magellanic Stream}
The region of sky where the high-negative velocity \OVI\ occurs is the same
region where the \HI\ velocities of the Magellanic Stream reach values between
$-$300 and $-$400\kms. However, the Magellanic Stream is a rather long, narrow
($\sim$5\deg--10\deg) feature at $l$=90\deg\ (see Fig.~\FHVCmap). At negative
velocities, only three sightlines intersect the Stream's \HI\ (NGC\,7469,
NGC\,7714 and PG\,2349$-$014). Tidal models for the Stream (Gardiner \& Noguchi
1996) predict that it fans out to about the $l$=80\deg--110\deg\ region at
$b$$\sim$$-$30\deg. The prediction for the velocities being $<$$-$300\kms\ is
fairly robust. However, to produce high-negative velocity gas over the wide area
that it is detected in \OVI, these models would need to be fundamentally
changed. Still, some of the gas in the more negative of the two high-negative
velocity \OVI\ components can be made to fit within the tidal models. For the
\OVI\ component near $-$150\kms, however, the 200\kms\ discrepancy implies that
it is probably unrelated to the Stream.
\par At positive velocities only one sightline intersects the \HI\ part of the
Stream (Fairall\,9), while several other pass close by (HE\,0226$-$4110,
NGC\,1705 and PKS\,0558$-$304). In each of these high-positive velocity \OVI\
absorption can be seen with velocities similar to that of the \HI. This suggests
that the Stream has an extended hot envelope. 

\subsubsection{Results -- Complex~C}
\par In the $-$200 to $-$100\kms\ velocity range, \OVI\ associated with HVC
complex~C can clearly be discerned in the northern sky (compare the region
around $l$=100\deg, $b$=45\deg\ in Figs.~\Flsrmap d and \Flsrmap e with
Fig.~\FHVCmap). This HVC has a low metallicity ($\sim$0.1 solar, see Wakker et
al.\ 1999, Richter et al.\ 2001b, Gibson et al.\ 2001), and seems to be a
tidally-stretched cloud that is falling toward the Milky Way.
\par Complex~C is seen in \HI\ emission in nine sightlines (Mrk\,279, Mrk\,290,
Mrk\,501, Mrk\,506, Mrk\,817, Mrk\,876, PG\,1259+593, PG\,1351+640 and
PG\,1626+554). In five of these \OVI\ is detected in the $-$200 to $-$150\kms\
channel. One (PG\,1259+593) gives an upper limit of 12.95 in this channel.
Strong absorption is seen in all nine sightlines in the $-$150 to $-$100\kms\
channel. The strong absorption toward H\,1821+643 in the $-$300 to $-$200\kms\
channel may also be related. That toward PG\,1626+554 in this channel appears to
be strong, but is only a 2.5$\sigma$ detection.
\par In the $-$200 to $-$150\kms\ channel, no \OVI\ absorption is seen in
sightlines adjacent to complex~C (except for a 4.5$\sigma$ detection in the wing
of the profile toward NGC\,3310). In the lower-velocity channel ($-$150 to
$-$100\kms), weak \OVI\ is seen in many directions (PG\,0804+761, Mrk\,116,
PG\,0953+414, Mrk\,421, Mrk\,209 and Mrk\,478), but this is just the
negative-velocity tail of the Galactic absorption. A stronger component occurs
toward Mrk\,106, but this sightline also crosses HVC complex~A. Strong \OVI\
further occurs toward H\,1821+643 and 3C\,382, but for these low-latitude
sightlines we conclude that the \OVI\ is associated with the Outer Arm. Only
toward NGC\,3310 is $N$(\OVI) high even though no high-velocity \HI\ is
detected. \par Weak absorption in the direction of complex~C also appears in the
100 to 150\kms\ channel. This component can clearly be discerned in the spectra
of Mrk\,817, PG\,1351+640 and PG\,1626+554 as an extended wing. Toward Mrk\,876
and PG\,1259+593 a weak extended wing is also detected, but it is less than
2$\sigma$ if integrated only in the 100 to 150\kms\ velocity range.
\par Sembach et al.\ (2002b) discuss the significance of the detection of \OVI\
associated with complex~C.

\subsubsection{Results -- High-Positive Velocity \OVI}
\par In all high-positive-velocity channel maps, there is much \OVI\ in the
northern third quadrant ($l$=240\deg--300\deg, $b$$>$30\deg). This \OVI\ extends
to much higher latitudes than the high-positive velocity gas seen in \HI, which
is concentrated at $b$$<$30\deg\ (see Fig.~\FHVCmap). Only one of our sightlines
samples such low latitudes (ESO\,265$-$G23). Although it is a Q=1 object, the
presence of strong high-velocity \OVI\ is clear.
\par In many cases the high-positive velocity \OVI\ manifests as an extended
wing, while in some it is clearly separated from the Galactic absorption. For
the case of 3C\,273.0 ($l,b$=290\deg,65\deg) Sembach et al.\ (2001b) suggested
that it represents outflowing material, as would be expected in the Galactic
Fountain picture; however, this sightline shows one of the weakest
positive-velocity wings. It is also possible that this gas is related to the
Magellanic Stream, for which models (Gardiner \& Noguchi 1996) predict that the
most accelerated gas lies in the region ($l$,$b$$\sim$270\deg,60\deg), i.e.\ in
the extension of the high-positive velocity gas seen in \HI. Other
interpretations are also possible, such as that this gas represents distant
Local Group material falling toward the Local Group barycenter. Sembach et al.\
(2002b) analyze each of these possibilities.
\par For two of the four sightlines with \OVI\ at $v$$>$300\kms\ (Mrk\,478 and
NGC\,4670) it is likely that the absorption occurs in intergalactic gas, as
these components are relatively narrow ($b$-values $\sim$30\kms) and isolated in
velocity and position.


\section{CONCLUSIONS}
We have presented information on a sample of 217 extragalactic objects and two
distant halo stars observed with \FUSE. We describe the process of calibration,
the alignment of the \OVI\ absorption lines and the construction of final
spectra. Most of the extragalactic objects are quasars and Seyfert galaxies with
relatively flat spectra, but some are nearby starburst galaxies. We fitted
continua to these spectra, and identified the contamination of the \OVIla\ line
by two \H2\ lines, by absorption intrinsic to the background target, and by
intergalactic gas. Of the original sample of 219 objects, 2 guest observer
objects were excluded because of overlapping science goals, 98 objects were too
faint to measure Galactic \OVI, and in 17 Galactic \OVI\ is contaminated by
other absorption, leaving us with a sample of 102 objects. For these we
separated low and high-velocity \OVI\ absorption, measured the \OVI\ equivalent
widths and column densities and studied the distribution of the errors. We reach
the following conclusions:
\par 1) To align \FUSE\ spectra, it is necessary to use \HI\ 21-cm emission
spectra in the target direction in order to determine the velocity of the peak
absorption as the \HI\ does not always peak at 0\kms\ -- it may peak at any
velocity between $\sim$$-$60 and 20\kms. Further, when using v1.8.7 of the
calibration pipeline, the \FUSE\ wavelength scale appears to have shifts of up
to $\sim$10\kms\ between different regions of the same spectrum, so that it is
necessary to align each absorption line individually. The wavelength calibration
of v2.0.5 of the pipeline is much better, but we conclude that a comparison with
\HI\ data is still necessary, and even then there may still be offsets up to
10\kms, due to both the intrinsic accuracy and the possibility that the gas
sampled by a broad \HI\ beam may differ from that in the narrow pencil beam
toward the background target.
\par 2) For bright objects (flux $>$8\tdex{-14}\fu) \FUSE\ can obtain good
spectra (Q$>$=3, S/N$>$9 per resolution element) in 15~ks. At flux levels of
4\tdex{-14}\fu\ this requires 20~ks, while Q=2 (S/N=5) requires only 6~ks. For
objects with a flux of 2\tdex{-14}\fu, a 10~ks integration time is needed to
obtain a spectrum with Q=2 (S/N=5), from which reasonable \OVI\ information can
be extracted after binning to 10 pixels, while good spectra (Q$>$=3, S/N$>$9)
require an exposure time $>$30~ks. For faint objects (flux $\sim$\dex{-14}\fu)
the minimum exposure time for detecting \OVI\ at Q$>$=1 (S/N$>$3) is 10~ks,
while Q$>$=2 (S/N$>$5) requires 25~ks or more and Q$>$=3 (S/N$>$9 per resolution
element) requires $>$80~ks. For fainter objects even very long observations do
not easily yield good spectra, because the background uncertainties start
playing a role. The highest S/N ratio that has been achieved is $\sim$30 per
resolution element, both for a very bright object (vZ\,1128, flux
60\tdex{-14}\fu, 31~ks) and a rather faint object (PG\,1259+593, flux
1.8\tdex{-14}\fu, 633~ks).
\par 3) Galactic \H2\ in the $J$=0 and $J$=1 states is detected in 80\% of the
102 directions with spectra of quality 1--4. The 19 non-detections cluster in
the regions $l$=0--120\deg, $b$$>$40\deg, $l$=210--60\deg, $b$$<$$-$40\deg. The
upper limit on $N$(\H2) can be as low as \dex{14}\cmm2\ in each rotational
level. In 62 out of 102 sightlines, lines of $J$=3 or higher are detected. \H2\
turns out to be ubiquitous in intermediate-velocity clouds. We analyze this in a
separate paper (Richter et al.\ 2002). In 23 cases \H2\ contaminates the \OVI\
absorption profile, but we can correct for it in all but one case (NGC\,3783).
\par 4) Contamination by intrinsic or intergalactic absorption occurs
occasionally. In 16 sightlines this makes it impossible to measure the Galactic
\OVI\ absorption.
\par 5) The Galactic \OVI, which we also refer to as ``thick disk'' absorption,
may extend as far $-$145 or 140\kms\ (Fig.~\Fvlim), but in 88 sightlines it is
confined to within $\pm$120\kms\ and in 61 to within $\pm$100\kms. The average
negative velocity limit is $-$90\kms, while the average positive velocity limit
is 90\kms.
\par 6) We present measurements of the \OVI\ equivalent widths in the Galactic
and high-velocity components, and calculate five contributions to the error: one
associated with the random noise fluctuations in the spectrum, one associated
with continuum fitting (placement), a fixed-pattern noise contribution, an error
associated with the choice of integration range, and an error related to
uncertainties in the determination of the parameters of contaminating \H2\
lines. The first two are combined into a statistical error, while the latter
three give a systematic error. Comparing two methods of determining the error
associated with the placement of the continuum (Fig.~\Ferrcorr) shows that the
$\pm$rms/3 method of estimating continuum errors tends to overestimate these for
flat continua, but gives reasonable answers for continua fit with a second-order
polynomial.
\par 7) In the subsample of 20 sightlines where it is possible to compare the
apparent optical depth profiles of the \OVIla\ and \OVIlb\ lines
(Fig.~\Foviratio), we find 17 components for which the \OVI\ absorption is not
saturated, since the ratio of column densities is unity, to within the 1$\sigma$
error. Of the remaining five components, minor saturation (ratio $\sim$1.2) may
occur in four, while clear saturation (ratio $\sim$1.6) occurs in only one
(Mrk\,421). From this we conclude that integration of the apparent optical depth
profile of the \OVIla\ line will yield a reliable column density in almost all
cases, and will still be correct to within 30\% in the maybe 5 total sightlines
where some saturation occurs.
\par 8) We clearly detect Galactic \OVIla\ absorption in 91 out of 102
directions. In eleven directions we can only set an upper limit, and for five of
these $N$(\OVI) must be relatively low. With even slightly more sensitive data
\OVI\ would probably have been found in the directions for which we now derive
non-detections. The largest \OVI\ equivalent width found is 429$\pm$12\maa,
toward PKS\,2005$-$489, while the lowest detected equivalent width is about
80\maa\ (toward Mrk\,1095 and NGC\,7714).
\par 9) High-velocity \OVIla\ absorption stronger than 65\maa\ is detected in 48
of the 102 sightlines (47\%) while high-velocity absorption stronger than
30\maa\ is found in 34 of the 49 sightlines with Q=3 or 4 (69\%). In 13
sightlines a high-velocity feature as small as 20\maa\ could have been detected,
and it is found in 12 of these (92\%).
\par 10) We show that both weak and strong high-velocity \OVI\ absorption
components occur at all velocities in the ranges $\sim$$-$400 to $-$100 and 100
to 400\kms\ (Fig.~\Fvelhist). The scatter plot of column density against
velocity suggests that many absorbers similar to those seen at high-velocity may
be blended with the Galactic absorption.
\par 11) We checked the sightlines that pass within a few degrees of Local Group
galaxies. None of the dwarf irregulars, ellipticals and spheroidals appear to
show associated \OVI, nor was any expected because of the small sizes of these
galaxies. Four sightlines are toward \HII\ regions in M\,33. Since several M\,33
OB stars have also been observed, it should be possible to study \OVI\ in M\,33.
Similarly, several M\,31 OB stars have been observed. One QSO near M\,31 is
bright enough that an improved spectrum is possible. However, a study of \OVI\
in M\,31 and M\,33 is complicated by the fact that high-negative velocity \OVI\
absorption (at velocities expected for \OVI\ in those two galaxies) is present
over a large part of the southern sky.
\par 12) For the case of M\,33, its rotation curve predicts gas velocities near
0\kms\ or larger along the sightlines to Mrk\,352, Mrk\,357 and PG\,0052+251,
which pass 90, 96 and 133\kpc\ from M\,33, respectively. No such absorption is
seen, limiting the extent of an \OVI\ halo around M\,33 to be $<$100\kpc, at a
3$\sigma$ detection limit of log $N$(\OVI)$\sim$14.0.
\par 13) A series of \OVI\ channel maps shows the imprint of differential
Galactic rotation on the low-velocity absorption: in the $-$100 to $-$50\kms\
channel, $N$(\OVI) is larger at longitudes $<$180\deg, while in the 50 to
100\kms\ channel it is larger at longitudes $>$180\deg.
\par 14) The $-$200 to $-$100\kms\ \OVI\ channel maps show that the HVC
complex~C is clearly detected in \OVI\ absorption. The positive-velocity side of
the Magellanic Stream ($l$$\sim$270\deg) is detected in the 150 to 300\kms\
channels although there is only one sightline where both \HI\ and \OVI\ are seen
(Fairall\,9). In all directions in the region $l$=20\deg--150\deg,
$b$$<$$-$30\deg\ \OVI\ is detected at high negative velocities $<$$-$200\kms,
while elsewhere in the sky only upper limits can be set. Conversely, in the 150
to 300\kms\ channels high-positive velocity \OVI\ is common in the region
$l$=180\deg--300\deg, $b$$>$20\deg, and only upper limits are set elsewhere in
the sky.


\newpage
\acknowledgements{
The work is based on data obtained for the Guaranteed Time Team by the
NASA-CNES-ESA \FUSE\ mission, operated by Johns Hopkins University. Financial
support to U.S.\ participants has been provided by NASA contract NAS5-32985.
B.P.W.\ was supported by NASA grants NAG5-9024, NAG5-9179, and NAG5-8967. K.R.S.
acknowledges financial support through NASA contract NAS5-32985 and grant
NAG5-3485. Some of the data presented in this paper was obtained from the
Multimission Archive at the Space Telescope Science Institute (\MAST); STScI is
operated by the Association of Universities for Research in Astronomy, Inc.,
under NASA contract NAS5-26555. Support for \MAST\ for non-\HST\ data is
provided by the NASA Office of Space Science via grant NAG5-7584. We thank
Veronique Buat, John Mulchaey, Trinh Xuan Thuan, Tim Heckman, Anuradha Koratkar,
Joel Bregman, William Keel, Thomas Brown, Brad Gibson, Smita Mathur and Jason
Prochaska for their permission to use their \FUSE\ guest investigator data. This
research has made use of the NASA/IPAC Extragalactic Database (\NED) which is
operated by the Jet Propulsion Laboratory, California Institute of Technology,
under contract with the National Aeronautics and Space Administration. We thank
Ricardo Morras for providing \HI\ spectra from the Villa Elisa survey for 14
southern targets. We thank Peter Kalberla for the \HI\ observations using the
Effelsberg telescope for 22 of our targets; the Effelsberg Telescope belongs to
the Max Planck Institute for Radio Astronomy in Bonn.
}


\newpage \hbox to \hsize{\hss{\bf APPENDIX}\hss}
In this appendix we give notes for all 119 objects for which spectral lines are
shown in Fig.~\Fspectra. All of these have a spectrum with a signal-to-noise
ratio $>$3 per resolution element near 1030\aa. These notes contain remarks on
special things to be aware of concerning the separation between the Milky Way
and high-velocity \OVI, concerning \H2\ contamination, as well as details about
the nearby galaxy groups intersected and any other special features.
\par Reference is often made to galaxy groupings from Tully's (1988) catalogue.
This catalogue gives the distance for each galaxy (based on direct
determinations), its radial velocity and group membership. These distances are
used to calculate the impact parameter of each target sightline from the angular
distance between the sightline and the galaxy. All galaxies with an impact
parameter of $<$200\kpc\ are explicitly mentioned. The correlation with galaxy
groups was done in order to identify possible contaminating intergalactic \Lyb\
absorption (see \Sect\Sgroupcontam). Where necessary, we give the name of the
galaxy groupings according to Tully (1988), followed by the average velocity and
velocity dispersion of the galaxies in the grouping.

\bigskip
\def\VirCld{Virgo Cluster ($v$=1360$\pm$720 \kms)}                          
\def\UMaCld{Ursa Major Galaxy Grouping ($v$=1360$\pm$540 \kms)}             
\def\UMaSouthSpur{Ursa Major Southern Spur ($v$=1760$\pm$530 \kms)}         
\def\ComSclCld{Coma-Sculptor Galaxy Grouping ($v$=420$\pm$300 \kms)}        
\def\LeoSpur{Leo Spur ($v$=620$\pm$160 \kms)}                               
\def\TriCld{Triangulum Spur ($v$=810\$pm$220 \kms)}                         
\def\PavAraCld{Pavo-Ara Galaxy Grouping ($v$=830$\pm$240)}                  
\def\LeoCld{Leo Galaxy Grouping ($v$=1390$\pm$390 \kms)}                    
\def\CrtCld{Crater Galaxy Grouping ($v$=1490$\pm$260 \kms)}                 
\def\LynCld{Lynx Galaxy Grouping ($v$=1850$\pm$280 \kms)}                   
\def\AntHydCld{Antlia-Hydra Galaxy Grouping ($v$=2090$\pm$500 \kms)}        
\def\CncLeoCld{Cancer-Leo Galaxy Grouping ($v$=2600$\pm$290 \kms)}          
\def\CarCld{Carina Galaxy Grouping ($v$=2430$\pm$350 \kms)}                 
\def\LepCld{Lepus Galaxy Grouping ($v$=2130$\pm$480 \kms)}                  
\def\VirLibCld{Virgo-Libra Galaxy Grouping ($v$=1820$\pm$490 \kms)}         
\def\CVnCld{Canes Venatici Galaxy Grouping ($v$=2360$\pm$330 \kms)}         
\def\CVnSpur{Canes Venatici Spur ($v$=1140$\pm$250 \kms)}                   
\def\DraCld{Draco Galaxy Grouping ($v$=1120$\pm$390 \kms)}                  
\def\ForEriCld{Fornax-Eridanus Galaxy Grouping ($v$=1580$\pm$490 \kms)}     
\def\CetAriCld{Cetus-Aries Galaxy Grouping ($v$=1810$\pm$510 \kms)}         
\def\DorCld{Dorado Galaxy Grouping ($v$=970$\pm$250 \kms)}                  
\def\TelGruCld{Telescopium-Grus Galaxy Grouping ($v$=2030$\pm$500 \kms)}    
\def\PavIndSpur{Pavo-Indus Spur ($v$=2850$\pm$210 \kms)}                    
\def\PsASpur{Pisces Austrinis Spur ($v$=2590$\pm$260 \kms)}                 
\def\PegCld{Pegasus Galaxy Grouping ($v$=2170$\pm$480 \kms)}                
\def\PegSpur{Pegasus Spur ($v$=1110$\pm$260 \kms)}                          
\def\SerCld{Serpens Galaxy Grouping ($v$=2090$\pm$290 \kms)}                
\def\BooCld{Bootes Galaxy Grouping ($v$=2650$\pm$310 \kms)}                 


\medskip\par 3C\,48.0
\par This sightline passes just 32\kpc\ from M\,33, and 164\kpc\ from M\,31.
However, the source is too faint (0.5\tdex{-14} \fu) to have sufficient S/N
ratio (1.3 per resolution element) after just 8.5 ks of integration. A longer
observation might reveal associated absorption. If present, such associated
absorption is expected at a velocity of $-$300\kms, considering the \HI\
velocity field of M\,33 (Deul \& van der Hulst 1987).



\medskip\par 3C\,232
\par This sightline passes through the halo of NGC\,3067, 10\kpc\ from its
center. A clear Ly$\beta$ line is seen at 1525\kms, as is redshifted \CII\
absorption, even though the S/N is only 1.5 per resolution element.

\medskip\par 3C\,249.1
\par The LiF2B channel of the second observation (P1071602) has no signal and
thus was not used. The sightline lies just 3\deg\ from the edge of complex~C as
seen in \HI, yet the \OVI\ absorption only extends out to $-$75\kms.
\par This sightline passes through the \UMaCld\ and the \CVnCld, as well as the
GH64 ($v$=1480\kms) and GH69 ($v$=3000\kms) groups. It lies 1\fdg1 (600\kpc\
impact parameter) from NGC\,3329 ($v$=2056\kms), and 2\fdg4 (55\kpc\ impact
parameter) from UGC6456 ($v$=96\kms, R=0.6\kpc), but no intergalactic Ly$\beta$
can be discerned near \OVI. Ly$\alpha$ is also absent at those velocities
(Savage et al.\ 2000).

\medskip\par 3C\,273.0
\par This is among the 10 sightlines with the highest S/N ratio (28 per
resolution element). A detailed investigation of this QSO was presented by
Sembach et al.\ (2001b). The velocity scale preferred here differs by 9\kms\
from that adopted by Sembach et al.\ (2001); see Sect.~\Svshift\ for a
discussion of this. Sembach et al.\ (2001b) report the same equivalent widths
(within the errors), except that they include the 125\kms\ component into the
Milky Way component, and they did not split the error into a statistical and
systematic component. The relatively large systematic error in the equivalent
width of this component is due to uncertain integration limits.
\par Both the \OVIla\ and the \OVIlb\ lines can be measured. In the velocity
range $-$70 to 105\kms\ the ratio N(1037)/N(1031) is 1.13$\pm$0.04, suggesting
that there might be some slight saturation.
\par Two H$_2$ components are clearly visible in the $J$=0, 1, 2 and 3 lines,
but for $J$=4 only one component is seen. The strongest H$_2$ component is
associated with the weak \HI\ at 25\kms\ (see Richter et al.\ 2002). This H$_2$
line has only a minor influence on the systematic error of the thick disk and
HVC \OVI\ components.
\par The features at 1029.161\aa\ and 1031.186\aa\ are Ly$\beta$ at z=0.00335
($v$=1005\kms) and z=0.00532 ($v$=1595\kms), which are associated with the Virgo
Cluster (Sembach et al.\ 2001). There are three small (diameter $<$10\kpc)
galaxies in this cluster with impact parameter $<$300\kpc\ (NGC\,4420, UGC\,7612,
and UGC\,7512). The feature at 1035.445\aa\ is \OVI\ in the Virgo cluster, at
$v$=1020\kms. The corresponding \OVIlb\ feature is blended with low-velocity
H$_2$ $J$=3.


\medskip\par 3C\,351.0
\par There is a Lyman limit system at z=0.22 in this sightline, which blocks out
all emission below 1112\aa. The sightline also passes just 6\kpc\ from the Draco
dwarf spheroidal (diameter 1.8\kpc), but if there had been flux at 1030\aa\ any
absorption associated with that galaxy ($v$=$-$30\kms) would be confused with
that in the Milky Way.

\medskip\par 3C\,382.0
\par This is one of three low-latitude sightlines in the sample toward which
\HI\ associated with the Outer Arm is seen (e.g., Habing 1966; Kepner 1970;
Hulsbosch \& Wakker 1988; Haud 1992). \OVI\ is seen at corresponding velocities,
extending to $-$130\kms. The \OVI\ measurement for the thick disk includes this
component. A separate measurement is also given for the velocity range over
which Outer Arm \HI\ is seen.
\par The feature at 1033.650\aa\ (500\kms\ on the \OVI\ velocity scale) is
intrinsic \CIII\ -- the corresponding Ly$\beta$, Ly$\gamma$ and Ly$\delta$ lines
are also seen. The features at 1033.090\aa\ and 1035.020\aa\ (340\kms\ and
900\kms\ on the \OVI\ velocity scale) remain unidentified. There are no
intrinsic Lyman lines at these velocities. Although there are no known galaxy
groups in this low-latitude (high extinction) direction, these features may be
Ly$\beta$ at $v$=2155\kms\ and 2715\kms. Unfortunately, the other Lyman lines
are then in a part of the spectrum that is too noisy to confirm this, and no
Ly$\alpha$ data are available.




\medskip\par ESO\,141$-$G55
\par The HVC component in this spectrum is similar to the one seen in the nearby
sightline toward PKS\,2005$-$489 (12\deg\ away).
\par The H$_2$ $J$=4 line somewhat confuses the measurement of the \OVI\ HVC
component, but it only increases the systematic error from 12 to 15\maa.
\par The sightline goes through the middle of the \TelGruCld, and several
galaxies lie within a few degrees (impact parameters 0.8--3 Mpc). No associated
Ly$\beta$ absorption can be seen, however. The high-velocity \OVI\ component at
175\kms\ cannot be Ly$\beta$, as the corresponding 200\maa\ Ly$\alpha$ line at
1223.72\aa\ is not visible in the \GHRS\ spectrum of this object.

\medskip\par ESO\,265$-$G23
\par This spectrum has low signal-to-noise ratio (3.2 per resolution element),
but the thick disk \OVI\ as well as the HVC \OVI\ at 260\kms\ are strong enough
to measure with confidence. Less than one degree away lies a large \HI\ cloud
with velocities centered near 260\kms, which is part of the leading arm of the
Magellanic Stream. After the cutoff date for our sample, new guest observer
data were obtained, increasing the exposure time from 5 to 45~ks. This shows
that the HVC component extends only to $\sim$310\kms, rather than the 345\kms\
suggested by the earlier observation. We therefore only integrated between 200
and 310\kms. HVC complex~WD is present in this sightline at 120\kms, but in this
velocity range no strong \OVI\ absorption is seen.
\par Intrinsic \CIII\ absorption (to go with the clearly detected intrinsic 
\OVI\ and Ly$\gamma$) would overlap the H$_2$ \l1031.191 $J$=3 line, but does
not appear to be present.

\medskip\par ESO\,350$-$IG38 (Haro\,11)
\par There is a broad absorption line near \OVIla, much of which is intrinsic
\SIII\l1012.510 absorption between 5600 and 6250\kms. In the same velocity range
absorption is also seen in the \OVIla, \CIII\l977.020, \NIII\l989.799, \CIIl,
\OIl, \SiIIla\ and Lyman lines. It is thus impossible to extract the
contribution of Galactic \OVI\ from this spectrum.

\medskip\par ESO\,572$-$G34
\par The \OVI\ profile in this direction is unusual in that both the Milky Way
and high-velocity components are very strong. Both are also clearly seen in the
\OVIlb\ line. This sightline lies just a few degrees from IRAS\,F11431$-$1810
and HE\,1115$-$1735, toward which the high-velocity \OVI\ is also clearly seen.
\par Both the \OVIla\ and the \OVIlb\ lines can be measured. In the velocity
range $-$70 to 100\kms\ the ratio N(1037)/N(1031) is 0.94$\pm$0.16, while in
the velocity range 100 to 240\kms\ this ratio is 0.95$\pm$0.16.
\par ESO\,572$-$G34 is part of the \CrtCld\ and has a systemic velocity of
1114\kms. The galaxy's Ly$\beta$ absorption is clearly visible at 1029.3\aa. The
sightline further passes just 0\fdg3 (114\kpc) from NGC\,4027 ($v$=1460\kms,
R=24\kpc), but no absorption is seen at its velocity.

\medskip\par Fairall\,9
\par This object is the only one in the sample that lies projected on the part
of the \HI\ Magellanic Stream where the Stream has positive velocities.
Associated \OVI\ absorption is clearly seen, but extends over a much wider
velocity range than the \HI. There is no clear separation between the \OVI\
absorption associated with the Milky Way and with the Stream; 100\kms\ was
chosen as the separation velocity. No contaminating H$_2$ is present, since both
the Milky Way and the  Magellanic Stream H$_2$ absorption only show lines up to
$J$=2.
\par The high-positive velocity absorption cannot be intergalactic Ly$\beta$: no
corresponding Ly$\alpha$ line is seen in the \GHRS\ spectrum.


\medskip\par H\,1821+643
\par This is one of three sightlines toward which the Outer Arm is seen in \HI.
\CII, \SiII\ and \FeII\ absorption at corresponding velocities is clearly
present, extending to $-$160\kms.
\par An earlier analysis of this sightline was presented by Oegerle et al.\
(2000), who already reported all the \OVI\ components that we list.
\par A strong H$_2$ $J$=3 line makes the measurement of the high-negative
velocity \OVI\ absorption uncertain. The absorption depth of 7 other $J$=3 lines
averages to 0.32$\pm$0.05, while the FWHM averages to 34$\pm$7\kms. The velocity
of the H$_2$ absorption is difficult to determine, but it clearly must be more
positive than $-$10\kms, the velocity of the peak \HI\ emission. Using the best
approximation to the H$_2$ $J$=3 absorption suggests that the absorption near
$-$210\kms\ is an about equal mixture of H$_2$ and \OVI, and that there are two
high-velocity \OVI\ components. A very weak one ($-$260\kms,
W=21$\pm$7$\pm$17)\maa\ ranges from $-$285 to $-$235\kms, while a stronger one
($-$190\kms, W=56$\pm$7$\pm$20) ranges from $-$225 to $-$160\kms. The uncertain
correction for H$_2$ is reflected in the large systematic error on the \OVI\
components.
\par A further complication is the planetary nebula K1-16, which lies less than
1 arcmin away. The strong saturated components in the \OVIla\ and \OVIlb\ lines
near 10\kms\ are very likely due to this object. The associated absorption was
removed from the calculation of the Milky Way \OVI\ equivalent width and column
density.


\medskip\par HE\,0226$-$4110
\par There are two observations for this object, the second of which was
executed using a focal-plane split. I.e., the spectrum was placed at one of four
different places on the detector during each of the 19 orbits. Fixed-pattern
noise therefore was reduced by a factor~4.
\par The \CII\ absorption does not reach zero flux, so there may be a small
background offset that has not been calibrated out properly.
\par This is one of 3 sightlines where the thick disk \OVI\ extends to more
negative velocities than $-$120\kms\ (to $-$140\kms), but in which there is no
convincing argument for measuring a separate high-velocity component.
\par There is absorption at a velocity of 165\kms\ which probably is \OVI, since
there appears to be a counterpart in the \OVIlb\ line.
\par Both the \OVIla\ and the \OVIlb\ lines can be measured. In the velocity
range $-$140 to 75\kms\ the ratio N(1037)/N(1031) is 0.91$\pm$0.17.
\par The sightline passes just 12\kpc\ (7\arcmin) from the center of the Fornax
dwarf spheroidal. However this galaxy is very small, having a diameter of just
0.5\kpc.
\par The sightline also lies about 8\deg\ from the main body of the Magellanic
Stream, and just 3\deg\ from a small HVC with Stream-like velocities.
\par The feature at 1034.659\aa\ (795\kms\ on the \OVI\ velocity scale) may be
intergalactic Ly$\beta$, though there are no confirming observations of
Ly$\alpha$.

\medskip\par HE\,0238$-$1904
\par The noisy nature of this spectrum makes it difficult to place the continuum
and to determine the extent of the Galactic and HVC absorption. The feature at
1033.055\aa\ (330\kms\ on the \OVI\ velocity scale) is just 2.9$\sigma$, and is
classified as ``unidentified''; it may not be real. If it were \OVI, the
corresponding absorption in the \OVIlb\ line is confused with a H$_2$ line.
\par When discernable, the other H$_2$ $J$=3 lines in the spectrum all have a
velocity of $\sim$0\kms. The mismatch between the predicted H$_2$ \l1031.191
line and the observed feature is probably due to the emission-like feature in
the LiF1A channel, which distorts the spectrum.

\medskip\par HE\,0450$-$2958
\par The LiF1A channel appears to show a weak \OVI\ line, but this is not
confirmed in the (noisier) LiF2B channel. The combined spectrum has low S/N, but
it is clear that the Milky Way \OVI\ component is rather weak. In fact, only a
3$\sigma$ upper limit of 115\maa\ or log N(\OVI)$<$13.96 can be set.
\par The sightline goes through the \ForEriCld\ (d=17 Mpc); UGCA\,95 and
UGCA\,97 (v=1290\kms) lie $\sim$1\deg\ away (300\kpc\ impact parameter). It also
passes through the \DorCld\ (d=10 Mpc), and several galaxies have impact
parameters of 450--800\kpc. However, no intergalactic Ly$\beta$ absorption can
be discerned.



\medskip\par HE\,1115$-$1735
\par The \OVI\ column density is low (3$\sigma$ upper limit of 105\maa\ or log
N(\OVI)$<$13.92), although this sightline is just a few degrees from
IRAS\,F11431$-$1810 toward which Galactic \OVI\ is very strong
(294$\pm$22\maa). On the other hand, the HVC component is clearly present and
about as strong as toward IRAS\,F11431$-$1810. There does not seem to be a
contaminating $J$=4 H$_2$ line, since no other $J$=4 lines can be discerned.

\medskip\par HE\,1228+0131
\par Toward this object there is strong Ly$\beta$ absorption at 1495, 1760 and
2335\kms, associated with the Virgo Cluster. Ly$\alpha$ counterparts for both
these lines are seen in the \STIS\ spectrum of this object. The first 14 Lyman
lines of this series can be clearly seen in the \FUSE\ spectrum. The sightline
passes within 300\kpc\ of a number of galaxies (NGC\,4517, UGC\,7685,
NGC\,4536).
\par The 1760\kms\ absorption distorts Galactic \OVIla, although the
positive-velocity wing (v$\sim$50--200\kms) must be Galactic \OVI.
Unfortunately, the \OVIlb\ line can not be used as an alternative, since the
contamination from H$_2$ $J$=1 cannot be determined; the apparent strengths of
the 1037.146 and 1038.156\aa\ lines are inconsistent with each other and with
other $J$=1 lines.
\par The nature of the feature at 1032.913\aa\ (290\kms\ on the \OVI\ velocity
scale) is unclear. It is unlikely to be high-velocity \OVI\ as there is no
counterpart in the \OVIlb\ line. It cannot be Ly$\beta$ since no Ly$\alpha$
absorption is visible.

\medskip\par HE\,1326$-$0516
\par Although the S/N ratio is low, the \OVI\ column density clearly is low. In
fact, only a 3$\sigma$ upper limit of 175\maa\ (log N(\OVI)$<$14.15) can be set.





\medskip\par HE\,2347$-$4342
\par This object is very faint (\dex{-15} \fu), but the 704 ks exposure time (the
longest done with \FUSE) allows the detection of the redshifted \HeII\ forest
(Kriss et al.\ 2001). This completely covers low-redshift absorption, however.



\medskip\par HS\,0624+6907
\par The individual channels for the two observations are rather noisy and are
not completely consistent. The Galactic absorption seems stronger in the two
LiF2B channels than in the two LiF1A channels.
\par This is one of three sightlines toward which the Outer Arm is seen in \HI\
emission (the other two are 3C\,382.0 and H\,1821+643). However, unlike in the
other cases, there is no \OVI\ absorption at the velocities of the Outer Arm.
in this direction.



\medskip\par HS\,1102+3441
\par This spectrum has one of the lowest S/N ratios in our final sample (3.4 per
resolution element). Still, it clearly shows a blended thick disk and HVC \OVI\
component. These were separated at a velocity of 95\kms, where the absorption is
minimal. In the \OVIlb\ line the HVC component is contaminated by a $J$=1 H$_2$
line. It is probably real, however, especially considering that in this part of
the sky ($l$=180\deg--240\deg, $b$=45\deg--70\deg) many sightlines show
absorption at high positive velocities.
\par The sightline passes through GH75 ($v$=1025\kms), but no absorption is
apparent at such velocities.
\par Note that \NED\ gives the name PG\,1102+347 for this source, but that there
was no such object in the original PG catalogue (Green et al.\ 1986).










\medskip\par HS\,1543+5921
\par This sightline passes 2\arcsec\ (300 pc) from the dwarf galaxy
SBS\,1543+593 ($v$=2698\kms, part of GH158). The flux of 0.75\tdex{-14} \fu\
(S/N ratio of 2 after the short 8.5 ks observation) is too low to detect
Galactic \OVI. Ly$\beta$ in SBS\,1543+593 is clearly present, however, while any
\OVI\ in that galaxy is confused with geocoronal \OI\ emission lines.






\medskip\par IRAS\,09149$-$6206
\par In this sightline strong and broad (470\kms\ wide) intrinsic \OVI\ and
\CIII\ lines can be discerned. Unfortunately, the \CIII\ overlies the
positive-velocity wing of the Galactic \OVI. It is not possible to determine how
much it contaminates the Galactic \OVI. We therefore did not include this
sightline in our sample.

\medskip\par IRAS\,F11431$-$1810
\par The positive-velocity \OVI\ absorption extends much farther than expected
for Galactic absorption. The H$_2$ $J$=4 \l1032.356 line does not seem to be
present, as no other $J$=4 lines can be seen. A choice was made to separate the
thick disk and HVC components at a velocity of 100\kms, which is the velocity at
which a clearer separation is seen between thick disk and high-velocity \OVI\ in
two neighboring sightlines: ESO\,572$-$G34 (3\deg\ away) and PG\,1302$-$102
(21\deg\ away).
\par This sightline passes $\sim$2\deg\ (1\Mpc) from the \CrtCld. The galaxies
in this small group that lie the closest to IRAS\,F11431$-$1810 have velocities
of 1000 to 1620\kms, so it is unlikely that some of the apparent Galactic \OVI\
absorption is inter-group Ly$\beta$. Unfortunately, the spectrum at 980\aa\ is
too noisy to check for Ly$\gamma$ and there are no HST spectra to check
for Ly$\alpha$.




\medskip\par MRC\,2251$-$178
\par The \OVI\ line in the LiF2B channel looks different from that in the LiF1A
channel. There are two associated Ly$\gamma$ absorption components ($v$=18550
and 18925\kms), centered on 1032.710 and 1033.943\aa\ (230 and 585\kms\ on the
\OVI\ velocity scale. These look smeared in the LiF2B channel. We therefore
decided to use only LiF1A data for this object. This system has a velocity of
$-$900\kms\ relative to the nominal velocity of the AGN. There may be a
foreground galaxy or associated foreground gas in the line of sight.
\par At first sight, the main Galactic absorption seems to extend to $-$145\kms\
and to have two components. However, we note that toward other sightlines in
this part of the sky (NGC\,7469, NGC\,7714, PHL\,1811, PKS\,2155$-$304) there is
a separate component between $-$100 and $-$160\kms, in addition to the
high-negative velocity component at v$<$$-$200 \kms. In these sightlines the
Galactic absorption usually extends only to about $-$70\kms. Therefore, the
absorption between $-$145 and $-$65\kms\ was measured separately and considered
a HVC component.
\par \CII\ absorption is seen at a velocity of $-$250\kms, with W=65$\pm$19\maa,
or N(\CII)$\sim$8\tdex{13}\cmm2. Assuming carbon is not depleted on dust, this
corresponds to N(\HI)=0.2\tdex{17}/Z\cmm2, with Z the metallicity in solar
units. The Green Bank spectrum sets a 3$\sigma$ upper limit of
$\sim$1\tdex{18}\cmm2\ on N(\HI).

\medskip\par Mrk\,9
\par For one of the three observations for this object, the detector
high-voltage was off for 5 out of 8 orbits. The remaining 2.5 ks represent just
10\% of the total integration time, and were discarded because they are
difficult to align.
\par The Galactic \OVI\ profile clearly shows two components, which is unusual.
\par This sightline crosses the edge of the \LynCld, whose galaxies have
velocities between 1380 and 2100\kms, and lie in a horseshoe shape around the
Mrk\,9 sightline. The closest galaxies in the group (NGC\,2460, $v$=1553\kms;
UGC\,4093, $v$=1646\kms; and UGC\,3826, $v$=1847\kms) lie 3\deg\ away (impact
parameter $\sim$1.2 Mpc). The Ly$\beta$ absorption associated with this group
would appear between $v$=$-$400 and 300\kms\ on the \OVI\ velocity scale;
Ly$\alpha$ observations will be necessary to exclude the probability that any of
the absorption in this velocity range is Ly$\beta$.

\medskip\par Mrk\,36 (Haro\,4)
\par The feature at $-$278\kms\ on the \OVI\ velocity scale is the H$_2$
\l1031.191 line at $-$55\kms\ in the IV-Arch (see Richter et al.\ 2002). Several
other H$_2$ $J$=3 lines with similar strengths can be seen.
\par Mrk\,36 lies in the \LeoSpur, but the galaxies closest on the sky lie
behind it.

\medskip\par Mrk\,54
\par There is intrinsic \OI\l988.773 and \SiIIlb\ absorption at approximately
1032.92 and 1034.26\aa\ (290 and 680\kms\ on the \OVI\ velocity scale). The wing
of intrinsic \OI\ may overlap Galactic \OVI\ absorption, making both the
continuum and the upper velocity limit too uncertain to measure Galactic \OVI.
We therefore did not include this object in our final sample.

\medskip\par Mrk\,59
\par This object is a bright \HII\ region representing the starburst in the
galaxy NGC\,4861 ($v$=847\kms).

\medskip\par Mrk\,79
\par The first of the three observations of this object was marred by many
detector burst events, but the \OVI\ features are consistent with those in the
second observation. During the third observation the high-voltage was off on
side 1 for 5 of the 7 orbits; we did not include this observation in the final
spectrum. In the final spectrum the continuum is difficult to determine, but
because the \OVI\ absorption is relatively strong, it is well-measured.

\medskip\par Mrk\,106
\par This spectrum has low S/N (3.6 per resolution element after combining both
LiF channels). The \OVI\ absorption in the individual channels looks different,
but the differences fall within the noise.
\par This is one of three sightlines projected on to HVC complex~A. Possible
associated absorption is seen between $-$150 and $-$100\kms, although it is only
a 3$\sigma$ feature and it does not overlap in velocity with the \HI\ in
complex~A.
\par There clearly is low-velocity H$_2$ in the $J$=2 level, but none of the
$J$=3 lines are visible, possibly because of the low S/N ratio.
\par The feature at 1033.918\aa\ (580\kms\ on the \OVI\ velocity scale) is most
likely intergalactic Ly$\beta$ at $v$=2395\kms\ that is associated with the
\LeoCld. The closest galaxy in this group is UGCA\,154 ($v$=2287\kms, 2\deg\
away, 1.2 Mpc impact parameter). The sightline also passes through the \LeoSpur\
and GH44 ($v$=930\kms), but no Ly$\beta$ absorption can be seen around these
velocities.


\medskip\par Mrk\,116
\par The continuum near the \OVIla\ line is somewhat difficult to determine, as
the absorption lies in the shoulder of the Ly$\beta$ absorption associated with
Mrk\,116. The decision on where to place the continuum was helped by knowing the
expected strength of the H$_2$ lines. This justifies placing the continuum
fairly high.
\par The H$_2$ $J$=4 line contaminates the positive-velocity edge of the thick
disk absorption, but it only increases the systematic error from 21 to 24\maa.
\par This is one of three sightlines projected onto HVC complex~A. No obvious
associated high-velocity feature is present in the combined spectrum.


\medskip\par Mrk\,205
\par This spectrum had a high background and the flux calibration is therefore
less reliable than usual. However, the central part of the \CIIl\ line lies at
zero flux, as expected.
\par \OVI\ is not detected, but because the spectrum is noisy, a 3$\sigma$ upper
limit of only 195\maa\ or log N\OVI)$<$14.19 can be set.
\par The feature at 1030.276\aa\ ($-$480\kms\ on the \OVI\ velocity scale) is
Ly$\beta$ in the halo of NGC\,4319 at $v$=1330\kms.

\medskip\par Mrk\,209 (Haro\,29)
\par There are two observations of this object. The individual LiF1A and LiF2B
channels are rather noisy, and the \OVI\ absorption features are slightly
different. The combined spectrum is fairly good, however.
\par This is one of 2 sightlines where the thick disk \OVI\ extends to more
positive velocities than 120\kms\ (to 130\kms), but in which there is no
argument for a separate high-velocity component.
\par The feature at 275\kms\ is \OVI\ absorption intrinsic to Mrk\,209, which
has $v$=281\kms.

\medskip\par Mrk\,279
\par This is among the 10 sightlines with the highest S/N ratio (27 per
resolution element). The flux of Mrk\,279 seems to vary over time. It was
11\tdex{-14} \fu\ on 1999 December 28, 9\tdex{-14} \fu\ on 2002 January 11 and
7\tdex{-14} \fu\ on 2002 January 28.
\par Mrk\,279 is one of 9 sightlines in which \OVI\ absorption associated with
HVC complex~C is seen. The HVC \OVI\ and \HI\ component overlap, although the
\HI\ profile is complicated, with two HVC, two IVC and four low-velocity
components clearly identifiable (each can be followed to neighboring
directions).
\par Both the \OVIla\ and the \OVIlb\ lines can be measured. In the velocity
range $-$80 to 100\kms\ the ratio N(1037)/N(1031) is 1.08$\pm$0.09.
\par The sightline crosses through the middle of the \CVnCld. The closest galaxy
that is part of this group is Mrk\,263 ($v$=1529\kms), at a distance of 1\fdg8
(775\kpc\ impact parameter). No intergalactic Ly$\beta$ is apparent at these
velocities. The \GHRS\ spectrum of this object goes down to 1222.6\aa, whereas
the possible Ly$\alpha$ line is expected at 1221.14\aa, so no check of
Ly$\alpha$ is possible.
\par The Ursa Minor dwarf spheroidal ($v$=17\kms, radius 1.4\kpc) lies 7\fdg3
away (13\kpc\ impact parameter), but any associated \OVI\ would be hidden in the
main Galactic absorption.

\medskip\par Mrk\,290
\par Although this is a spectrum with low S/N, absorption associated with HVC
complex~C is clearly present at the 4.7$\sigma$ level, extending over a velocity
range similar to that of the \HI. The Galactic and HVC component were separated
at a velocity of $-$70\kms, based on the appearance of the \HI\ profile. There
does not appear to be H$_2$ in the $J$=4 state, and the extended
positive-velocity wing is more likely to be \OVI. On the other hand, the
negative-velocity edge of the HVC absorption is contaminated by H$_2$ $J$=3, but
the amount of contamination is minor.
\par The sightline passes through the \CVnCld\ and GH152 ($v$=1050\kms).
NGC\,5963 lies 1\fdg5 (350\kpc) away. No Ly$\beta$ can be seen near these
velocities, however.

\medskip\par Mrk\,304
\par The positive-velocity edge of the high-negative velocity \OVI\ component is
contaminated by the H$_2$ $J$=3 line. This increases the systematic error from
38 to 46\maa.
\par The features at 1033.564\aa\ and 1034.000\aa\ (475 and 605\kms\ on the
\OVI\ velocity scale) may be Ly$\beta$ at 2290 and 2420\kms, though no other
Lyman lines are seen. No \GHRS\ or \STIS\ spectrum exists that allows a check.
The sightline passes through the middle of the \PegCld. The nearest two galaxies
in this group (NGC\,7280, $v$=2090\kms\ and UGCA\,429, $v$=2158\kms) lie 3
degrees away (1.4~Mpc impact parameter).

\medskip\par Mrk\,335
\par The high-negative velocity \OVI\ component centered at $-$305\kms\ is
clearly present, as is the secondary component just to the side of the H$_2$
\l1031.191 line. Because the parameters of the H$_2$ line are uncertain, the
systematic error on the equivalent width and column density of the high-negative
velocity components reflects a possible variation from 25 to 35\kms\ in the
width, from $-$20 to 0\kms\ in the velocity as well as a $\pm$20\% amplitude
variation. This increases the systematic error from $\sim$8 to $\sim$20\maa.
\par The separation between the HVC and Galactic absorption is based on the fact
that in most nearby sightlines Galactic absorption extends to about $-$75\kms\
(see \Sect4.1.3 point d).
\par The absorption at 1032.446\aa\ (150\kms\ on the \OVI\ velocity scale;
W=47$\pm$8\maa) is intergalactic Ly$\beta$ at $v$=1965\kms. Its Ly$\alpha$
counterpart can be found in the \GHRS\ spectrum of Mrk\,335 and was listed by
Penton et al.\ (2000) as a feature at $v$=1965\kms\ with W=229$\pm$30\maa. A
second Ly$\alpha$ feature at 2290\kms\ is too weak to have a detectable
Ly$\beta$ counterpart. The sightline to Mrk\,335 passes through the edge of the
\PegCld, which corresponds to GH175. The nearest galaxies in this group
(NGC\,7817, $v$=2532\kms\ and NGC\,7798, $v$=2621\kms) lie 0\fdg8 and 1\fdg7
away (impact parameters 430 and 970\kpc). The Ly$\alpha$\ and Ly$\beta$
absorption lines are probably associated with this galaxy group.
\par The weak features at 1028.659\aa\ and 1034.897\aa\ ($-$950 and 865\kms\ on
the \OVI\ velocity scale; 20$\pm$7 and 15$\pm$5\maa, respectively) may be
Ly$\beta$ and \OVIla\ at 865\kms, associated with the \PegSpur. The
corresponding Ly$\alpha$ lies outside the wavelength range of the \GHRS\
spectrum, which goes down to $v$(Ly$\alpha$)=1420\kms. The resolution of the
\FOS\ spectrum of this object is insufficient to separate Galactic Ly$\alpha$
and geocoronal Ly$\alpha$ from the probable Ly$\alpha$ at 865\kms.

\medskip\par Mrk\,352
\par Side 2 was misaligned in this early observation, so only LiF1A was used to
measure \OVI.
\par The H$_2$ $J$=3 line strongly contaminates the \OVI\ HVC components. Its
presence increases the systematic error on the equivalent width of the HVC \OVI\
components from 23 to 47\maa\ (for the $-$295\kms\ component) and from 25 to
61\maa\ (for the $-$180\kms\ component).
\par This sightline passes only about 1\deg\ from the 2\tdex{18}\cmm2\ contour
of HVC WW466 (the HVC near M33; Wright 1974), which has velocities of about
$-$365\kms. The high-velocity \OVI\ component may be associated with that cloud,
although similar high-velocity \OVI\ is seen in every sightline in this part of
the sky, whether or not the sightline passes close to a \HI\ HVC. It is possible
that there is Ly$\beta$ absorption corresponding to \HI\ outside the 21-cm
contour. However, the higher Lyman lines cannot be checked since there is no SiC
data. The optical depth of any \CIIl\ associated with WW466 is $<$0.2.
\par M\,33 lies 7\fdg4 (90\kpc) away; see \Sect\SLocGrp\ for further discussion.

\medskip\par Mrk\,357
\par The LiF1A and LiF2B spectra differ somewhat, especially in the depth of the
HVC feature.
\par This is one of two sightlines where intermediate-velocity H$_2$
contaminates the \OVI\ line (PG\,1351+640 being the other). After removing the
H$_2$ \l1031.191 absorption, we interpret the HVC absorption as showing two
components, centered at $-$280 and $-$185\kms. However, because of the H$_2$
contamination the systematic uncertainty on both components is rather large.
\par The sightline passes only 3\deg\ from the outer contour of HVC WW466 (the
HVC near M33, Wright 1974), which has velocities similar to the \OVI-HVC
component. There is not enough signal in the SiC channels to search for
associated Lyman lines. The optical depth of \CIIl\ is $<$0.2.
\par M\,33 lies 7\fdg9 (96\kpc) away; see \Sect\SLocGrp\ for further discussion.

\medskip\par Mrk\,421
\par The high-positive velocity wing on the \OVI\ profile is not clearly
separated from the Galactic component. In several nearby directions such a wing
is also seen, but with a clearer separation: at 95\kms\ toward HS\,1102+3441
(4\deg\ away), at 100\kms\ toward PG\,0947+396 (14\deg\ away), and at 115\kms\
toward PG\,1116+215 (18\deg\ away). Toward Mrk\,421 the wing extends rather far
and the separation was placed at 100\kms. The wing is rather weak, however:
37$\pm$11$\pm$29\maa.
\par Both the \OVIla\ and the \OVIlb\ lines can be measured. In the velocity
range $-$130 to 100\kms\ the ratio N(1037)/N(1031) is 1.60$\pm$0.18, making
this one of two sightlines with substantial saturation (the other is Mrk\,876).
\par The galaxy group GH75 ($v$=1025\kms) and the \LeoSpur\ are sampled by this
sightline, but no intergalactic absorption is apparent.



\medskip\par Mrk\,478
\par The \OVI\ absorption in this direction is clearly offset from 0\kms\ and is
centered at $-$20\kms.
\par The sightline passes through the \CVnSpur. A \STIS\ spectrum of Mrk\,478
shows Ly$\alpha$ absorption at $v$=1560\kms. A corresponding weak Ly$\beta$
feature is seen at 1031.119\aa\ ($v$=$-$235\kms\ on the \OVI\ velocity scale) in
the \FUSE\ spectrum.
\par The 79$\pm$17\maa\ feature at 1033.238\aa\ (380\kms\ on the \OVI\ velocity
scale) may be \OVI. It cannot be Ly$\beta$ at 2170\kms, as there is no
corresponding Ly$\alpha$ line in the \STIS\ spectrum of Mrk\,478. There is a
32$\pm$20\maa\ feature in the 1037 line that may correspond to it. There is also
a feature at 1027.179\maa\ that may be Ly$\beta$ at 425\kms. More exposure time
is needed to positively identify the feature.

\medskip\par Mrk\,487
\par There is a component in the \CII\ line at $-$140\kms, although there is no
\HI\ detected at this velocity. However, both to the north and to the south, HVC
complex~C is detected at similar velocities in 21-cm emission.

\medskip\par Mrk\,501
\par This is one of 9 sightlines where \OVI\ absorption is detected at
velocities where complex~C is seen in \HI. However, in this direction, the IVC
complex~K is also present. It is not clear at which velocity to separate the
Galactic and HVC component. A choice was made to cut at a velocity of
$-$100\kms, which is also about where the complex~C and K components can be
separated in \HI. 
\par The sightline also passes through the \DraCld, as well as through grouping
\#70, whose galaxies have $v$=2320$\pm$400\kms. No inter-group Ly$\beta$ is
seen. 
\par Note the feature at 1035.596\aa\ (60$\pm$15\maa) It cannot be Ly$\beta$
since there is no corresponding Ly$\alpha$ line in the \GHRS\ spectrum. It is
most likely \CII\ at a velocity of $-$215\kms. The nearest \HI\ with similar
velocities are parts of complex~C that lie about 5\deg\ away
(N(\HI)$\sim$2\tdex{18}\cmm2). If it is \CII\ in complex~C, the corresponding
\HI\ column density would also be about 2\tdex{18}\cmm2.

\medskip\par Mrk\,506
\par This is one of 9 sightlines with \OVI\ associated with HVC complex~C. The
velocity extents of the \OVI\ and the \HI\ are similar. The HVC and Galactic
components were separated at $-$100\kms, because the \OVI\ components seem to be
split there.
\par The negative-velocity edge of the $-$145\kms\ \OVI\ component is
contaminated by the H$_2$ $J$=3 line, increasing the systematic error from 34 to
39\maa.
\par There is a feature at 1032.549\aa\ (180\kms\ on the \OVI\ velocity scale),
which appears to be present in both the LiF1A and LiF2B channels, and measures
as a 2.0$\sigma$ feature. It cannot be H$_2$ $J$=4. It is at the right velocity
to be intrinsic \SiIIlb, but since no intrinsic Ly$\beta$ is seen this is
unlikely. Possibly, the feature is weak Ly$\beta$ at 1995\kms, although there
are no known galaxy groups in this direction. It is unlikely that it is
high-positive velocity \OVI\ because it is narrow ($b$=18\kms). This feature may
not be real.

\medskip\par Mrk\,509
\par This is among the 10 sightlines with the highest S/N ratio (26 per
resolution element). The flux seems to vary over time, being 12\tdex{-14} \fu\
on 1999 November 2, but 5\tdex{-14} \fu\ on 2000 September 5.
\par The high-negative velocity HVC component is split in two parts by Galactic
H$_2$, but after removing this, it becomes a fairly clear single component lying
between $-$390 and $-$180\kms\ with a secondary component between $-$180 and
$-$100\kms. However, the systematic error is increased from 11 to 31\maa.
Similarly, on the positive-velocity side, the wing becomes clearer after removal
of the H$_2$ $J$=4 line. The component at $v$=150\kms\ is a 7.0$\sigma$
detection.
\par Both the \OVIla\ and the \OVIlb\ lines can be measured. In the velocity
range $-$40 to 100\kms\ the ratio N(1037)/N(1031) is 0.99$\pm$0.05.
\par \CII\ absorption is seen at a velocity of $-$295\kms, with W=53$\pm$10\maa,
or N(\CII)$\sim$6\tdex{13}\cmm2. Assuming carbon is not depleted on dust, this
corresponds to N(\HI)=0.2\tdex{17}/Z\cmm2, with Z the metallicity in solar
units. The Green Bank spectrum sets a 3$\sigma$ upper limit of
$\sim$2\tdex{18}\cmm2\ on N(\HI). A similar component is seen in the
\CII\l1334.532 line (Sembach et al.\ 1999).



\medskip\par Mrk\,618
\par This sightline shows unusually narrow \OVI\ absorption, similar to that
seen in the two objects nearest to it (Mrk\,1095 and HE\,0450$-$2958).

\medskip\par Mrk\,734
\par This is one of two sightlines where very little negative-velocity \OVI\
absorption can be discerned. However, the \OVIlb\ line seems weaker than
expected from the \OVIla\ line. Some of this may be because of low S/N, as there
is a difference between the LiF1A and LiF2B channels -- the 1037 line is more
visible in LiF2B. The noise is sufficiently high that the differences are
(barely) acceptable within the errors: W(1031)=282$\pm$30$\pm$30\maa, while
W(1037)=65$\pm$50$\pm$14\maa, where a value of 120--170\maa\ is expected. There
also seems to be strong positive-velocity \OVI, but the corresponding 1037 line
is hidden in the H$_2$ $J$=1 line.
\par The H$_2$ $J$=4 line contaminates the edges of both the thick disk and the
HVC \OVI\ absorption, but it can be determined well enough that the systematic
errors increase only sightly.
\par The sightline passes through the \LeoSpur, the \LeoCld\ as well as GH78
($v$=1200\kms). Three galaxies in the \LeoSpur\ have an impact parameter of
$\sim$200\kpc\ (NGC\,3627, $v$=623\kms, NGC\,3623 ($v$=693\kms) and NGC\,3593
($v$=510\kms). The absorptions at 1027.460\aa\ and 1028.353\aa\ are Ly$\beta$ at
$v$=510\kms\ and $v$=770\kms; the corresponding Ly$\gamma$ and Ly$\delta$ lines
are weak but probably present. The feature at 1034.506\aa\ is most likely \OVIla\
at 750\kms, associated with one of the Ly$\beta$ absorbers. No absorption is
seen in the more distant groups.

\medskip\par Mrk\,771
\par There may be a slight calibration problem for this spectrum, as the flux
does not go down to zero in the \CIIl\ line. Also, the LiF1A and LiF2B spectra
differ substantially -- the \OVI\ absorption is much stronger in the LiF1A
channel. However, because of the low S/N ratio this difference still lies within
the noise. The most serious problem, however, is that the \OVIla\ line is
probably contaminated by Ly$\beta$. In the \STIS\ spectrum of this object three
Ly$\alpha$ absorbers can be seen, at velocities of 1170, 1875 and 2545\kms, with
equivalent widhts of $\sim$250\maa. Depending on the $b$-value the predicted
strength of the corresponding Ly$\beta$ lines is $\sim$100\maa. The Ly$\beta$
line at 1875\kms\ is expected at 1032.138\aa, or 60\kms\ on the \OVIla\ velocity
scale. It is thus likely that \OVIla\ is contaminated by Ly$\beta$. That this is
the case is also suggested by the fact that the \OVIlb\ line is a factor two to
three weaker than expected. Fortunately, this line is uncontaminated, and the
continuum can be determined well enough that we decided to use \OVIlb\ to
measure N(\OVI). This is the only sightline for which this was possible.

\medskip\par Mrk\,817
\par This is among the 10 sightlines with the highest S/N ratio (29 per
resolution element). It is also one of 9 sightlines in which complex~C is seen
in \OVI\ absorption. The separation between the HVC and Galactic components is
not clear, however. A cut was made at a velocity of $-$90\kms, which is at the
minimum of the \HI\ emission, and is where the \OVI\ profile shows a slight
rise. This is probably real, as the S/N ratio is high.
\par The flux of Mrk\,817 changes from one observation to the next. It was
9\tdex{-14} \fu\ on 2002 February 17, 6\tdex{-14} \fu\ on 2000 December 23, and
13\tdex{-14} \fu\ on 2001 February 18. These fluctuations are probably intrinsic.
\par Both the \OVIla\ and the \OVIlb\ lines can be measured. In the velocity
range $-$90 to 60\kms\ the ratio N(1037)/N(1031) is 1.28$\pm$0.07, showing that
some saturation is present.
\par The extended ledge at positive velocities is intergalactic Ly$\beta$
absorption associated with the \CVnCld\ (more specifically, with GH144,
$v$=2320\kms). The sightline passes just 0\fdg6 (330\kpc\ impact parameter) from
UGC\,9391 $v$=2097\kms). Penton et al.\ (2000) report Ly$\alpha$ absorption in a
\GHRS\ spectrum, with velocities of 1933 and 2097\kms, equivalent widths of
29$\pm$13 and 135$\pm$15\maa, and $b$-values of 34$\pm$13 and 40$\pm$4\kms. The
corresponding Ly$\beta$ lines are then expected at 1032.336 and 1031.897\aa, or
119 and 282\kms\ on the \OVI\ velocity scale, with equivalent widths of 4$\pm$3
and 21$\pm$3\maa. The feature between 145 and 320\kms\ has an equivalent width
of 66$\pm$10\maa, and clearly is Ly$\beta$. The weaker intergalactic absorption
would be too weak too discern, implying that the feature in the wing of the
\OVI\ line between 60 and 145\kms\ must be high positive-velocity \OVI.

\medskip\par Mrk\,829
\par The Galactic \OVI\ absorption sits in the wing of broad intrinsic Ly$\beta$
absorption. This galaxy is part of the \CVnSpur.

\medskip\par Mrk\,876
\par This is one of 9 sightlines toward which \OVI\ associated with HVC
complex~C is seen. Both the \HI\ and the \OVI\ have an extended wing out to
$-$200\kms. The cut between the HVC and the Galactic component was placed at
$-$100\kms, as there are no objective criteria to use.
\par The H$_2$ $J$=3 line contaminates the negative-velocity edge of the HVC
absorption, but it has only a minor influence on the measurement of the
equivalent width. The weak positive-velocity feature is more strongly
contaminated by the H$_2$ $J$=4 line, and it may not be real.
\par At first sight, the \OVIlb\ line appears uncontaminated, while there
appears to be some saturation in the \OVI\ lines. However, it turns out that the
\STIS\ spectrum shows a Ly$\alpha$ line at a velocity of 3457\kms. This has an
equivalent width of 245\maa, and a fitted FWHM of 63\kms. With those parameters
the corresponding Ly$\beta$ should be a 46\maa\ feature lying between $-$30 and
30\kms\ on the \OVIlb\ velocity scale. In this velocity range the equivalent
width of \OVIla\ is 113\maa, while that of \OVIlb\ is 95\maa. The latter could
very well be the sum of 46\maa\ due to Ly$\beta$ and 49\maa\ due to \OVI. We
therefore conclude that there is no evidence for \OVI\ saturation, but that
instead the \OVIlb\ line is contaminated by intergalactic Ly$\beta$.
\par The sightline passes through the \DraCld, but no intergalactic Ly$\beta$ is
seen at its velocities. The Ursa Minor dwarf spheroidal ($v$=17\kms, radius
1.4\kpc) lies 6\fdg6 away (12\kpc\ impact parameter), but any associated \OVI\
would be hidden in the main Galactic absorption.
\par The feature at 1035.175\aa\ cannot be intergalactic Ly$\beta$ at 2725\kms,
as there is no Ly$\alpha$ absorption at that velocity in the \STIS\ spectrum of
Mrk\,876. It is also unlikely to be interstellar \CII\ at $-$330\kms, as there
is no known \HI\ with similar velocities within tens of degrees. It is most
likely intergalactic \OVI\ at 945\kms, with a small amount of H$_2$ $J$=4
included. Other H$_2$ $J$=4 lines of similar strength are much weaker, though
not absent. There is a possible 2$\sigma$ (20\maa) counterpart in the night-only
LiF1A channel at 1040.744\aa, which corresponds to 905\kms\ on the \OVIlb\
velocity scale. However, even in the night data this feature may be affected by
geocoronal OI* emission. There is Ly$\alpha$ absorption at $v$=935\kms\ in the
\STIS\ spectrum of Mrk\,876. The corresponding Ly$\beta$ is heavily contaminated
by the 1028.986\aa\ H$_2$ $J$=3 line, but the feature near this wavelength has a
much broader base then expected if it were just H$_2$, especially when compared
to the other $J$=3 lines.
\par The large (37\kpc\ at 25 mag surface brightness), relatively isolated SB
galaxy NGC\,6140 ($v$=1147\kms; part of the \DraCld) has an impact parameter of
just 260~kpc. It is likely that the 940\kms\ Ly$\alpha$, Ly$\beta$ and \OVI\
absorptions are associated with this galaxy.


\medskip\par Mrk\,926
\par This spectrum has an S/N of 3 per resolution element, and is the noisiest
included in the final sample. It is still clear that the Galactic \OVI\
absorption is weak. \OVI\ absorption can be seen between $-$130 and $-$65\kms,
while between $-$65 and 100\kms\ only a 3$\sigma$ upper limit of 215\maa\ can be
set (log N(\OVI)$<$14.23). High-velocity \OVI\ absorption extends from $-$395 to
$-$65\kms\ and can be split into two components. This absorption is rather clear
and similar to that seen in many other nearby sightlines.
\par The 3$\sigma$ feature at 1032.757\aa\ (240\kms\ on the \OVI\ velocity
scale) is more difficult to identify. It might be \OVI, but the 1037 line is too
noisy to confirm this. It might be Ly$\beta$, but no verifying Ly$\alpha$
spectrum exists (the sightline passes through the \PsASpur). However, it is most
likely a noise feature in the LiF2B channel, as it is less clear in LiF1A.
\par The feature near 1034.748\aa\ (820\kms\ on the \OVI\ velocity scale) is
unidentified. It is no Ly$\beta$ as there is no Ly$\alpha$ absorption at
corresponding velocities in the \STIS\ spectrum.
\par \CII\ absorption is seen at a velocity of $-$240\kms, with
W=120$\pm$45\maa, or N(\CII)$\sim$3\tdex{14}\cmm2. Assuming carbon is not
depleted on dust, this corresponds to N(\HI)$>$0.9\tdex{17}/Z\cmm2, with Z the
metallicity in solar units. The Green Bank spectrum sets a 3$\sigma$ upper limit
of $\sim$6\tdex{17}\cmm2\ on N(\HI).

\medskip\par Mrk\,1095 (Akn\,120)
\par This sightline shows narrow Galactic \OVI, similar to that seen in the
neighboring directions toward Mrk\,618 and HE\,0450$-$2958.
\par The feature at 1030.578\aa\ ($-$390\kms\ on the \OVI\ velocity scale) is
obvious and strong in the LiF2B channels of both observations, but may be absent
in the LiF1A channels. The measured $b$-value is very low (14\kms). This casts
doubt on its reality, even though in the combined spectrum it nominally is a
3.4$\sigma$ detection. If this feature were intergalactic Ly$\beta$ at
$v$=1420\kms, the corresponding Ly$\alpha$ line would have an equivalent width
of 140\maa. Unfortunately, it is not possible to check this, as this line would
be at \l=1221.420\aa, whereas the \GHRS\ spectrum of Mrk\,1095 only extends to
1222.4\aa\ (see Penton et al.\ 2000, who use the name Akn\,120). The \FOS\
spectrum may show a feature at the right wavelength, but is rather noisy.
However, the absence of a known galaxy group with similar velocities argues
against an interpretation as Ly$\beta$. We list it as ``unidentified''.


\medskip\par Mrk\,1383
\par In principle the \OVI\ line could be contaminated by Ly$\delta$ absorption
intrinsic to Mrk\,1383, which is expected at $v$(\OVI)=$-$17\kms. However, no
other Lyman lines are seen at the redshift of Mrk\,1383.
\par Both the \OVIla\ and the \OVIlb\ lines can be measured. In the velocity
range $-$100 to 100\kms\ the ratio N(1037)/N(1031) is 1.23$\pm$0.08, showing that
some saturation is present.
\par The feature at 1032.356\aa\ (125\kms\ on the \OVI\ velocity scale) is
clearly significant and unusual. It is seen in both observations. The sightline
passes through the \VirLibCld, with several galaxies lying 2--3\deg\ away
(1--1.5\Mpc\ impact parameter). Even though this sightline also passes through
GH145 ($v$$\sim$1880\kms), this feature cannot be Ly$\beta$ at 1940\kms, as there
is no corresponding Ly$\alpha$ feature at 1223.529\aa\ in the \STIS\ spectrum of
this target. On the other hand, there is a hint of a counterpart in the \OVIlb\
line, so it is probably high-velocity \OVI.
\par 


\medskip\par Mrk\,1502
\par The strength of the high-velocity component differs somewhat in the LiF1A
channels of the two observations. The LiF2B channel of the first observation
contains no flux and was not used.
\par The negative-velocity edge of the HVC \OVI\ component is contaminated by
H$_2$ $J$=3, which increases the systematic error from 29 to 37\maa.
\par Intrinsic Ly$\gamma$ would be at $v$(\OVI)=21\kms, but no other intrinsic
Lyman lines are seen.

\medskip\par Mrk\,1513
\par The positive-velocity edge of the HVC \OVI\ component is contaminated by
the H$_2$ $J$=3 line, which results in increasing the systematic error from 28
to 33\maa.
\par Both the \OVIla\ and the \OVIlb\ lines can be measured. In the velocity
range $-$75 to 80\kms\ the ratio N(1037)/N(1031) is 1.02$\pm$0.22.
\par The feature at 1033.346\aa\ (415\kms\ on the \OVI\ velocity scale) is
\CIII\l977.020 at z=0.05765 ($v$=17280\kms), an associated system in which
Ly$\alpha$ to Ly$\zeta$ and \OVI\ are also seen. The velocity of this system
relative to Mrk\,1513 is $-$1600\kms. As this is a relatively small velocity
difference with respect to an AGN, we cannot rule out that this absorption
system is associated with Mrk\,1513, rather than an intergalactic cloud.
\par The sightline passes through the edge of the \PegCld, but no intergalactic
absorption is visible within 1200\kms\ of \OVIla.

\medskip\par MS\,0700.7+6338
\par The sightline passes $\sim$1 Mpc from the edge of the \LynCld, but no
obvious associated Ly$\beta$ can be seen.


\medskip\par NGC\,588
\par The LiF1A and LiF2B spectra appear to show some differences. However, the
differences lie within the noise.                                 
\par This object is an \HII\ region in M\,33. The spectrum is therefore an
amalgam of the radiation from many O stars. However, the region near \OVIla\
looks fairly clean -- it is probably dominated by the hottest and brightest
stars. Absorption at Milky Way velocities is very weak -- the 3$\sigma$ upper
limit is 110\maa\ (log N(\OVI)$<$14.07). This is consistent with the weak Milky
Way lines seen in the nearby sightlines to Mrk\,352 (7\fdg1 away), Mrk\,357
(7\fdg8 away) and PG\,0052+251 (9\fdg8 away). The high-negative velocity \OVI\
components are also seen toward those sightlines. The LSR velocity of the \HI\
in M\,33 at the position of NGC\,588 is $-$220\kms\ (Deul \& van der
Hulst 1987), but the \OIl, \ArIl\ and \SiIIla\ absorption lines are centered at
$-$180\kms, while the \CIIl\ absorption extends to $-$230\kms.
\par A-priori one might favor an interpretation in which the \OVI\ components at
$-$370 (W=122$\pm$27$\pm$18\maa) and especially the one at $-$210\kms\
(W=301$\pm$31$\pm$25\maa) are associated with M\,33. However, similar components
are seen toward Mrk\,352 (at $-$295 and $-$180\kms, W=103 and 123\maa), toward
Mrk\,357 (at $-$280 and $-$185\kms, W=131 and 80\maa) and toward PG\,0052+251
(at $-$335 and $-$195\kms, W=153 and 77\maa), as well as toward all other
sightlines in this part of the sky. Comparing these, it is clear that the
$-$370\kms\ component toward NGC\,588 has an equivalent width that is similar to
that of the other very-high negative velocity components. The $-$210\kms\
component, however, is much stronger and this component is probably dominated by
absorption associated with M\,33. It is not possible, however, to disentangle
the relative contributions from M\,33 and the more extended high-negative
velocity \OVI.

\medskip\par NGC\,592
\par This is an \HII\ region in M\,33, but unlike NGC\,588, the combined O-star
spectrum does not provide a simple continuum, and this sightline is not included
in the final sample.

\medskip\par NGC\,595
\par The LiF1A and LiF2B spectra appear to show some differences. However, they
lie within the noise.                                 
\par Like NGC\,588, this object is an \HII\ region in M\,33. The spectrum is
therefore an amalgam of that of many O stars. The region near \OVIla\ looks
fairly clean. Only a 3$\sigma$ upper limit of 126\maa\ (log N(\OVI)$<$14.00) can
be set on the Milky Way absorption. This is consistent with the weak Milky Way
lines seen in the nearby sightlines to Mrk\,352 (7\fdg1 away), Mrk\,357 (7\fdg8
away) and PG\,0052+251 (9\fdg8 away). Two strong high-negative velocity \OVI\
components are also detected. See the notes for NGC\,588 for further comments.

\medskip\par NGC\,604
\par This is the brightest of the four \HII\ regions in M\,33 that were observed
by \FUSE. However, the continuum placement near 1032\aa\ is unclear, as there is
a sharp upturn at slightly longer wavelengths. This target was therefore not
included in the final sample.


\medskip\par NGC\,985
\par The feature at 1031.215\aa\ ($-$205\kms\ on the \OVI\ velocity scale) is
broader than expected if it is solely H$_2$, but after removal of the H$_2$ line
it is too weak to measure, so we decided that it is probably not real.
\par The sightline passes through the middle of the \CetAriCld, and 35 arcmin
(190\kpc\ impact parameter) from NGC\,988, which has $v$=1504\kms. However,
there is no inter-group Ly$\beta$, nor are there Ly$\alpha$ features near
1222.18\aa\ in the \STIS\ spectrum of NGC\,985.

\medskip\par NGC\,1068
\par This object is bright, and in the \OVIla\ region the flux is even higher
because of the wing of strong \OVI\ emission (peak flux 8.5\tdex{-13} \fu)
intrinsic to NGC\,1068. The continuum still seems fairly easy to determine in
the region near 1032\aa.
\par There is a feature centered at 1033.463\aa\ (445\kms\ on the \OVI\ velocity
scale). This is probably \OVI\ associated with NGC\,1068, although its
counterpart is confused with the low-velocity Galactic \OIl\ absorption.
NGC\,1068 is part of the \CetAriCld, but none of the galaxies near NGC\,1068
that are part of this group has a velocity as low as 500\kms, so no
intergalactic Ly$\beta$ is expected.


\medskip\par NGC\,1399
\par The Ly$\beta$ absorption associated with NGC\,1399 is very strong, and
reduces the S/N ratio in the continuum near the \OVIla\ line to about 3.5 per
resolution element. However, it can clearly be established that the Galactic
\OVI\ absorption is weak. The S/N ratio near the \OVIlb\ line is $\sim$8, and a
good measurement is possible. The two measurements are in good agreement, giving
W(\OVIla)=124$\pm$39$\pm$21\maa, and W(\OVIlb)=70$\pm$23$\pm$36\maa. Both are
$\sim$3$\sigma$ detections, but the column density ratio is $\sim$1. In Table~2
the \OVIla\ measurement is listed, but in the channels maps of Fig.~10 the
\OVIlb\ column densities are used, since it has a better S/N ratio.
\par This galaxy is part of the \ForEriCld, and several other galaxies have an
impact parameter of $<$150\kpc. However, they have similar velocities and it is
not clear whether they are in front or behind NGC\,1399.




\medskip\par NGC\,1705
\par This is among the 10 sightlines with the highest S/N ratio (23 per
resolution element). Heckman et al.\ (2001) present a detailed analysis of this
sightline. The continuum placement is somewhat difficult. We decided that the
flux at 1034.5\aa\ represents the continuum level. The implication of this
choice is that there is absorption at all velocities between the Galactic and HVC
components at 0 and +325\kms. On the other hand, this velocity range may show
absorption associated with NGC\,1705. Unfortunately, the continuum is too
complicated to check this using the \OVIlb\ line. Heckman et al.\ (2001) chose
the continuum to be much lower, and do not list the intermediate (180\kms)
component. This would imply a strong wiggle and a sharp upturn at 1034\aa. This
could be justified by arguing that the \OVI\ in NGC\,1705 has a P-Cygni profile.
It would yield an equivalent width of 180$\pm$9\maa\ between $v$=$-$70 and
105\kms\ for the Milky Way (rather than 239$\pm$15 between $-$120 and 120\kms)
and of 120$\pm$8$\pm$9\aa\ between $v$=265 and 400\kms\ for the HVC component
(rather than 200$\pm$9\kms). Half the difference between these values and the
values derived from the higher-placed continuum is included in the systematic
error. 
\par The HVC absorption between 245 and 435\kms\ is associated with HVC WW487
(Wakker \& van Woerden 1991), which was also detected in several other
absorption lines by Sahu \& Blades (1997).
\par The feature at 1033.816\aa\ (550\kms\ on the \OVI\ velocity scale) is \OVI\
associated with NGC\,1705.



\medskip\par NGC\,3310
\par In this sightline the H$_2$ lines near \OVI\ look very broad. Weaker $J$=2
and $J$=3 lines show that there is H$_2$ absorption associated with both the
low- and the intermediate-velocity gas in the sightline. The strongest H$_2$
absorption is associated with the IV-Arch, and this is used to align the
spectrum. Just using the \SiIIla\ and \ArIl\ lines would lead to an incorrect
alignment.
\par This is one of 3 sightlines where the thick disk \OVI\ extends to more
negative velocities than $-$120\kms\ (to $-$135\kms), but in which there is no
obvious separate high-velocity component. This sightline passes through the Ursa
Major window and HVCs complexes~A and C lie just a few degrees away. The
negative-velocity wing may be associated with \OVI\ around the edges of these
HVC complexes.


\medskip\par NGC\,3504
\par Although the S/N ratio in this spectrum is low (3.3 per resolution
element), the Galactic \OVI\ line clearly is very weak. The 3$\sigma$ limit for
the $-$70 to 70\kms\ velocity range is 190\maa\ (log N(\OVI)$<$14.19). Just one
degree away lies Mrk\,36, toward which W(\OVI)= 232$\pm$57\maa.
\par The intrinsic Ly$\beta$ line in NGC\,3504 is clearly visible, and centered
at about 1370\kms. The galaxy's spectrum further shows a depression in the
region between 1030 and 1040\aa. This makes the flux near the Galactic \OVI\
absorption rather low, and the resulting upper limit rather high.



\medskip\par NGC\,3690
\par Although this galaxy is bright, and Galactic \OVI\ absorption is clearly
present, it cannot be measured since the continuum placement is too uncertain,
partly because intrinsic, redshifted \SiIIla\ overlaps the negative-velocity
wing of the \OVI.

\medskip\par NGC\,3783
\par Although this is a bright object with a high S/N spectrum, the contamination
by both H$_2$ and intrinsic lines is too severe to allow a reliable measurement
of the Galactic \OVI\ line, even though it is clearly present.


\medskip\par NGC\,3991
\par The spectra in the LiF1A and LiF2B channels look somewhat different, but
especially the LiF2B channel is rather noisy. The combined spectrum was still
used.
\par The short wavelength wing of the Ly$\beta$ absorption associated with
NGC\,3991 extends to near the \OVIla\ line, but only marginally distorts the
continuum there. A major problem with this sightline is that the \SiIIla\ line
associated with NGC\,3991 lies adjacent to the negative-velocity edge of the
\OVIla\ absorption. Fitting other intrinsic lines (\CIIl, \OIl, \ArIl, \SiIIlb)
shows that the feature centered at $-$108\kms\ is \SiIIla\ in NGC\,3991. This
feature was fitted and removed before measuring the Galactic N(\OVI).

\medskip\par NGC\,4151
\par The Galactic \OVI\ line sits in a relatively clean part of the spectrum,
which shows strong emission lines associated with NGC\,4151. The components seen
in \OVI\ can also be identified in the \CIV\ profile, which was analyzed by
Brandt et al.\ (2001).
\par There is a second \FUSE\ spectrum of this object (P2110201), which we did
not include in our analysis, as it has a shorter integration time. However, this
observation shows that the feature at 1032.520\aa\ (175\kms\ on the \OVI\
velocity scale) is time-variable and thus associated with NGC\,4151.

\medskip\par NGC\,4214
\par This object is bright and the observation is relatively long, resulting in
a spectrum with high S/N ratio (12 per resolution element). Strong Galactic
\OVI\ absorption can be seen, as well as strong \OVI\ and Ly$\beta$ associated
with NGC\,4214. However, the continuum placement in the 1030--1034\aa\
wavelength region is too uncertain. Placing it low would yield
W(\OVI)=165$\pm$25\maa, whereas placing it high gives W(\OVI)=265$\pm$20\maa.
Both of these continuum placements can be defended. Because of this large
uncertainty NGC\,4214 was excluded from the final sample.









\medskip\par NGC\,4649
\par The Galactic \OVI\ line lies in the wing of the strong Ly$\beta$ line
associated with the galaxy itself, which is a galaxy in the Virgo cluster. This
substantially reduces the S/N ratio, but the Galactic \OVI\ line is still
clearly visible and can easily be measured.

\medskip\par NGC\,4670 (Haro\,9)
\par This passes within 1\fdg5 of the North Galactic Pole: $b$=88\fdg6.
\par Intrinsic Ly$\beta$ absorption can clearly be seen, centered at 1040\kms.
\par Both the \OVIla\ and the \OVIlb\ lines can be measured. In the velocity
range $-$80 to 80\kms\ the ratio N(1037)/N(1031) is 0.88$\pm$0.17.
\par The feature at 1033.155\aa\ is \OVI\ absorption at $v$=355\kms. There is
clearly a counterpart in the other line of the doublet. The velocity of this
\OVI\ doublet relative to NGC\,4670 is $-$720\kms. Since NGC\,4670 is a dwarf
galaxy (albeit one with vigorous star formation), it seems unlikely that this
\OVI\ absorption is associated with that galaxy. In no other direction in this
part of the sky is \OVI\ absorption seen at such high positive velocities (see
Fig.~10l). There is also absorption at this velocity in the Ly$\beta$,
Ly$\gamma$ and Ly$\epsilon$ lines (Ly$\delta$ is blended with Galactic \OI).


\medskip\par NGC\,5236 (M\,83)
\par This galaxy is a nearby, large starburst (velocity 516\kms, diameter
16.1\kpc\ or 1.2 arcmin at its distance of 4.7 Mpc). The \FUSE\ aperture covers
only part of the galaxy. The flux is probably due to a mixture of many O stars.
Very strong \OVI\ absorption is seen between $-$200 and 740\kms, but the Galactic
contribution is unclear. This sightline was therefore excluded from the final
sample.

\medskip\par NGC\,5253 (Haro\,10)
\par This small (3.4\kpc\ diameter), nearby (3.2 Mpc) starburst was observed
using the MDRS aperture. A strong, wide Ly$\beta$ line can be seen, which is so
wide that the continuum near \OVIla\ becomes uncertain. \OVI\ absorption near
400\kms\ that is associated with NGC\,5253 is also seen, as is absorption at all
velocities between 100 and 450\kms. Thus, not only is the continuum difficult to
place, it is further unclear which absorption is Galactic and which is not.
NGC\,5253 was therefore excluded from the final sample.

\medskip\par NGC\,5461
\par This object is an \HII\ region in M\,101. Broad \OVI\ absorption can be
discerned, which is a mixture of Galactic \OVI\ and \OVI\ in NGC\,5461. In
addition, the S/N ratio is low and the placement of the continuum is difficult.
This sightline was therefore excluded from the final sample.

\medskip\par NGC\,5548
\par The \OVI\ absorption looks narrower in the LiF2B channel than in the LiF1A
channel, but both channels were combined to create the final spectrum.



\medskip\par NGC\,7469
\par The strong high-negative velocity HVC component is split in two by the
Galactic H$_2$ $J$=3 absorption, making the equivalent width measurement
difficult. However, some of the absorption near $-$210\kms\ is \OVI, rather than
H$_2$. After removal of the H$_2$ line, the high-velocity component can be split
in two components. This direction lies within the tip of the \HI\ Magellanic
Stream, and there is \HI\ at $-$333 \kms. Thus, some of the \OVI\ HVC component
at $-$305\kms\ may be associated with the Stream.
\par The systematic error on the equivalent width and column density of the
high-negative velocity components reflects the uncertainty in the H$_2$
parameters. This increases the systematic errors from 9 to 21\maa\ and from 17
to 40\maa\ for the $-$305 and $-$185\kms\ \OVI\ components, respectively.
\par Based on the strength of the H$_2$ $J$=0 and $J$=1 lines (they have damping
wings, and N(\H2)$>$\dex{19}\cmm2), the HD line at 1031.912\aa\ may be present.
At the wavelengths of the HD 3--0 R(0) through 8--0 R(0) lines at 1066.271,
1054.433, 1042.847, 1031.912, 1021.456 and 1011.457\aa\ (Dabrowski \& Herzberg
1976) features with equivalent widths of 21$\pm$8, 12$\pm$6, 29$\pm$11,
19$\pm$7, 15$\pm$7 and 15$\pm$7\maa\, respectively, can be measured. It is
therefore likely that the sharp feature centered at 1031.912\aa\ is HD. This
feature was removed before calculating for the Galactic \OVI\ parameters.
\par The sightline passes through the \PegSpur\ and the edge of the \PegCld.
The absorption between $-$375 and $-$120\kms\ cannot be intergalactic Ly$\beta$
since the corresponding Ly$\alpha$ absorption is not seen in the \FOS\ spectrum
of NGC\,7469. A strong Ly$\alpha$ line is found at $v$=3070\kms. The
corresponding Ly$\beta$ absorption mostly overlaps the low-velocity \CIIl\
line, and it causes the sloped negative-velocity wing on that line.
\par \CII\ absorption is seen at a velocity of $-$335\kms, with
W=185$\pm$10\maa. An \HI\ component with N(\HI)=3.3\tdex{18}\,\cmm2\ is seen in
the Effelsberg spectrum with a FWHM of 27\kms. This is compatible with the \CII\
absorption for gas with Magellanic abundances.

\medskip\par NGC\,7496
\par Strong intrinsic Ly$\beta$ absorption centered at $v$=1370\kms\ in NGC\,7496
hides the Galactic \OVI\ line.

\medskip\par NGC\,7673
\par Absorption in NGC\,7673 is seen in the Ly$\beta$, Ly$\gamma$, Ly$\delta$,
Ly$\epsilon$ and \CIIl\ lines, centered near the nominal velocity of 3408\kms.
This places the intrinsic \SiIIla\ line at 1032.302\aa, or 112\kms\ on the \OVI\
velocity scale. The Ly$\beta$ damping wings indicate that
N(\HI)$\sim$1.4\tdex{21}\cmm2. Most of the low-ion lines appear to be very
broad; they seem to be several hundred \kms\ wide (e.g., the \CIIl\ line seems
to range from 3000 to 3600\kms) \SiIIlb\ is also present, and strong. The
\SiIIla\ line will have an optical depth that is 10 times lower, on the order of
0.1--0.5. The upshot of the presence of the strong, wide intrinsic low-ion lines
is that the continuum near the Galactic \OVIla\ line is very unreliable and thus
we do not include this sightline in the final sample.

\medskip\par NGC\,7714
\par This direction lies within the tip of the Magellanic Stream and there is
\HI\ at $v$=$-$316\kms. The high-velocity \OVI\ component lies at slightly less
negative velocities. It may or may not be associated with the Magellanic Stream.
\par Absorption in NGC\,7714 is seen in the Ly$\beta$, \CII, \SiIIla,
\FeII\l1144.938 and \FeII\l1063.176 lines, centered at a velocity of 2725\kms.
This places the intrinsic \SiIIla\ line at 1029.975\aa\ ($-$565\kms\ on the
\OVI\ velocity scale). This line therefore does not contaminate the Galactic
\OVI.

\medskip\par PG\,0052+251
\par The Galactic \OVI\ component is very weak in this direction. The 3$\sigma$
upper limit is 115\maa\ (log N(\OVI)$<$13.97). This sightline crosses a HVC at
$-$121\kms\ (WW478), but no \OVI\ absorption is seen at those velocities. The
sightline also lies only about 2\deg\ from cloud WW466 (the HVC near M33, Wright
1974), which has a velocity of $-$365\kms. High-velocity \OVI\ absorption at
similar velocities is clearly seen. This 4.6$\sigma$ component may be associated
with that cloud, or the \OVI\ and/or \HI\ may be Local Group gas. A secondary
3.5$\sigma$ component is also seen at $-$195\kms.
\par \CII\ absorption is seen at a velocity of $-$285\kms, with W=78$\pm$29\maa,
or N(\CII)$\sim$1\tdex{14}\cmm2. Assuming carbon is not depleted on dust, this
corresponds to N(\HI)=0.3\tdex{17}/Z\cmm2, with Z the metallicity in solar
units. The Green Bank spectrum sets a 3$\sigma$ upper limit of
$\sim$3\tdex{18}\cmm2\ on N(\HI).

\medskip\par PG\,0804+761
\par There are two observations for this object. The flux on 1999 October 5 was
9\tdex{-14} \fu, while on 2000 January 4 it was 14\tdex{-14} \fu. This change
may be due to a problem with early \FUSE\ spectra, but it also may be due to an
intrinsic change in the flux.
\par The redshift of PG\,0804+761 is given as 0.100 by {\it NED}, based on the
original data from Green et al.\ (1986). However, as Richter et al.\ (2001a)
discuss, the actual redshift is probably 0.102.
\par The feature at 1033.867\aa\ (564\kms\ on the \OVI\ velocity scale) is
intrinsic Ly$\epsilon$. Ly$\beta$ and Ly$\gamma$ are detected (Richter et al.\
2001a), with equivalent widths that are compatible with this interpretation.
\par This sightline passes about 0\fdg4 (170\kpc\ impact parameter) from
UGC\,4238 ($v$=1714\kms), which is part of the \UMaCld. No associated Ly$\beta$
absorption is visible.

\medskip\par PG\,0832+251
\par In spite of low S/N (3.5 per resolution element) this sightline gives a
fairly clean detection of Galactic \OVI.
\par The sightline passes through the \CncLeoCld. The feature at 1033.106\aa\
(345\kms\ on the \OVI\ velocity scale) is probably Ly$\beta$ at 2160\kms.
Observations of Ly$\alpha$ are required to confirm this interpretation. 

\medskip\par PG\,0832+675
\par This is one of two distant halo stars in the sample (d=8.1~kpc, z=4.7~kpc;
Ryans et al.\ 1997, spectral analysis by Hambly et al.\ 1996). The star is
classified as a post-AGB star, with T=23,000 K. It is one of three sightlines
toward HVC complex~A. A relatively large number of unidentified, but probably
stellar, features can be seen. Since there is no \CIIl, \SiIIla\ or \OIl\
absorption associated with complex~A, the star sets a lower distance limit to
that HVC, and a HVC \OVI\ component is not expected.

\medskip\par PG\,0844+349
\par The wing at velocities up to 250\kms\ seems to connect smoothly to the
Galactic component, but after removing the H$_2$ $J$=4 line a fairly clear break
can be discerned (see the apparent column density panel in Fig.~1).
\par The feature at 1035.380\aa\ (1000\kms\ on the \OVI\ velocity scale) is
intrinsic Ly$\gamma$ absorption ($v$=19372\kms). Many Lyman lines as well as
\CIII\ and \OVI\ are seen at this velocity.
\par The sightline passes near the edge of the \LeoSpur. The features at
1033.210\aa\ and 1033.494\aa\ (375 and 455\kms\ on the \OVI\ velocity scale)
have strengths of 35$\pm$9 and 100$\pm$13\maa. If these were intergalactic
Ly$\beta$ at 2190 and 2270\kms, there should be corresponding Ly$\alpha$
absorption at 1224.55 and 1224.88\aa\ with a total equivalent width on the order
of 300\maa. In the \FOS\ spectrum of PG\,0844+349 there is a 600\maa\ feature,
but it lies at 1226.17\aa, or 2600\kms. However, this is completely absent in
the short (600~sec) low-resolution (150\kms\ bins) \STIS\ observation of
PG\,0844+349. There is also no room for a 300\maa\ feature near 2200\kms.
\par Therefore, an alternative interpretation for the 1033.210 and 1033.494\aa\
features is that they are intergalactic \OVI. For the higher-velocity feature
the \OVIlb\ line overlaps the Galactic \OIl\ line and is impossible to recover.
For the lower-velocity feature there may possibly be an \OVIlb\ counterpart, but
the continuum placement is uncertain, and the possible counterpart is centered
at 405, rather than 375\kms. On balance, we decided to classify the 375 and
455\kms\ features as possibly being \OVI. Better data for the \FUSE\ SiC
channels as well as a high-resolution \STIS\ observations are required to
exclude the possibility of Ly$\beta$.


\medskip\par PG\,0947+396
\par The LiF1A and LiF2B spectra differ substantially in this sightline, but
they both show the high-positive velocity \OVI\ component. The combined spectrum
was used.
\par The \HI\ spectrum in this direction is complex and shows two
intermediate-velocity components, at $-$66 and $-$48\kms. The first of these is
associated with the IV16 core. However, there appears to be no \OVI\ at the
velocities of the IV \HI\ components.
\par The sightline lies just 2\deg\ from PG\,0953+414. The Galactic \OVI\ column
density differs little, but the HVC is 1.7 times stronger toward PG\,0947+396.
\par The sightline passes through the \LeoSpur\ and the \LeoCld. Weak (90\maa)
absorption associated with the latter group may be seen at 1030.811\aa\
(1485\kms, $-$325\kms\ on the \OVI\ velocity scale). This feature is weak enough
that the corresponding Ly$\alpha$ line (expected to be $\sim$250\maa) does not
show up clearly in the \FOS\ spectrum.
\par The broad feature between 1033.430 and 1035.178\aa\ (440 to 940\kms\ on the
\OVI\ velocity scale) has low significance (260$\pm$120\maa). It is barely
visible in the LiF1A data, but adding in LiF2B strengthens it somewhat. We
decided it may not be real and classify it as ``unidentified''.

\medskip\par PG\,0953+414
\par This is among the 10 sightlines with the highest S/N ratio (23 per
resolution element). The flux varied from 7\tdex{-14} \fu\ on 1999 December 30
to 4\tdex{-14} \fu\ on 2000 May 4. This may be an intrinsic variation.
\par There is no clear separation between the Galactic component and the
high-velocity wing. In several nearby directions such a wing is also seen, but
with a clearer separation: at 100\kms\ toward PG\,0947+396 (2\deg\ away), at
95\kms\ toward HS\,1102+3441 (15\deg\ away) and at 120\kms\ toward PG\,0844+349
(15\deg\ away). Based on this, a velocity of 100\kms\ was chosen to separate the
HVC and Galactic \OVI\ absorption.
\par Both the \OVIla\ and the \OVIlb\ lines can be measured. In the velocity
range $-$100 to 100\kms\ the ratio N(1037)/N(1031) is 0.99$\pm$0.011.
\par The identification of the feature at 1034.139\aa\ (645\kms\ on the \OVI\
velocity scale) remains uncertain. It is visible in both of the LiF1A and the
LiF2B channels, and thus appears real. In the \STIS\ spectrum there is a weak
Ly$\alpha$ absorber at 2555\kms with W=50\maa. The corresponding Ly$\beta$
feature is expected at 735\kms\ on the \OVI\ velocity scale, but with a strength
of just 6\maa. The sightline also contains many intergalactic absorbers. For a
system with z=0.05876 the \CIIIl\ line might lie at 1035.487\aa\ (730\kms\ on
the \OVI\ velocity scale), but only Ly$\alpha$ (and no \OVI) is found in this
system. For a system with z=0.14258 \CII\l903.9616 might lie at 1035.481\aa\
(270\kms\ on the \OVI\ velocity scale), but nothing is present there. For none
of the other systems with strong Ly$\alpha$ do strong absorption lines end up
near 1034.139\aa.
\par The most likely explanation for the 1034.139\aa\ feature is therefore that
it is \OVI\ at 645\kms. If just the two LiF1A channels are combined it measures
as 52$\pm$16\maa, though if the LiF2B channels are added in this is reduced to
40$\pm$12\maa. There appears to be a corresponding \OVIlb\ feature at 640\kms,
which measures as 30$\pm$14\maa\ in the LiF1A channels, and as 27$\pm$15\maa\
in the combined LiF1A+LiF2B data.


\medskip\par PG\,1001+291
\par The \OVI, \SiIIla, \ArIl\ and \CIIl\ lines in the LiF2B channel look rather
different than those in the LiF1A channel, so only LiF1A was used to measure
\OVI.
\par This is one of several sightlines near $l$=180\deg\ in which a
high-positive velocity wing is seen. It was separated from the Galactic
absorption by choosing to cut at a velocity of 100\kms. This is similar to the
velocity at which a fairly clear separation exists in neighboring sightlines (at
100\kms\ toward PG\,0947+396, 11\deg\ away, at 95\kms\ toward HS\,1102+3441,
14\deg\ away, and at 120\kms\ toward PG\,0844+349, 17\deg\ away).
\par Although this sightline passes through the \LeoSpur, the \LeoCld\ and GH51
($v$=1600\kms), and there are two galaxies nearby (UGC\,5340, $v$=441\kms,
impact parameter 165\kpc, and UGCA\,201=Haro\,23, $v$=1402\kms, impact parameter
170\kpc), there is no Ly$\beta$ absorption associated with the intersected
groups.
\par Nevertheless, there are some unidentified features in the \FUSE\ and \FOS\
spectra of this object. The strong line at 1028.96\aa\ may be Ly$\beta$ at
$v$=945\kms, rather than H$_2$ $J$=3, as all other H$_2$ $J$=3 lines are much
weaker. Further, in the \FOS\ spectrum of this target (Savage et al.\ 2000)
there are features near 1217.84 and 1223.37\aa\ (at velocities of 535 and
1900\kms\ if they are Ly$\alpha$), but these have no apparent Ly$\beta$
counterpart.

\medskip\par PG\,1004+130
\par This is one of 3 sightlines where the thick disk \OVI\ extends to more
negative velocities than $-$120\kms\ (to $-$130\kms), but in which there is no
clear case for a separate high-velocity component. Both the \OVIla\ and the
\OVIlb\ lines can be measured. In the velocity range $-$90 to 80\kms\ the ratio
N(1037)/N(1031) is 1.16$\pm$0.16.
\par Many of the H$_2$ $J$=3 lines are seen to be broad (average FWHM of
70\kms), which suggests that H$_2$ is present in the weak intermediate-velocity
\HI\ components. The $J$=4 \l1032.350 line appears present but weak. However, no
independent width measurement can be obtained from other $J$=4 lines, so a FWHM
of 70 \kms\ (equal to that found for the $J$=3 lines) was assumed to remove it
from the Galactic \OVI\ absorption.
\par The sightline goes through the \LeoSpur, the \LeoCld\ and the \CncLeoCld.
The feature at 1030.136\aa\ ($-$520\kms\ on the \OVI\ velocity scale) is
Ly$\beta$ at 1290\kms, as there clearly is a matching Ly$\alpha$ feature in the
\STIS\ spectrum of this object.
\par There are several features whose identification remains problematic. These
lie at 1032.701, 1033.048, 1034.122 and 1034.644\aa\ (225, 325, 640 and 790\kms\
on the \OVI\ velocity scale). They cannot be \OVI, as they clearly do not have
counterparts in the \OVIlb\ line. They are very unlikely to be Ly$\beta$, as no
strong Ly$\alpha$ absorption is seen at corresponding velocities. The most likely
possibility seems to be that they are several instances of \OIII\l832.927 at a
redshift of 0.2398 to 0.2422, as the redshift of PG\,1004+130 itself is 0.2400.
The velocities and relative strengths of these features are inconsistent with an
interpretation as redshifted \OII\ll832.757, 833.329, 834.466. The Galactic
\OVI\ absorption seems unaffected by redshifted \OIII, since the N$_a$(v)
profiles of the \OVIla\ and \OVIlb\ lines agree very well.


\medskip\par PG\,1048+342
\par Several galaxies in the \LeoCld\ lie close to this sightline: NGC\,3442
($v$=1713\kms, 20 arcmin, impact parameter 160\kpc), NGC\,3430 ($v$=1533\kms,
1\deg, impact parameter 490\kpc), NGC\,3395 and NGC\,3396 ($v$=1595 and
1649\kms, 1\fdg1, impact parameter 525\kpc). Ly$\beta$ absorption from the group
is seen at about 1031.6\aa\ ($v$=1720\kms, $-$90\kms\ on the \OVI\ velocity
scale), which overlaps the Galactic \OVI\ and makes it impossible to measure.
That this feature is Ly$\beta$ follows from the combination of facts that it is
very strong, while no \OVIlb\ counterpart can be seen.



\medskip\par PG\,1116+215
\par This is one of the many sightlines near $l$=180\deg\ with high-positive
velocity \OVI. In this case the Galactic and HVC component are clearly separated
and the HVC component (also visible in the \OVIlb\ line) shows internal
structure. Two separate HVC components were measured. There is no contamination
by H$_2$ $J$=4 apparent.
\par Both the \OVIla\ and the \OVIlb\ lines can be measured. In the velocity
range $-$55 to 90\kms\ the ratio N(1037)/N(1031) is 1.09$\pm$0.13, while in the
velocity range 150 to 230\kms\ this ratio is 1.08$\pm$0.16.
\par This is one of the sightlines for which assuming that the \HI\ is centered
around 0\kms\ would yield an erroneous alignment; the strongest \HI\ component
is at $-$40\kms. Assuming that it is at 0\kms\ might lead to the conclusion that
there is high-velocity \CII, but not \CII*.
\par Although the sightline passes through the edge of the \LeoCld, the feature
at 19\kms\ cannot be intergalactic Ly$\beta$ -- the corresponding Ly$\alpha$
line is clearly absent in the \STIS\ spectrum. Ly$\alpha$ lines are detected at
velocities of 960 and 1480\kms, but these are so weak that the corresponding
Ly$\beta$ lines are undetectable. The two features at 1028.813 and 1029.171\aa\
($-$905 and $-$800\kms\ on the \OVI\ velocity scale) are intergalactic
\CII\ll903.6235, 903.9616 in a system at z=0.138 in which many lines are found.


\medskip\par PG\,1211+143
\par The spectrum in the LiF2B channel differs in many ways from that in the
LiF1A channel. For the \OVIla\ line there is a broad $\sim$200\kms\ wide
depression which is centered around 1035\aa. The LiF2B and SiC2A channels also
differ. Since the LiF1A channel appears to behave normally, only that channel
was used for the measurements.
\par Similar to the neighbouring direction toward Mrk\,734, the \OVI\
absorption at negative velocities is weak.
\par The sightline passes through the Virgo Cluster, with NGC\,4212 ($v$=$-$163)
and NGC\,4189 ($v$=2039\kms) having impact parameters of 105 and 185\kpc,
respectively. The 3$\sigma$ feature centered at 1032.938\aa\ (295\kms\ on the
\OVI\ velocity scale) is probably Ly$\beta$ at 2110\kms\ associated with the
\VirCld, as there is a Ly$\alpha$ line at 2115\kms\ in the \STIS\ spectrum.
On the other hand, the feature at 1033.393\aa\ (425\kms\ on the \OVI\ velocity
scale) cannot be Ly$\beta$. Furthermore, although there are 13 Ly$\alpha$
absorbers visible in the \STIS\ spectrum, none of these is at a redshift such
that other Lyman lines, \OVI, \CIII\ or low-ionization lines fall near 1033\aa.
There is a possibility that it is \OVI, but the corresponding \OVIlb\ line
would be hidden by low-velocity \OIl\ absorption. We therefore classify it as
``unidentified''.
\par The feature at 1035.203\aa\ is intergalactic Ly$\gamma$ at 19315\kms\
(z=0.0680). Ly$\alpha$ to Ly$\eta$, \CIII\ and \OVI\ are found in this system.
\par The redshift of PG\,1211+143 is given as 0.08090 in {\it NED}. However,
this is based on a fit to an asymmetric \CIV\ profile at 1650\aa. From the
intrinsic Lyman lines it is clear that the redshift actually is 0.08040.

\medskip\par PG\,1216+069
\par There is a Lyman series at $v$=1880\kms\ in this spectrum, associated with
the Virgo Cluster. The Ly$\alpha$ line can be seen in the \FOS\ spectrum (Savage
et al.\ 2000). The Ly$\beta$ line in this series overlaps the Galactic \OVI\
component. This makes it impossible to measure the Galactic \OVIla\ absorption.
The Galactic \OVIlb\ is not clearly seen either. It is still possible to discern
a high-velocity \OVI\ component in the \OVIla\ and \OVIlb\ lines (near 290\kms),
whose presence would be consistent with other similar components seen in
neighbouring directions (3C273.0, HE\,1228+0131, PG\,1116+215, Mrk\,734).

\medskip\par PG\,1259+593
\par This is the sightline with the highest S/N ratio (30 per resolution
element), and the second longest \FUSE\ observation (633 ks).
\par It is one of 9 sightlines which show \OVI\ absorption associated with HVC
complex~C, and one of the clearest complex~C components. Richter et al.\ (2001b)
analyze the low-ionization absorption lines in complex~C and the IV-Arch. There
is also a high-velocity \OVI\ wing at positive velocities (140\kms), although
this is a weak non-gaussian component (optical depth $<$0.03) that integrates
to a 2.3$\sigma$ detection. This cannot be intergalactic Ly$\beta$ since there
is no corresponding Ly$\alpha$ absorption at 1223.699\aa\ in the \STIS\ spectrum
of PG\,1259+593.
\par Both the \OVIla\ and the \OVIlb\ lines can be measured. In the velocity
range $-$60 to 100\kms\ the ratio N(1037)/N(1031) is 1.29$\pm$0.11, showing that
some saturation is present.
\par The sightline passes through the \UMaCld\ and the \CVnCld. The feature at
1033.578\aa\ (480 \kms\ on the \OVI\ velocity scale) is inter-group Ly$\beta$ at
2295\kms. The Ly$\alpha$ line is clearly seen at 1224.902\aa\ in the \STIS\
spectrum of PG\,1259+593.
\par The galaxy (UGC\,8146; $v$=669\kms) in the closer Ursa Major Galaxy
Grouping lies just 22 arcmin away (90\kpc\ impact parameter), and associated
Ly$\alpha$ and Ly$\beta$ absorption is seen at corresponding velocities. There
may even be \OVI\ absorption at similar velocities. Three weak ($\sim$25\maa)
features appear at 1034.124, 1034.386 and 1034.714\aa\ (640, 715 and 810\kms\ on
the \OVI\ velocity scale). The corresponding \OVIlb\ absorption is confused with
\OI* emission lines, but in the smoothed orbital-night-only data there may be a
2$\sigma$ feature.

\medskip\par PG\,1302$-$102
\par This object has a wavy continuum, so that it is necessary to fit a 5th
order polynomial. This makes the continuum near the \OVIlb\ line too uncertain
to confidently compare the \OVIla\ and \OVIlb\ column densities.
\par The sightline passes through the \VirCld, but although there are several
features that might be Ly$\beta$, none can be confirmed. In particular, the
identification of the feature at 1030.237\aa\ ($-$490\kms\ on the \OVI\ velocity
scale) remains tentative. If it is Ly$\beta$ at 1320\kms\ associated with the
Virgo Cluster, there should be a 500\maa\ Ly$\alpha$ line at 1221.022\aa, but
this is not seen clearly in the \FOS\ spectrum (Savage et al.\ 2000). The
corresponding Ly$\gamma$ line is blended with \CIIIl, which falls at
$v$=1375\kms\ on the Ly$\gamma$ velocity scale. The \CIII\ absorption seems to
have a wing at $v$(Ly$\gamma$)$\sim$1300\kms, however, which is where the
intergalactic Ly$\gamma$ would be expected.
\par The feature at 1032.804\aa\ (255\kms\ on the \OVI\ velocity scale) is most
likely high-velocity \OVI, as there appears to be a corresponding feature in the
\OVIlb\ line, and in many sightlines in the neighbourhood of PG\,1302$-$102
high-positive velocity \OVI\ is also found (e.g.\ HE\,1115$-$1735,
IRAS\,F11431$-$1810, 3C273.0). This feature might be Ly$\beta$ at $v$=2070\kms,
but from the \FOS\ spectrum it is not clear whether there is corresponding
Ly$\alpha$. A \STIS\ spectrum of this target has been taken, but is not yet
public data.
\par The apparent structure in the high-velocity component is caused by noise
peaks in the shorter observations.
\par This spectrum is particularly rich in Lyman systems. There are systems at
z=0.09397 and 0.09484 (both in Ly$\alpha$ to Ly$\eta$, both showing \CIII), at
z=0.14537 (Ly$\alpha$, Ly$\beta$, as well as a weak \CII\l903.9616 at
1035.482\aa, 1033\kms\ on the \OVI\ velocity scale) and at z=0.19160 (Ly$\alpha$
through Ly$\zeta$. Another system near z=0.097 is less clear in just Ly$\alpha$
and Ly$\beta$. In none of these systems do \HI, \CIII, or low-ion lines fall
near Galactic \OVI.

\medskip\par PG\,1307+085
\par This sightline passes through the outer edge of the \VirCld, and 190\kpc\
from UGC\,8091 ($v$=165\kms). However, no associated intergalactic absorption
can be discerned.


\medskip\par PG\,1351+640
\par This is one of two sightlines where intermediate-velocity H$_2$ $J$=3
contaminates the \OVI\ spectrum (Mrk\,357 being the other). It is also one of 9
sightlines in which \OVI\ absorption associated with complex~C is seen. The
Galactic and HVC \OVI\ components are separated at a velocity of $-$100\kms, as
there are no clear components visible.
\par The amount of contamination due to H$_2$ $J$=4 is unclear. Other $J$=4
lines of similar strength are either confused with other H$_2$ lines or in a
part of the spectrum with low S/N ratio. Lines in clearer regions are a factor 2
weaker and absent. The tabulated Galactic \OVI\ column density assumes that
there is no contamination. A cut was placed at 100\kms\ to separate the weak
(3$\sigma$) positive-velocity wing from the Galactic absorption.
\par The sightline passes through the \CVnCld, and there are two galaxies within
1\deg (UGC\,8894, $v$=1943\kms, and Mrk\,277, $v$=1828\kms), but both have an
impact parameter of 320\kpc\ and there does not appear to be Ly$\beta$
absorption associated with the galaxy group.
\par The absorptions centered at 1029.991 and 1030.760\aa\ ($-$560 and
$-$340\kms\ on the \OVI\ velocity scale) are Ly$\delta$ in two absorption
systems at 25330 and 25575\kms\ (z=0.08449 and 0.08531), which are associated
with PG\,1351+640. Corresponding Ly$\beta$, Ly$\gamma$, Ly$\epsilon$ \CIII\ and
\OVI\ are also visible. Considering the extent of the Ly$\gamma$ and
Ly$\epsilon$ lines, the Ly$\delta$ line should extend to $\sim$1030.83\aa, or
$-$320\kms\ on the \OVI\ velocity scale. It does not contaminate the Galactic
\OVI\ absorption.

\medskip\par PG\,1352+183
\par The sightline passes 20 arcmin (135\kpc) from UGC\,8839 ($v$=971\kms),
which lies at the very edge of the \VirLibCld. The feature at 1029.933\aa\
($-$580\kms\ on the \OVI\ velocity scale) may be Ly$\beta$ at 1230\kms\
associated with UGC\,8839.

\medskip\par PG\,1402+261
\par This sightline passes through the \CVnSpur, but no intergalactic absorption
is seen. The \OVI\ profile is simple and clear.

\medskip\par PG\,1404+226
\par This spectrum has a S/N ratio near the limit (3.2 per resolution element).
\par The feature at 1030.516\aa\ ($-$410\kms\ on the \OVI\ velocity scale) is
intrinsic Ly$\epsilon$ absorption at z=0.09886.
\par The sightline passes 90\kpc\ from UGC\,9128 ($v$=196\kms), but no apparent
associated absorption can be discerned near \OVIla.

\medskip\par PG\,1411+442
\par This sightline passes about 5\deg\ north of HVC complex~C. The continuum
placement and the negative-velocity extent of the thick disk \OVI\ do not seem
to be adversely affected by the presence of the absorption centered at
1029.861\aa\ ($-$600\kms\ on the \OVI\ velocity scale), which is intrinsic
Ly$\delta$ absorption at z=0.08436.
\par This sightline passes near the edges of the \DraCld, the \CVnSpur\ and the
\CVnCld, while UGC\,9240 ($v$=281\kms) lies 130\kpc\ away. No associated
intergalactic Ly$\beta$ is visible within 1200\kms\ of \OVIla.
\par The feature at 1034.674\aa\ may be intergalactic \OVI\ at 800\kms\ rather
than intergalactic Ly$\beta$ at 2620\kms, because the Ly$\alpha$ absorption is
not clear at this velocity in the \FOS\ spectrum (although it has low S/N
ratio). A possible counterpart for \OVI\ is seen in the \OVIlb\ line near
1040.455\aa, although the spectrum is noisy and this feature is offset in
velocity by 20\kms. Associated Ly$\delta$ at $v$=25320\kms\ is found at about
1030.5\aa\ ($-$600\kms\ on the \OVI\ velocity scale; $-$1600\kms\ relative to
PG\,1411+442).
\par This is also one of a close triplet of sightlines in our sample.
SBS\,1415+437 lies 46 arcmin away, and PG\,1415+451 is at 65 arcmin (the angular
distance between the latter two is 85 arcmin). This triplet clearly illustrates
the large changes that can occur in W(\OVI) on small angular scales. Toward
PG\,1411+442 a deep (65\% absorption) \OVI\ profile ranges from $-$95 to 90\kms,
with W= 303$\pm$30\maa. Toward PG\,1415+451 the \OVI\ profile (55\% absorption)
ranges from $-$80 to 100\kms, with W= 200$\pm$42\maa. Toward SBS\,1415+437 the
\OVI\ is much shallower (35\% absorption) and ranges from $-$50 to 100\kms, with
W= 145$\pm$39\maa.

\medskip\par PG\,1415+451
\par The LiF1A and LiF2B spectra differ for this object. LiF1A shows a fairly
narrow central \OVI\ absorption with an extended positive-velocity wing, whereas
LiF2B is shallower, and the positive-velocity wing is not as clear. The combined
spectrum was still used.
\par The feature centered at 210\kms\ is a 3.0$\sigma$, and appears to be real.
It is likely that this is Ly$\gamma$ at z=0.0618 (18525\kms), as there are
matching features at the wavelengths corresponding to Ly$\alpha$, Ly$\beta$ and
possibly Ly$\delta$. All these features are $\sim$2--3$\sigma$, but they match
if log N(\HI)=14.9$\pm$0.6 and $b$=12$^{+17}_{-6}$ \kms.
\par This sightline passes near the edges of the \DraCld, the \CVnSpur\ and the
\CVnCld. UGC\,9240 ($v$=281\kms) lies 90\kpc\ away. No intergalactic Ly$\beta$
is visible. See also the notes to PG\,1411+442 about the triplet of sightlines
of which PG\,1415+451 is a part.



\medskip\par PG\,1444+407
\par The feature at 1034.913\aa\ (870\kms\ on the \OVI\ velocity scale) is
Ly$\beta$ at 2685\kms, as there is a Ly$\alpha$ counterpart of the proper
strength in the \FOS\ spectrum. There is a small group of nine galaxies nearby:
the \BooCld, although the closest galaxy (I\,Zw\,97; $v$=2669\kms) lies 2\deg\
away (1.4\Mpc\ impact parameter).
\par The 90\maa\ feature at 1031.096\aa\ ($-$240\kms\ on the \OVI\ velocity
scale) is a 3.0$\sigma$ detection. It is unlikely to be \OVI\ since there is no
counterpart in the \OVIlb\ line, which would fall between the Galactic \CIIl\
and \CII*\l1037.018 lines. If it is intergalactic Ly$\beta$ at 1570\kms, the
corresponding Ly$\alpha$ line at 1222.035\aa\ is not clearly visible in the
\FOS\ spectrum of this object, though its presence is not excluded.



\medskip\par PG\,1626+554
\par This is one of 9 sightlines toward which \OVI\ associated with HVC
complex~C is seen. The absorption near $-$210\kms\ mimics a H$_2$ $J$=3 line,
but no evidence is seen for any of the other $J$=3 lines. Similarly, no evidence
is seen for any $J$=4 H$_2$ lines, which implies that the feature at 140\kms\ is
high-positive-velocity \OVI. The $-$150\kms\ component is the strongest of the 9
complex~C components (W=220$\pm$37\maa).
\par This sightline passes just 13\kpc\ from the Draco dwarf spheroidal
($v$=$-$31\kms), which has a 1.8\kpc\ diameter. Strong corresponding absorption
is not evident, but weak absorption would be hidden in the Galactic profile.


\medskip\par PG\,2349$-$014
\par The Galactic \OVI\ absorption is very weak; the 3$\sigma$ upper limit for
the $-$70 to 70\kms\ velocity range is 120\maa\ (log N(\OVI)$<$13.97). Two clear
high-negative velocity components are present, dissected by a very strong H$_2$ 
$J$=3 line. This is one of three sightlines where high-negative velocity \HI\
($v$=$-$300\kms) in the tip of the Magellanic Stream is also detected. The
relation with the high-velocity \OVI\ is unclear.
\par The identity of the feature at 1032.537\aa\ (180\kms\ on the \OVI\ velocity
scale) is unclear. It measures as 2.6$\sigma$, but it seems real. It is most
likely Ly$\beta$ at 1990\kms, even though the nearest galaxies with similar
velocities have an impact parameter of 3~Mpc. In the low-resolution (300\kms)
\STIS\ snapshot a possible Ly$\alpha$ match can be discerned.    
\par \CII\ absorption is seen at a velocity of $-$290\kms, with
W=245$\pm$37\maa. An \HI\ component with N(\HI)=3.0\tdex{18}\,\cmm2\ is seen in
the Green Bank spectrum, with a FWHM of 49\kms. This is compatible with the
\CII\ absorption for gas with Magellanic abundances.


\medskip\par PHL\,1811
\par This sightline has a Lyman limit system at z=0.08088 (985\aa). There are
also absorption systems at z=0.07339, 0.07774, 0.07895, 0.13220, 0.13536 and
0.17645. None of these systems yield lines that can interfere with the Galactic
\OVI\ absorption; the wavelength range of the z$\sim$0.08 system is redshifted
to 1031--1033\aa\ is 954--962\aa\ and the only strong interstellar line in this
range is \NI\l954.104. In the system at z=0.08088, this line would be at
$v$=$-$210\kms\ on the \OVI\ LSR velocity scale. However, in that case the
\OIlb\ line should be very strong (the optical depth ratio \OIlb/\NI\l954.104 is
$\sim$100), but only a weak \OIlb\ line is seen..
\par {\it NED} lists the redshift of this object as 0.190, even though the
original publication (Leighly et al.\ 2001) lists 0.192. 
\par \CII\ absorption is seen at a velocity of $-$195\kms, with
W=200$\pm$24\maa. The Leiden-Dwingeloo Survey \HI\ spectrum sets an upper limit
on N(\HI) at this velocity of $\sim$4\tdex{18}\cmm2. Assuming that the
$b$-value is large, this is compatible with the \CII\ absorption for gas with
Magellanic abundances, even though this sightline lies 15\deg\ away from where
the Stream is detected in \HI\ 21-cm emission.

\medskip\par PKS\,0405$-$12
\par This QSO has a flat continuum and high S/N ratio. The weak Galactic \OVI\ is
clearly seen, as is a weak positive-velocity wing.
\par Both the \OVIla\ and the \OVIlb\ lines can be measured. In the velocity
range $-$40 to 50\kms\ the ratio N(1037)/N(1031) is 0.96$\pm$0.21.
\par The sightline passes through the outskirts of the \DorCld, but no Ly$\beta$
is seen. There is a Lyman limit system at z=0.167, in which many metal lines can
be seen.

\medskip\par PKS\,0558$-$504
\par This is one of 2 sightlines where the thick disk \OVI\ extends to more
positive velocities than 120\kms\ (to 135\kms), but in which there is no clear
case for a separate high-velocity component. Both the \OVIla\ and the \OVIlb\
lines can be measured. In the velocity range $-$100 to 100\kms\ the ratio
N(1037)/N(1031) is 1.13$\pm$0.15.
\par The sightline also passes through the \DorCld, but no intergalactic
Ly$\beta$ can be discerned.
\par This sightline passes only about 2\deg\ from HVC WW425, a cloud with
$v$=260\kms, which may be part of the Magellanic Stream. The high-positive
velocity \OVI\ feature may be associated with that HVC.

\medskip\par PKS\,2005$-$489
\par A HVC component similar to the one at 155\kms\ is seen toward
ESO\,141$-$G55, which lies 12\deg\ away. Judging from the shape of the wing of
the \OVIlb\ line this feature does seem to be high-velocity \OVI. It is not
clear where to cut the HVC and Galactic components, but there seems to be a
discontinuity near 120\kms.
\par Both the \OVIla\ and the \OVIlb\ lines can be measured. In the velocity
range $-$50 to 120\kms\ the ratio N(1037)/N(1031) is 0.94$\pm$0.07.
\par The sightline passes within 400\kpc\ of a subgroup of 5 galaxies in the
\TelGruCld. Intergalactic Ly$\beta$ may be present. However, if such a line
contaminates the \OVI\ line, the corresponding Ly$\gamma$ line is too weak to
discern. There is a Lyman series at $v$=5075\kms\ in this spectrum, in which
Ly$\beta$, Ly$\gamma$, Ly$\delta$ and \CIII\ are seen.

\medskip\par PKS\,2155$-$304
\par This is among the 10 sightlines with the highest S/N ratio (27 per
resolution element). There are three observations, with differing fluxes. On
1999 October 23/24 the flux was 17\tdex{-14} \fu, while on 2001 June 18 the flux
was 11\tdex{-14} \fu. This change is probably intrinsic to PKS\,2155$-$304.
\par There is no H$_2$ present in this sightline, so the absorption between
$-$280 and $-$85\kms\ is all \OVI. The separation between the HVC and Galactic
absorption is based on the fact that in most nearby sightlines Galactic
absorption extends to about $-$85\kms\ (see \Sect4.1.3 point d).
\par Both the \OVIla\ and the \OVIlb\ lines can be measured. In the velocity
range $-$70 to 120\kms\ the ratio N(1037)/N(1031) is 1.05$\pm$0.07.
\par The sightline passes through the \PsASpur, but no intergalactic Ly$\beta$
is visible.
\par \CII\ absorption is seen at a velocity of $-$130\kms, with W=48$\pm$7\maa,
or N(\CII)$\sim$5\tdex{13}\cmm2. Assuming carbon is not depleted on dust, this
corresponds to N(\HI)=0.2\tdex{17}/Z\cmm2, with Z the metallicity in solar
units. The Green Bank spectrum sets a 3$\sigma$ upper limit of
$\sim$2\tdex{18}\cmm2\ on N(\HI).











\medskip\par SBS\,0335$-$052
\par The Galactic \OVI\ absorption is very weak. Only a 3$\sigma$ upper limit of
120\maa\ (log N(\OVI)$<$13.98) can be set. The intrinsic Ly$\beta$ absorption
(centered at 4050\kms\ or 1039.5\aa) provides a fairly smooth continuum. Low
Galactic \OVI\ column densities are also seen in several other nearby sightlines
(NGC\,1068, Mrk\,618, Mrk\,1095 and PKS\,0405$-$12).

\medskip\par SBS\,1415+437
\par On the LiF1 side, only 3 of the 12 exposures have signal. Thus, the LiF2B
channel has a better S/N ratio than LiF1A.
\par This sightline passes near the edges of the \DraCld, the \CVnSpur\ and the
\CVnCld. UGC\,9240 ($v$=281\kms) lies 110\kpc\ away. No intergalactic Ly$\beta$
is visible near these velocities. See also the notes to PG\,1411+442 about the
triplet of sightlines of which SBS\,1415+437 is a part.

\medskip\par Tol\,0440$-$381
\par The emission-like feature at the short wavelength side of the \OVI\
absorption is troublesome. It occurs in both the LiF1A and the LiF2B channels,
and is present in both the orbital day and orbital night data. It is not at the
redshift of an absorption system. If it were intrinsic to Tol\,0440$-$381 it is
emission at about 990\aa, in the wavelength region near \NIII, \OI\ and \SiII,
which makes this unlikely. In any case, this feature makes the extent of the
Galactic \OVI\ absorption too difficult to determine and we did not include this
sightline in the final sample.                                

\medskip\par Tol\,1247$-$232
\par Near the \OVI\ line the continuum placement for this source is slightly
problematic. Over the 1010--1050\aa\ range the continuum is on average rather
flat. However, on the short wavelength side of the \OVI\ line the level is
higher than on the long wavelength side. On larger scales, the continuum is
closer to that on the positive-velocity side, so the high continuum near
$-$300\kms\ was ignored in the fitting. This may lead to an underestimate of
W(\OVI).
\par The high-velocity \OVI\ component was separated from the thick disk
absorption at a velocity of 100\kms, even though there is no clear boundary
visible in the spectrum. The separation velocity is based on other sightlines in
this region of the sky that have high-velocity absorption (the closest are
ESO\,572$-$G34, 13\deg\ away, and Mrk\,1383, 35\deg\ away).                                 

\medskip\par Tol\,1924$-$416
\par The LiF1A and LiF2B spectra look different, although this may mostly be due
to low S/N ratio in the LiF2B channel. The main difference is that the H$_2$
$J$=3 line seems absent in LiF1A, but strong in LiF2B.
\par There does not appear to be high-velocity \OVI\ in this direction, even
though such absorption is seen toward all other sightlines within about 30\deg\
(ESO\,141$-$G55, PKS\,2005$-$489 and Mrk\,509).                                
\par Intrinsic \SiIIla\ absorption at $v$=2825\kms\ can be seen at 1030.322\aa\
($-$465\kms\ on the \OVI\ velocity scale)


\medskip\par Ton\,1187
\par The LiF1A and LiF2B spectra are both noisy and show some differences, but
these are still within the noise.
\par This is the only sightline in which HVC complex~M is clearly visible in
\HI\ emission (at \vlsr=$-$104\kms). There does not appear to be \OVI\
absorption at this velocity, however.


\medskip\par Ton\,S180
\par The LiF2B channel has a much lower apparent flux than LiF1A, most likely
because it was misaligned in this early observation. Only LiF1A was used.
\par The intrinsic Ly$\gamma$ absorption line at $v$=18580\kms\ is expected to
lie at 1032.814\aa\ (260\kms\ on the \OVI\ velocity scale). However, there is
no matching Ly$\beta$ absorption, although intrinsic \OVI\ at $v$=18700\kms\ is
clearly present.
\par This sightline passes 110\kpc\ from NGC\,247 ($v$=190\kms, distance 2.1
Mpc), 145\kpc\ from UGCA\,15 ($v$=331\kms) and 200\kpc\ from NGC\,253
($v$=260\kms), all of which lie in the \ComSclCld. The Ly$\beta$ line shows a
positive-velocity wing out to $\sim$200\kms.
\par Considering the points above, the 48$\pm$12\maa\ feature at 1032.787\aa\
(250\kms\ on the \OVI\ velocity scale) appears to be high-velocity \OVI\ rather
than an intrinsic absorption line or intergalactic Ly$\beta$. A weak (2$\sigma$)
matching \OVIlb\ line appears present, with equivalent width 22$\pm$13\maa
(i.e., the strength is about that expected from the possible \OVIla\ feature).
No other sightlines in this part of the sky show high-positive velocity \OVI.
\par \CII\ absorption is seen at a velocity of $-$130\kms, with W=52$\pm$18\maa,
or N(\CII)$\sim$6\tdex{13}\cmm2. Assuming carbon is not depleted on dust, this
corresponds to N(\HI)=0.2\tdex{17}/Z\cmm2, with Z the metallicity in solar
units. The Green Bank spectrum sets a 3$\sigma$ upper limit of
$\sim$2\tdex{18}\cmm2\ on N(\HI).

\medskip\par Ton\,S210
\par The LiF2B channel of the first observation was misaligned and shows no
flux.
\par This sightline passes within 15 arcmin of (or even through) a HVC which has
v=$-$190\kms. Sembach et al.\ (2002a) discuss this sightline. \CIIl\ absorption
is detected at $-$170\kms, and is probably associated with the outer envelope of
the HVC. The high-negative velocity \OVI\ component centered at $-$185\kms\ may
trace an outer ionized envelope of this HVC, or it may be part of a more general
distribution of high-velocity \OVI\ seen in this part of the sky (it is possible
that both the high-velocity \HI\ and \OVI\ originate in the same manner).
\par Both the \OVIla\ and the \OVIlb\ lines can be measured. In the velocity
range $-$90 to 95\kms\ the ratio N(1037)/N(1031) is 0.98$\pm$0.10.
\par \CII\ absorption is seen at a velocity of $-$175\kms, with
W=230$\pm$13\maa. The Effelsberg spectrum sets an upper limit of
3\tdex{18}\cmm2\ for \HI\ at this velocity. For gas with Magellanic abundances
this is compatible with the \CII\ absorption, if the $b$-value is large.


\medskip\par UGC\,12163 (Akn\,564)
\par During the 61.6~ks observation, the source was misaligned on detector side
1 for the first 8 of the 32 exposures, effectively reducing the exposure time to
40~ks.
\par The Galactic \OVI\ is very weak in this direction. In the velocity range
between $-$100 and 100\kms\ the absorption integrates to 107$\pm$32\maa,
corresponding to a 3.3$\sigma$ detection. A high-negative velocity \OVI\
component is also present, whose positive-velocity edge is contaminated by H$_2$
$J$=3.
\par The absorption features at 1034.990 and 1035.677\aa\ (890 and 1090\kms\ on
the \OVI\ velocity scale) are most likely interstellar \CIIl\ at $-$390 and
$-$190\kms. The Ly$\beta$ and Ly$\gamma$ lines show extended negative-velocity
wings, while there is a separate feature in Ly$\delta$. This gas is probably
associated with the Magellanic Stream. In \HI\ emission the tip of the Stream is
about 10\deg\ away, and points in the direction of UGC\,12163. The equivalent
widths of the \CII\ features are 160$\pm$34 and 98$\pm$27\maa, corresponding to
N(\CII) $\sim$\dex{14}\cmm2. This is compatible with the upper limit for N(\HI)
of $\sim$5\tdex{18}\cmm2.
\par The sightline passes through the \PegCld, but no Ly$\beta$ is visible.

\medskip\par VII\,Zw\,118
\par Although the continuum seems well defined, the comparison between the 1031
and 1037 N$_a$(v) profiles indicates either a problem or the presence of
saturation. However, the H$_2$ column density is large enough that the H$_2$
$J$=1 \l1037.146 line has damping wings that may shift the continuum down near
the \OVIlb\ line.
\par The feature at 1034.104\aa\ (630\kms\ on the \OVI\ velocity scale) is
intergalactic Ly$\beta$ at $v$=2450\kms. The corresponding Ly$\alpha$ line can
be seen in the \STIS\ spectrum. However, no galaxy groups are intersected by
this sightline. The nearest group ($>$3\deg\ or 1.5~Mpc impact parameter away)
is the \LynCld. The nearest galaxies in this group have $v$=1900\kms, much less
than the 2450\kms\ of the intergalactic absorption. MS\,0700.7+6338 lies 1\deg\
closer to the Lynx Galaxy Grouping, but no Ly$\beta$ is seen in its spectrum.

\medskip\par vZ\,1128
\par This is the sightline with the second highest S/N ratio (29 per resolution
element).
\par vZ\,1128 is a star in the globular cluster M\,3, which lies at a distance
of 9.7~kpc ($z$=9.5~kpc).
\par The alignment of this spectrum relative to the \HI\ is slightly
problematic. Howk et al.\ (2002) use v2.0.5 of the pipeline and argue that the
shape of the \ArI\ and \OI\ lines fits the \HI\ quite well. This implies that
the \SiII\ line is shifted, and that there is \CII\ absorption at velocities up
to 30\kms\ more negative than the \HI. We use the same alignment as Howk et al.\
(2002), but this shifts the spectrum by $-$20\kms\ relative to a determination
that would be based on just the shape of the \ArI\ and \SiII\ absorption as seen
in the v1.8.7 version of the spectrum. This sight line and that for 3C\,273.0
illustrate the substantial problems associated with determining the velocity
offsets in \FUSE\ spectra.                                  
\par Both the \OVIla\ and the \OVIlb\ lines can be measured. In the velocity
range $-$130 to 80\kms\ the ratio N(1037)/N(1031) is 0.97$\pm$0.05.



\newpage
\def\fgnumber#1{\bigskip\bigskip Figure #1. }

\fgnumber{\Fovishift.} Four examples showing the shift required to align the
apparent column density profiles of the \OVIla\ and \OVIlb\ lines. For each of
the four objects, the two left panels shows the $N_a(v)$ profiles using the
nominal wavelength calibration, the two right panels show them after shifting
\OVIlb\ by 10\kms. The top panel give the \OVIla\ line is shown as a thick line,
the \OVIlb\ line as a thin line, while the bottom panel present the ratio of the
two profiles, slightly smoothed by a gaussian with a FWHM of 2 pixels. Over the
uncontaminated velocity range defined by the heavy vertical lines on the
spectra, the ratio clearly shows an unphysical slope before applying the
velocity correction.

\fgnumber{\Fcompare.} Comparison of the flux observed with \FUSE\ to the flux
expected from an extrapolation of the \IUE\ spectrum to 1030\aa. Star symbols
are for QSOs, closed circles for Seyfert galaxies and open circles for other
kinds of galaxies. The horizontal and vertical lines in the top panel are drawn
at a level of 2\tdex{-14}\fu; the diagonal line shows the 1--1 relationship.

\fgnumber{\Fmagflux.} Comparison of the visual and blue magnitudes from the
V\'eron-Cetty \& V\'eron catalogue (2000) with the flux observed near 1030\aa,
separately for QSOs, Seyferts and galaxies. The symbols indicate the reddening,
E(B$-$V), which is almost always $<$0.07 mag. A reddening of 0.04 mag
corresponds to a reduction in flux of a factor $\sim$1.6, a shift of 0.2 in the
log. The correlation coefficient, $\rho$, is indicated in the top right corner
of each panel. The horizontal lines indicate a flux of 5\tdex{-15}\fu, below
which the measured fluxes are unreliable.

\fgnumber{\Fvlim.} Histograms of the number of objects for which the \OVI\
absorption extends out to a velocity $v$$_{\rm min}$ or $v$$_{\rm max}$. Thin
lines include all sightlines, thick lines exclude 26 sightlines where the
separation of Milky Way and high-velocity absorption is difficult and eleven
upper limits. Top panel: velocity limits for the Milky Way component. Middle
panel: most-negative velocity edge for the HVC component. Bottom panel:
most-positive velocity edge for the HVC component.

\fgnumber{\Ferrhist.} Histograms of the equivalent widths and their errors. The
top set of panels is for Milky Way \OVI\ absorption (mostly between
$\pm$120\kms). The bottom set of panels is for high-velocity \OVI. The quality
factor Q for a row of histograms is indicated at the right hand edge. The four
columns of four panels give, from left to right, the equivalent width, the error
associated with random noise (\sig{noise}), the error associated with fitting
the continuum (\sig{cfit}) and the error associated with choosing an integration
range (\sig{vlim}).

\fgnumber{\Ferrcorr.} Correlations between the random-noise error (\sig{noise}),
the continuum-fit error (\sig{cfit}), the error in the continuum fit calculated
by shifting the continuum up or down by \sig{noise}/3 (\sig{rms/3}) and the
velocity-limits error (\sig{vlim}). Plus signs indicate local continuum fits
with polynomial order 1, filled circles are for local continuum fits with order
2; open circles are for the 15 fits with higher orders. The solid lines show
least-squares fits, separately for the fits with order 1 and order 2.

\fgnumber{\Foviratio.} The upper panel shows the ratio (and its error) for the
column densities derived from the two \OVI\ lines, in the velocity range where
the \OVIlb\ absorption can be reliably measured. The error includes the effect
of random noise and continuum placement uncertainty, but not the contributions
to the systematic error. The vertical axis gives the rank of the object after
sorting on ratio. In the lower panel the ratios are plotted against the column
density derived from the \OVIla\ line.

\fgnumber{\Fvelhist.} Top panels (a, h); scatter plots of \OVI\ column density
against average component velocity, with symbol sizes proportional to the
$b$-value. Panels b to g (left) show the number of high-velocity \OVI\ absorbers
in 25\kms\ intervals, for a series of equivalent width ranges. Panels i to m
(right) give the distribution of the low-velocity Galactic \OVI\ absorbers, in
4\kms\ intervals.

\fgnumber{\Flsrmap a.} \OVI\ column density for 10 sightlines (and 23
significant non-detections), integrated from \vlsr=$-$500 to $-$300\kms, in an
aitoff projection of galactic coordinates, with the galactic Anti-Center in the
middle. The dots indicate all sightlines toward which a measurement was
obtained, surrounded by a 12\deg\ radius colored area. Stronger mottling
indicates sightlines with lower S/N ratios. Upper limits are shown by a colored
area 2\fdg5 in diameter. See Sect.~\Smaps\ for more details.

{\fgnumber{\Flsrmap b.} \OVI\ column density for 18 sightlines (and 20 significant non-detections), integrated from \vlsr=$-$300 to \vlsr=$-$200\kms.}

{\fgnumber{\Flsrmap c.} \OVI\ column density for 20 sightlines (and 17 significant non-detections), integrated from \vlsr=$-$200 to \vlsr=$-$150\kms.}

{\fgnumber{\Flsrmap d.} \OVI\ column density for 32 sightlines (and 12 significant non-detections), integrated from \vlsr=$-$150 to \vlsr=$-$100\kms.}

{\fgnumber{\Flsrmap e.} \OVI\ column density for 56 sightlines (and  4 significant non-detections), integrated from \vlsr=$-$100 to \vlsr=$-$50\kms.}

{\fgnumber{\Flsrmap f.} \OVI\ column density for 79 sightlines,                                     integrated from \vlsr=$-$50 to \vlsr=$-$0\kms.}

{\fgnumber{\Flsrmap g.} \OVI\ column density for 83 sightlines,                                     integrated from \vlsr=0     to \vlsr=50\kms.}

{\fgnumber{\Flsrmap h.} \OVI\ column density for 54 sightlines (and  4 significant non-detections), integrated from \vlsr=50    to \vlsr=100\kms.}

{\fgnumber{\Flsrmap i.} \OVI\ column density for 26 sightlines (and 14 significant non-detections), integrated from \vlsr=100   to \vlsr=150\kms.}

{\fgnumber{\Flsrmap j.} \OVI\ column density for 16 sightlines (and 17 significant non-detections), integrated from \vlsr=150   to \vlsr=200\kms.}

{\fgnumber{\Flsrmap k.} \OVI\ column density for 13 sightlines (and 20 significant non-detections), integrated from \vlsr=200   to \vlsr=300\kms.}

{\fgnumber{\Flsrmap l.} \OVI\ column density for  3 sightlines (and 25 significant non-detections), integrated from \vlsr=300   to \vlsr=500\kms.}

\fgnumber{\FHVCmap.} Map of the deviation velocities of the \HI\ high-velocity
gas. Data from Hulsbosch \& Wakker (1988) and Morras et al.\ (2000). The
deviation velocity is the difference between the observed LSR velocity and the
maximum velocity that can be easily understood in a simple model of Galactic
differential rotation (v$_{\rm rot,}\odot$=220\,\kms, R$_\odot$=8.5\kpc, R$_{\rm
MW}$=26\kpc, z$_{\rm ISM}$=2\kpc\ at R$_\odot$, increasing to 6\kpc\ R$_{\rm
MW}$, see Wakker 1991). The names of the major complexes are indicated. Contour
levels are at 0.05, 0.5 and 1~K brightness temperature, or $\sim$2, 20 and
40\tdex{18}\cmm2.

\end{document}